\documentclass[]{aa}

\usepackage{graphicx}
\usepackage{txfonts}
\usepackage{natbib}
\usepackage{url}
\usepackage{multirow,bigdelim}
\usepackage{rotating}
\usepackage{longtable}
\usepackage[singlelinecheck=off]{caption}

\begin{document}

\newcommand{\zabs}{\ensuremath{z_{\rm abs}}}
\newcommand{\zem}{\ensuremath{z_{\rm em}}}
\newcommand{\zqso}{\ensuremath{z_{\rm QSO}}}
\newcommand{\zgal}{\ensuremath{z_{\rm gal}}}
\newcommand{\HH}{\mbox{H$_2$}}
\newcommand{\HD}{\mbox{HD}}
\newcommand{\DD}{\mbox{D$_2$}}
\newcommand{\CO}{\mbox{CO}}
\newcommand{\dla}{damped Lyman\,$\alpha$}
\newcommand{\Dla}{damped Lyman\,$\alpha$}
\newcommand{\lya}{\ensuremath{{\rm Ly}\,\alpha}}
\newcommand{\lyb}{Ly\,$\beta$}
\newcommand{\Ha}{H\,$\alpha$}
\newcommand{\Hb}{H\,$\beta$}
\newcommand{\lyg}{Ly\,$\gamma$}
\newcommand{\ArI}{\ion{Ar}{i}}
\newcommand{\CaII}{\ion{Ca}{ii}}
\newcommand{\CI}{\ion{C}{i}}
\newcommand{\CII}{\ion{C}{ii}}
\newcommand{\CIV}{\ion{C}{iv}}
\newcommand{\ClI}{\ion{Cl}{i}}
\newcommand{\ClII}{\ion{Cl}{ii}}
\newcommand{\CrII}{\ion{Cr}{ii}}
\newcommand{\CuII}{\ion{Cu}{ii}}
\newcommand{\DI}{\ion{D}{i}}
\newcommand{\FeI}{\ion{Fe}{i}}
\newcommand{\FeII}{\ion{Fe}{ii}}
\newcommand{\HI}{\ion{H}{i}}
\newcommand{\MgI}{\ion{Mg}{i}}
\newcommand{\MgII}{\ion{Mg}{ii}}
\newcommand{\MnII}{\ion{Mn}{ii}}
\newcommand{\NI}{\ion{N}{i}}
\newcommand{\NII}{\ion{N}{ii}}
\newcommand{\NV}{\ion{N}{v}}
\newcommand{\NiII}{\ion{Ni}{ii}}
\newcommand{\OI}{\ion{O}{i}}
\newcommand{\OII}{\ion{O}{ii}}
\newcommand{\OIII}{\ion{O}{iii}}
\newcommand{\OVI}{\ion{O}{vi}}
\newcommand{\PII}{\ion{P}{ii}}
\newcommand{\PbII}{\ion{Pb}{ii}}
\newcommand{\SI}{\ion{S}{i}}
\newcommand{\SII}{\ion{S}{ii}}
\newcommand{\SiII}{\ion{Si}{ii}}
\newcommand{\SiIV}{\ion{Si}{iv}}
\newcommand{\TiII}{\ion{Ti}{ii}}
\newcommand{\ZnII}{\ion{Zn}{ii}}
\newcommand{\AlII}{\ion{Al}{ii}}
\newcommand{\AlIII}{\ion{Al}{iii}}
\newcommand{\Ho}{\mbox{$H_0$}}
\newcommand{\angstrom}{\mbox{{\rm \AA}}}
\newcommand{\abs}[1]{\left| #1 \right|} 
\newcommand{\avg}[1]{\left< #1 \right>} 
\newcommand{\kms}{\ensuremath{{\rm km\,s^{-1}}}}
\newcommand{\cmsq}{\ensuremath{{\rm cm}^{-2}}}
\newcommand{\ergs}{\ensuremath{{\rm erg\,s^{-1}}}}
\newcommand{\ergsa}{\ensuremath{{\rm erg\,s^{-1}\,{\AA}^{-1}}}}
\newcommand{\ergscm}{\ensuremath{{\rm erg\,s^{-1}\,cm^{-2}}}}
\newcommand{\ergscma}{\ensuremath{{\rm erg\,s^{-1}\,cm^{-2}\,{\AA}^{-1}}}}
\newcommand{\msyr}{\ensuremath{{\rm M_{\rm \odot}\,yr^{-1}}}}
\newcommand{\nhi}{n_{\rm HI}}
\newcommand{\fhi}{\ensuremath{f_{\rm HI}(N,\chi)}}
\newcommand{\refs}{{\bf (refs!)}}
\newcommand{\jonze}{J1135$-$0010}

\newcommand{\iap}{Institut d'Astrophysique de Paris, CNRS-UPMC, UMR7095, 98bis bd Arago, 75014 Paris, France\label{iap}}
\newcommand{\arizona}{Steward Observatory, University of Arizona, Tucson, AZ 85721, USA\label{arizona}}
\newcommand{\portsmouth}{Institute of Cosmology and Gravitation, University of Portsmouth, UK \label{portsmouth}}
\newcommand{\uchile}{Departamento de Astronom\'ia, Universidad de Chile, Casilla 36-D, Santiago, Chile\label{uchile}}
\newcommand{\chicagoa}{Department of Astronomy and Astrophysics, University of 
Chicago, 5640 South Ellis Avenue, Chicago, IL 60637, USA \label{chicagoa}}
\newcommand{\chicagob}{Enrico Fermi Institute, University of Chicago, 5640 South
 Ellis Avenue, Chicago, IL 60637, USA \label{chicagob}}
\newcommand{\florida}{\mbox{Astronomy Department, University of Florida, 211 Bryant Space Science Center, PO\,Box\,112055, 
Gainesville, FL\,32611-2055, USA}\label{florida}}

\hyphenation{ESDLA}
\hyphenation{ESDLAs}

   \title{A connection between extremely strong Damped Lyman-$\alpha$ Systems and Lyman-$\alpha$ Emitting Galaxies 
         at small impact parameters}
   \titlerunning{A connection between extremely strong DLAs and LAEs}

   \author{
        P.~Noterdaeme       \inst{\ref{iap}}
\and    P.~Petitjean        \inst{\ref{iap}}
\and    I.~P\^aris          \inst{\ref{uchile}}
\and    Z.~Cai              \inst{\ref{arizona}}
\and    H.~Finley           \inst{\ref{iap}}
\and    J.~Ge               \inst{\ref{florida}}
\and    M.~M.~Pieri         \inst{\ref{portsmouth}}
\and    D.~G.~York          \inst{\ref{chicagoa}}
          }

   \institute{    
\iap\  -- \email{noterdaeme@iap.fr}
\and \uchile
\and \arizona
\and \florida
\and \portsmouth
\and \chicagoa
             }

   \date{}

 \abstract{
  We present a study of $\sim$100 high redshift ($z \sim $~2-4)  extremely strong damped Lyman\,$\alpha$ systems 
(ESDLA, with $N(\HI)\ge 0.5 \times 10^{22}$~\cmsq) 
 detected in quasar spectra from the Baryon Oscillation Spectroscopic Survey (BOSS) of the Sloan Digital Sky Survey 
 (SDSS-III) Data Release 11. 
 We study the neutral hydrogen, metal, and dust content of this elusive population of absorbers and confirm our previous finding that 
 the high column density end of the $N(\HI)$ frequency distribution has a relatively shallow slope with
power-law index $-$3.6, similar to what is seen from 21-cm maps in nearby galaxies.
 The stacked absorption spectrum indicates a typical metallicity $\sim$1/20$^{\rm th}$ solar, 
similar to the mean metallicity of the overall DLA population.
The relatively small velocity extent of the low-ionisation 
lines suggests that ESDLAs do not arise from large-scale flows of neutral gas. 
The high column densities involved are in turn more similar to what is seen in DLAs associated 
with gamma-ray burst afterglows (GRB-DLAs), which are known to occur close to star forming regions. 
This indicates that ESDLAs arise from lines of sight passing at very small impact parameters from the 
host galaxy, as observed in nearby galaxies. This is also supported by simple theoretical considerations and 
recent high-$z$ hydrodynamical simulations.
 We strongly substantiate this picture by the first statistical detection of \lya\ emission with 
 $\avg{L_{\rm ESDLA}(\lya)} \simeq (0.6\pm0.2) \times 10^{42}~\ergs$ in the core of ESDLAs (corresponding 
 to about 0.1\,$L^{\star}$ at $z\sim 2-3$), obtained through 
 stacking the fibre spectra (of radius  1\,$\arcsec$ corresponding to $\sim$\,8\,kpc at $z \sim 2.5$).
 Statistical error on the \lya\ luminosity are of the order of $0.1 \times 10^{42}~\ergs$ but we 
caution that the measured \lya\ luminosity may be overestimated by $\sim 35\%$ 
due to sky light residuals and/or FUV emission from the quasar host and that we have neglected 
flux-calibration uncertainties. We estimate 
a more conservative uncertainty of $0.2 \times 10^{42}~\ergs$.

 The properties of the \lya\ line (luminosity distribution, velocity width and velocity offset compared to systemic redshift) are 
very similar to that of the population of Lyman-$\alpha$ emitting galaxies (LAEs) with $L_{\rm LAE}(\lya) \ge 10^{41} \ergs$ detected 
in long-slit spectroscopy or narrow-band imaging surveys. 
By matching the incidence of ESDLAs with that of the LAEs population, we estimate the high column density gas radius 
to be about $r_{\rm gas}=2.5$~kpc, i.e., significantly smaller than that corresponding to the BOSS fibre aperture, 
and making fibre losses likely negligible. Finally, the average measured Ly\,$\alpha$ luminosity indicates a star-formation rate 
consistent with the 
Schmidt-Kennicutt law, SFR (\msyr)~$\approx 0.6/f_{\rm esc}$, where $f_{\rm esc} < 1$ is the \lya\ escape 
fraction. 
Assuming the typical escape fraction of LAEs, $f_{\rm esc} \sim 0.3$, the Schmidt-Kennicutt law implies a galaxy radius of 
about $r_{\rm gal} \approx 2.5$~kpc. Finally, we note that possible overestimation of the \lya\ emission would result 
in both smaller $r_{gas}$ and $r_{gal}$.
 Our results support a close association between LAEs and strong DLA host galaxies.
}
 
  \keywords{Quasars: absorption-lines - Galaxies: high-redshift, ISM, star formation}
   \maketitle
%

\section{Introduction}

In the past two decades, astronomers have found several efficient observational 
strategies to detect and study galaxies in the early Universe. Each strategy targets 
a subset of the overall population of galaxies, which is then named after the selection 
technique.
Lyman-break galaxies \citep[LBGs,][]{Steidel96} are selected in broad-band imaging using 
colour cuts around the Lyman-limit at 912~{\AA}. Because of this selection, LBGs probe 
mostly bright massive galaxies with strong stellar continuum \citep[e.g.][]{Steidel03, 
Shapley03, Shapley11}. 
Since hydrogen recombination following ionisation by young stars produces \lya\ emission, 
this line can also be used to detect star-forming galaxies at high-redshift, where it is conveniently 
redshifted in the optical domain. 
\lya\ emitting galaxies (more generally called Lyman-$\alpha$ emitters: LAEs, \citealt[][]{Cowie98, Hu98}) 
are detected using narrow-band filters tuned to the wavelength of \lya\ 
\citep[e.g.][]{Rhoads00, Ouchi08, Ciardullo12}, long-slit spectroscopy \citep{Rauch08,Cassata11} or 
integral field spectroscopy \citep[e.g.][]{Petitjean96,Adams11}.
Because their selection is independent of 
the stellar continuum, these galaxies are often faint in broad-band imaging and likely represent 
low-mass systems with little dust attenuation \citep{Gawiser07}. 
Several studies have attempted to relate these two populations 
in a single picture by studying how the \lya\ emission line properties are related to 
the galaxy stellar populations \citep[e.g.][]{Lai08,Kornei10}.
Additionally, infrared observations together with detections of molecular emission have opened a new 
and very promising way to study galaxies at high redshift \citep[e.g.][]{Omont96, Daddi09}.

Another and very different technique to detect high-redshift galaxies is based on the absorption 
they imprint on the spectra of bright background sources, such as quasi-stellar objects (QSOs) or gamma ray burst (GRB) 
afterglows. These detections depend only on the gas cross-section and are thus independent 
of the luminosity of the associated object.
Large surveys have demonstrated that Damped Lyman-$\alpha$ systems (DLAs, see \citealt{Wolfe05}), characterised 
by $N(\HI) \ge 2\times 10^{20}$~cm$^{-2}$, contain $\ge$80\% of the neutral gas immediately 
available for star formation \citep{Peroux03, Prochaska05, Prochaska09, Noterdaeme09dla, Noterdaeme12c,Zafar13}.
  
Constraints on the star-formation activity associated with DLAs can be obtained by measuring the metal abundances in the gas 
\citep[e.g.][]{Prochaska03, Petitjean08} and their evolution with cosmic time 
\citep[e.g.][]{Rafelski12}. The excitation of different atomic and/or molecular species provides indirect constraints on 
instantaneous surface star-formation rates \citep{Wolfe03, Srianand05, Noterdaeme07lf, Noterdaeme07}.
\citet{Prochaska97,Prochaska98} tested a variety of models and concluded that the DLA kinematics, 
as traced by the profiles of low-ionisation metal absorption lines, could be characteristic of rapidly 
rotating discs. This interpretation is, however, problematic in the cold dark matter models that predict 
low rotation speeds \citep{Kauffmann96}. Alternatively, \citet{Ledoux98} showed that merging protogalactic 
clumps can explain the observed profiles, as expected in the now prevailing hierarchical models of galaxy 
formation \citep[see e.g.][]{Haehnelt98}. \citet{Schaye01} proposed that large scale outflows would also 
give rise to DLAs when seen in absorption against a background QSO and that the outflows would have 
sufficiently large cross-section to 
explain a significant fraction of DLAs. It has also been proposed that the fraction of neutral gas in cold 
streams of gas infalling onto massive galaxies is non-negligible at high redshift, where this is an important 
mode of galactic growth \citep[e.g.][]{Moller13}. This gas potentially gives rise to DLAs 
\citep{Fumagalli11} with moderate column densities.

Although the chemical and physical state of the gas in DLAs is relatively well understood, we still 
know little about the properties (mass, kinematics, stellar content) of the associated galaxy population. 
Since the total cross-section of DLAs is much larger than that of starlight-emitting regions in observed 
galaxies, a large fraction of DLAs potentially arises from atomic clouds in the halo or circumgalactic 
environments, as supported by high-resolution galaxy formation simulations \citep[e.g.][]{Pontzen08}.
Direct detection of galaxies associated with DLAs (hereafter ``DLA-galaxies'') is needed to address 
these issues. This has appeared to be a very difficult task, mainly due to the faintness of the 
associated galaxies and their unknown location (i.e. impact parameter) with respect to the quasar line of sight.
Thankfully, substantial progress has been made in the past few years, owing to improved selection strategies and 
efficient instrumentation on large telescopes \citep{Bouche12, Fynbo10, Fynbo11, Noterdaeme12a, Peroux11}. 
Although still rare, these observations show that it is possible to relate the properties of the gas to 
star formation activity in the host galaxy \citep[e.g.][]{Krogager12}. For example, large scale kinematics 
have recently been invoked to link the absorbing gas with star-forming regions 
located 10-20~kpc away \citep[e.g.][]{Bouche13,Fynbo13,Krogager13,Kashikawa13}. 

Here, we aim to study the link between star formation and the absorbing gas within or very close to 
the host galaxy. Our rationale is that this can be achieved by selecting DLAs with very high column densities of 
neutral hydrogen, which will be closely connected both spatially and physically to star forming regions in galaxies, since a Schmidt-Kennicutt law is expected to apply to quasar absorbers \citep[e.g.][]{Chelouche10}.
This idea is also supported by 21-cm maps of nearby galaxies \citep[e.g.][]{Zwaan05,Braun12} and existing 
observations of impact parameters for high-$z$ DLA galaxies that decrease with 
increasing column density \citep{Krogager12}, an effect which 
is also seen in simulations \citep[e.g.][]{Pontzen08,Yajima12,Altay13b}.

Until recently, very high column density DLAs, with $\log N(\HI)\sim 22$, were very rare occurrences 
(\citealt{Guimaraes12}, \citealt{Noterdaeme12a}, see also \citealt{Kulkarni12}), but the steadily increasing 
number of quasar spectra obtained by the Sloan Digital Sky Survey \citep[SDSS,][]{York00} and more recently 
by the Baryon Oscillation Spectroscopy Survey \citep[BOSS,][]{Dawson13} component of SDSS-III \citep[][]{Eisenstein11} 
opens the possibility to study such a population.

We present our DLA sample in Sect.~\ref{sample} and its column density distribution in Sect.~\ref{fhi}. We then 
study the metal content of our DLA sample and compare it to the population of DLAs 
associated with GRB afterglows (Sect.~\ref{metals}). In Sect.~\ref{colours}, we analyse the colour 
distortions that DLAs induce 
on their background QSOs. The rest of the paper explores 
the \lya\ emission 
detected using stacking procedures and discusses the nature of DLA galaxies. Throughout the paper, we use 
standard $\Lambda$CDM cosmology with $\Ho=70$~\kms\,Mpc$^{-1}$, $\Omega_{\Lambda}=0.7$ and $\Omega_{\rm m}=0.3$.

\section{Sample \label{sample}}

DLAs were detected with a fully-automatic procedure based 
on profile recognition using correlation analysis \citep[see][]{Noterdaeme09dla}. 
In \citet{Noterdaeme12c}, we applied this technique to about 65\,000 quasar spectra  
from the SDSS-III BOSS Data Release 9 \citep{Paris12}. 
Here, we extend this search to nearly 140\,000 quasar spectra from Data Release 11 
(that includes DR9 \citep{Ahn12} and DR10 \citep{Ahn13}, 
to be released together with DR12 in December 2014. 
The reduction of data obtained with the BOSS spectrograph \citep{Smee13} mounted on the SDSS telescope 
\citep{Gunn06} is described in \citet{Bolton12}.
Quasar spectra 
featuring broad absorption lines were rejected from the sample after a systematic visual inspection \citep[see][]{Paris12,Paris13}.
The automatic detection procedure provides the redshift and neutral hydrogen column density for each DLA candidate. Although we focus on systems with 
$N(\HI) \ge 5 \times 10^{21}$~\cmsq, we carefully checked all candidates with column densities down to 
0.2~dex below this limit. We paid particular attention to the Lyman series and the low-ionisation metal 
lines in order to remove possible blends or misidentifications. 
Whenever we felt it necessary, we 
refined the redshift measurement based on the low-ionisation metal lines and refitted the DLA 
profile\footnote{We note 
that the data quality is not good enough to observationnally test the theoretical assymetry 
of the strong damped \lya\ and \lyb\ profiles recently calculated by \citet{Lee13}.}. 
In a few cases, data are of poor quality and there is possibility of mis-indentification 
or large uncertainty on the $N(\HI)$-measurement. This should affect a small fraction of 
our sample and would have no effect on any statistical result presented in the paper.

We then selected intervening DLAs with $\log N(\HI) \ge 21.7$ (hereafter called extremely 
strong DLAs or ``ESDLAs'') avoiding proximate DLAs with velocities less than 5000~\kms\ from the QSO 
as these may be physically associated with the quasar environment. Our sample then consists 
of 104 {\sl intervening} ESDLAs with absorption redshifts in the range $\zabs \simeq $~2 - 4.3 
(see Table~\ref{tab:esdla} and Fig.~\ref{fig:esdla}). 
We note that even at $\Delta{_v}<3000~\kms$, the properties of DLAs are generally 
consistent with an origin external to the QSO host and may simply sample over-dense environments 
\citep{Ellison10}. Our conservative higher velocity cut-off should ensure a very low probability
 for a given ESDLA 
to be physically associated with the QSO and increasing this cutoff to 10\,000~\kms\ would 
only reject a further four ESDLAs with almost no consequence on the number derived in the paper.
Strong proximate DLAs from BOSS are studied in \citet{Finley13}.

\section{$N(\HI)$-frequency distribution \label{fhi}}

\citet{Petitjean93} showed early that different physical processes shape the $N(\HI)$ frequency distribution. 
Above $\log N(\HI) \sim 20$, i.e. the regime probed by 
DLAs, the gas is neutral and the slope reflects the average projected distribution of the gas in and around 
high redshift galaxies \citep[e.g.][]{Prochaska09}. The very high end of this distribution is of particular 
interest since local processes including molecular hydrogen formation and UV radiation or outflows from 
star-formation activity will influence its shape (\citealt{Rahmati13}, \citealt{Altay13}). However, this 
regime was until recently poorly constrained due to the small cross-section of the high column density gas. 
The availability of the larger DR9 data-set of DLAs allowed to statistically probe the very high column density 
end \citep{Noterdaeme12c}, which we now extend using DR11 data.

\begin{figure}
\centering
\includegraphics[bb = 60 175 500 570,clip=,width=\hsize]{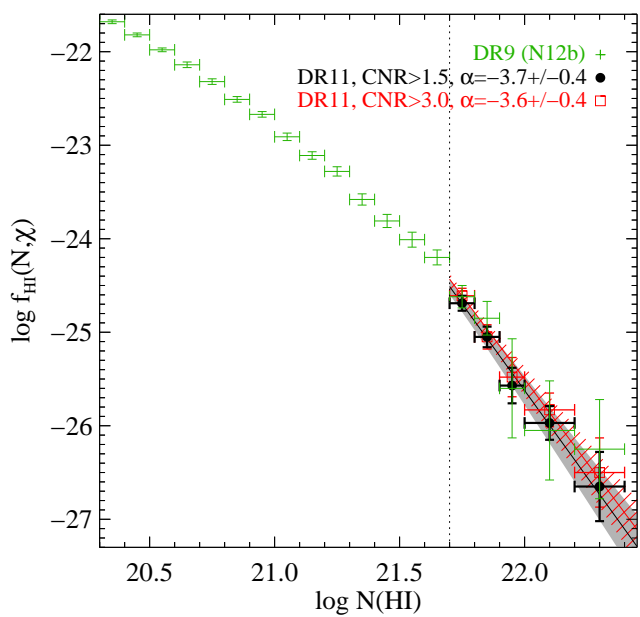} 
\caption{The DLA $N(\HI)$-distribution function. The high column density end of the distribution studied 
here is shown in black for the overall DR11 sample and a subset with higher continuum-to-noise ratio (red). 
Green points show the results from \citet{Noterdaeme12c} based on DR9. \label{fig:fhiall}}
\end{figure}

Fig.~\ref{fig:fhiall} presents the $N(\HI)$ frequency distribution derived from our sample. We 
confirm our previous result \citep{Noterdaeme12c} that the distribution extends to very high column 
densities ($\log N(\HI)=22.35$) with a moderate power-law slope, although steeper than what is 
seen at lower column 
densities  \citep[see][for a discussion on the shape of the high redshift $N(\HI)$-distribution 
over the range $10^{12}-10^{22}$~cm$^{-2}$]{Prochaska14}. 
Including all the lines of sight searched for ESDLAs (with continuum-to-noise ratio\footnote{The continuum-to-noise ratio is averaged over the \lya-forest, 5000~\kms\ 
redwards (resp. bluewards) of the \lyb\ (resp. \lya) emission line. See \citet{Noterdaeme09dla} and 
\citet{Noterdaeme12c} for more details.} CNR~$>$~1.5), 
we derive a power-law slope $\alpha = -3.7 \pm 0.4$. Restricting the sample to lines of sight with twice the 
minimum CNR value (i.e. CNR~$> 3$), we get $\alpha = -3.6 \pm 0.4$. There is a small 
systematic normalisation offset between the two distributions (resp. -24.51 and -24.46 at $\log N(\HI)=21.7$), 
but it remains within errors. 
This indicate little, if any, S/N-dependent bias.
If actually present, any bias would indicate slightly lower completeness at low data quality rather 
than poor identification because the normalisation increases slightly when considering CNR~$>3$. Indeed, 
restricting to CNR~$>5$ does not introduce significant change either in the slope ($\alpha = -3.6 \pm 0.4$) or in the normalisation (-24.43 at $\log N(\HI)=21.7$).

\begin{figure}
\includegraphics[bb=74 178 491 397,width=\hsize]{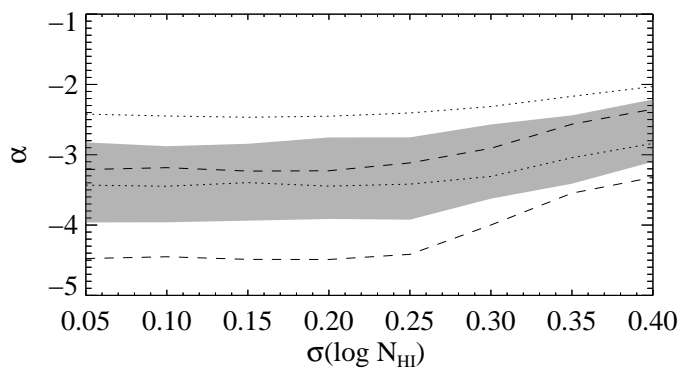} 
\caption{Measured power-law slope as a function of input slope and measurement uncertainty. 
The different regions (resp. dotted, grey area, dashed) shows the 1\,$\sigma$ range around the mean 
value for respectively $\alpha_i=$-3;-3.5 and -4.
\label{fig:slopeio}}
\end{figure}

We note that measurement uncertainties combined with limited sample size can in principle 
introduce a bias in the power-law slope measurement \citep{Koen09}. To test this, we 
simulate a DLA population with column density distributions of power-law slopes 
$\alpha_i=-3,-3.5,-4$, to which we added measurement uncertainties 
(1\,$\sigma$ level ranging from 0.05 to 0.40~dex by step 0.05~dex, see Fig.~\ref{fig:slopeio}).
We reconstruct the resulting ``observed'' distribution function for 104 ESDLAs randomly 
selected from this population and repeat this exercise 50 times for each input slope and 
measurement uncertainty.
We find that the slope tends to be shallower with increasing measurement uncertainties 
(see Fig.~\ref{fig:slopeio}). While this test remains simplistic, it supports our neglecting 
systematic errors as the 1\,$\sigma$ uncertainty on $\log N(\HI)$ measured in SDSS spectra is typically 
less than 0.3~dex (\citealt{Noterdaeme09dla}, see also Fig.~\ref{fig:esdla}), although it 
can be larger in a few cases.

While the large number of quasars discovered  by BOSS allows us to constrain for the first time the 
slope of the $N(\HI)$ frequency distribution at the very high column density end, the sample size 
remains too small to perform statistically sound studies on subsamples (e.g. as a function of the 
absorber's redshift). Nonetheless, 
no strong evolution in the slope of the distribution function is seen in the redshift range 
$z=2-4$. Finally, as noted by \citet{Noterdaeme12c}, the slope of the 
distribution is close to that derived in nearby galaxies from opacity-corrected 21-cm emission maps 
\citep{Braun12}.

\section{Absorption properties \label{metals}}

\subsection{Metal equivalent widths}

The equivalent width of metal absorption lines was obtained automatically for each system by locally normalising the 
quasar continuum around each line of interest and subsequently modelling the absorption lines with a Voigt profile. 
For non-detections, an upper-limit was derived from the noise around the expected line position, 
assuming an optically thin regime. Upper-limits were also set for absorptions clearly blended with a line 
from another system.
It is important to note here that most of the lines are saturated
even if the apparent optical depth is small because of the low spectral resolution
(R$\sim$2000). 
Although the EWs of absorption systems can provide some statistical indication of their metallicity 
and/or velocity widths, it is necessary to measure EWs from optically thin lines to accurately derive 
the metallicity. The S/N of SDSS spectra generally does not allow meaningful metallicity estimates 
for individual systems.

\begin{figure}
\centering
\includegraphics[bb = 60 84 580 780,clip=,width=\hsize]{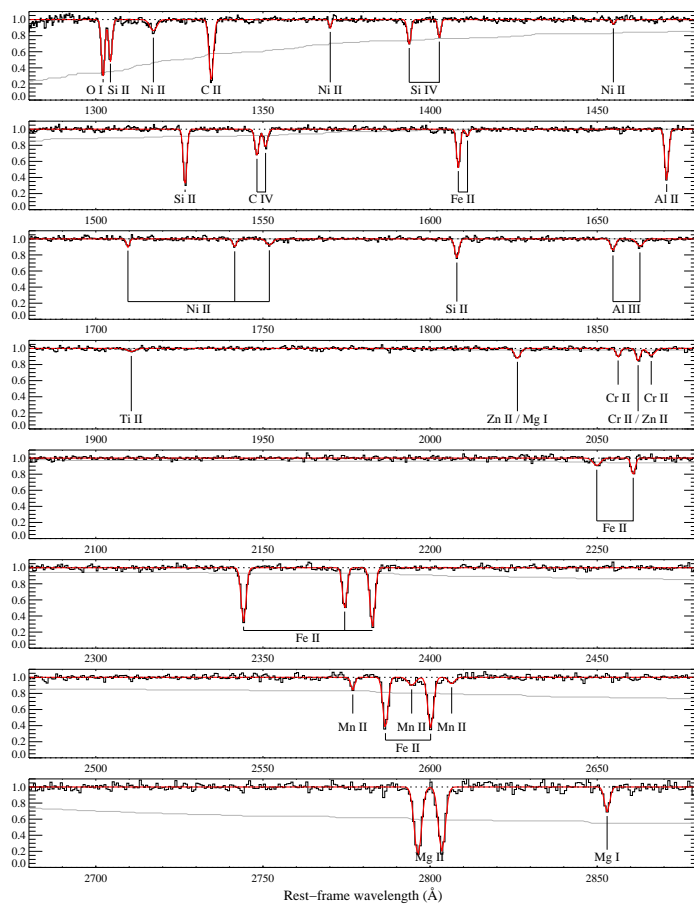}
\caption{Median ESDLA absorption spectrum redwards of the \lya\ emission. The grey line shows the 
fraction of the total sample contributing to the stacked spectrum at a given wavelength. 
\label{fig:absstack}}
\end{figure}

In order to derive the typical metal content of ESDLAs and avoid blending with \lya\ forest lines, we produced 
an average absorption spectrum by stacking the portion of the normalised spectra redwards the \lya\ forest. 
To do this, we first shifted the spectra to the DLA-rest frame and rebinned the data onto a common grid, keeping the 
same pixel size (constant in velocity space) as the original data. For each pixel $i$, the stacked spectrum is taken 
as the median of the normalised fluxes measured at $\lambda_i$. Note that we do not apply any weighting, which means that each DLA contributes equally to the stacked, as long as it covers the considered wavelength. 
Residual broad-band 
imperfections in the normalisation (resulting from combining imperfections in individual spectra in 
different wavelength ranges) were then corrected by re-normalising the stacked spectrum using a 
median-smoothing filter to get the pseudo-continuum. 
The resulting stacked spectrum reaches a S/N ratio per pixel of about 50 (see Fig.~\ref{fig:absstack}), 
which sets the typical 1\,$\sigma$ detection limit to about 0.02~{\AA} for an unresolved line at the 
BOSS spectral resolution and sampling.

Thirty-eight absorption features are detected between 1300 and 2900~{\AA} (rest-frame). These arise mostly 
from low-ionisation species (e.g. \FeII, \SiII, \ZnII, \CrII, \MgI) but also from high-ionisation species (\SiIV, \CIV). 
Weak lines such as \FeII$\lambda\lambda$2249,2260 are clearly detected even though they are well below the detection 
limit for any individual spectrum. \FeII$\lambda$1611, \NiII$\lambda$1454 and the \TiII$\lambda$1910 doublet are 
also possibly detected, but at less than the 3$\sigma$ level. 
We proceed to measure the average equivalent widths of the different absorption features through Gaussian fitting (see Table~\ref{tab:ew}). Interestingly, these are the same lines that are detected in a stack by 
\citet{Khare12} using the whole DLA catalogue by \citet{Noterdaeme09dla}, which reaches much higher 
signal-to-noise ratio. 
However, the equivalent widths of weak lines are found ten times higher here than in \citet{Khare12}, 
consistent with the ten times higher \HI\ column densities. This already indicates that the abundances 
in ESDLAs should not be significantly different from that of the overall DLA population.

\begin{figure}
\centering
\includegraphics[bb=74 178 491 397,width=\hsize]{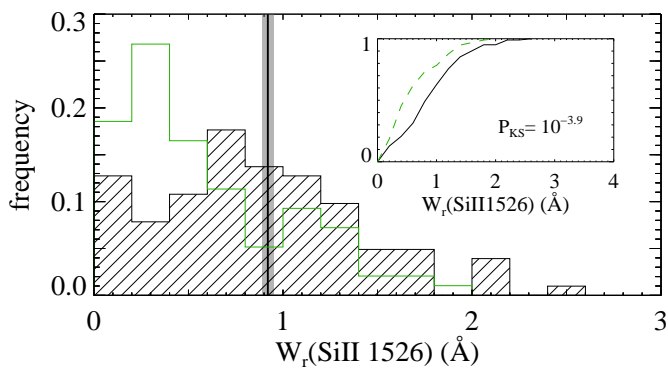}
\caption{Distributions of the \SiII\,$\lambda$1526 equivalent width measured in ESDLAs (hashed black histogram) and in a sample 
representative of the overall population of DLAs \citep[green][]{Jorgenson13}. The vertical line 
corresponds to the equivalent width (the 1\,$\sigma$ error is represented by the grey area) measured in the stacked 
ESDLA spectrum. The inset shows the corresponding cumulative distributions (solid black: ESDLAs, dashed green: 
\citealt{Jorgenson13}). \label{fig:si}}
\end{figure}

Similarly, in Fig.~\ref{fig:si}, we compare the distribution of \SiII$\lambda$1526 equivalent widths measured 
in individual ESDLAs to that of a sample drawn from SDSS without any criterion on $N(\HI)$ \citep{Jorgenson13}. 
There is a deficit of small EWs ($\la 0.6~{\AA}$) in ESDLAs compared to the overall DLA population, 
with an almost zero probability that this is due to chance coincidence. We note that this is not due to 
incompleteness at low EWs, as only one system has \SiII$\lambda$1526 undetected at 3\,$\sigma$ (a weak line 
in a low S/N spectrum). This suggests that ESDLAs do not reach drastically lower abundances than 
the rest of the DLA population.

\begin{table}
\caption{Equivalent widths of metal absorption lines \label{tab:ew}}
\centering
\begin{tabular}{c c c c}
\hline
\hline
{\large \strut} Transition                  &  W$_r$     &   $\log N$                 & $b_{eff}$            \\
                            &  ({\AA})    &    (\cmsq)                 & (\kms)            \\

\hline                                           
\OI\,$\lambda$1302           &  0.90 $\pm$ 0.05  &   -                            &  -                    \\
     \hline 
\CII\,$\lambda$1334          &  1.25 $\pm$ 0.04  &   -                            &  -                     \\
     \hline                                                  
\SiII\,$\lambda$1304         &  0.67 $\pm$ 0.05  &  \multirow{3}{*}{16.0 $\pm$ 0.1\tablefootmark{a}}   & \multirow{3}{*}{41 $\pm$ 2} \\
\SiII\,$\lambda$1526         &  0.92 $\pm$ 0.03  &   & \\
\SiII\,$\lambda$1808         &  0.37 $\pm$ 0.03  &   &     \\
     \hline                                                                                                                          
\FeII\,$\lambda$1608         &  0.63 $\pm$ 0.04  &  \multirow{9}{*}{15.5 $\pm$ 0.1\tablefootmark{a} }  & \multirow{9}{*}{39 $\pm$ 1} \\
\FeII\,$\lambda$1611         &  0.10 $\pm$ 0.04  &  & \\
\FeII\,$\lambda$2249         &  0.25 $\pm$ 0.04  &  & \\
\FeII\,$\lambda$2260         &  0.31 $\pm$ 0.05  &  & \\
\FeII\,$\lambda$2344         &  1.24 $\pm$ 0.05  &  & \\
\FeII\,$\lambda$2374         &  0.86 $\pm$ 0.05  &  & \\
\FeII\,$\lambda$2382         &  1.37 $\pm$ 0.05  &  & \\
\FeII\,$\lambda$2586         &  1.20 $\pm$ 0.07  &  & \\
\FeII\,$\lambda$2600         &  1.35 $\pm$ 0.07  &  & \\
     \hline                                                                                             
\NiII\,$\lambda$1317         &  0.36 $\pm$ 0.06  &  \multirow{6}{*}{14.2 $\pm$ 0.3\tablefootmark{b}}   & \multirow{6}{*}{-}          \\
\NiII\,$\lambda$1370         &  0.09 $\pm$ 0.02  &   & \\
\NiII\,$\lambda$1454         &  0.06 $\pm$ 0.03  &   & \\
\NiII\,$\lambda$1709         &  0.12 $\pm$ 0.04  &   & \\
\NiII\,$\lambda$1741         &  0.14 $\pm$ 0.04  &   & \\
\NiII\,$\lambda$1751         &  0.18 $\pm$ 0.04  &   & \\
     \hline                                                                                                                          
\MgI\,$\lambda$2852          &  0.61 $\pm$ 0.13  &  12.7 $\pm$ 0.1 \tablefootmark{b}& -\\
\hline                                                 
\CrII\,$\lambda$2056         &  0.16 $\pm$ 0.04  &  \multirow{2}{*}{13.8  $\pm$ 0.2\tablefootmark{b}} & \multirow{2}{*}{-}\\
\CrII\,$\lambda$2066         &  0.23 $\pm$ 0.06  & & \\
\hline                                                 
\ZnII+\MgI\,$\lambda$2026    &  0.26 $\pm$ 0.05  &    \multirow{2}{*}{13.1  $\pm$ 0.1\tablefootmark{c}} & \multirow{2}{*}{-} \\
\ZnII+\CrII\,$\lambda$2062   &  0.25 $\pm$ 0.04  &   & \\
   \hline                                                                                               
\MnII\,$\lambda$2576         &  0.23 $\pm$ 0.06  &  \multirow{3}{*}{13.2 $\pm$ 0.1\tablefootmark{b}}    & \multirow{3}{*}{-}\\
\MnII\,$\lambda$2594         &  0.27 $\pm$ 0.04  &  & \\
\MnII\,$\lambda$2606         &  0.18 $\pm$ 0.07  &  &  \\
\hline                                                 
\TiII\,$\lambda$1910         &  0.10 $\pm$ 0.03  &   13.2 $\pm$ 0.2 \tablefootmark{b} & - \\
          \hline                                                                                                                     
\MgII\,$\lambda$2796         &  2.30 $\pm$ 0.12  &   - & - \\
\MgII\,$\lambda$2803         &  2.10 $\pm$ 0.12  &   - & - \\
          \hline                                                                                                                                         
\AlII\,$\lambda$1670         &  0.89 $\pm$ 0.04  &   - &  - \\
          \hline                                                                                                                                         
\AlIII\,$\lambda$1854        &  0.25 $\pm$ 0.03  &  -  & - \\
\AlIII\,$\lambda$1862        &  0.23 $\pm$ 0.04  &  -  & - \\
          \hline                                                                                                  
\SiIV\,$\lambda$1393         &  0.42 $\pm$ 0.02  &   - & - \\
\SiIV\,$\lambda$1402         &  0.26 $\pm$ 0.03  &   - & - \\
      \hline                                                                                                                                             
\CIV\,$\lambda$1548          &  0.49 $\pm$ 0.03  &   - & - \\
\CIV\,$\lambda$1550          &  0.34 $\pm$ 0.04  &   - & - \\

      \hline                                                                                                                                             
\OVI\,$\lambda$1031\tablefootmark{d}          &  0.44 $\pm$ 0.22  &   - & - \\
\OVI\,$\lambda$1037\tablefootmark{d}          &  0.27 $\pm$ 0.15  &   - & - \\

\hline
\end{tabular}
\tablefoot{
\tablefoottext{a}{Derived using curve-of-growth analysis}
\tablefoottext{b}{Derived from equivalent width in the optically thin regime}
\tablefoottext{c}{Column density of \ZnII. See text for details.}
\tablefoottext{d}{Located in the \lya\ forest.}

}
\end{table}

\subsection{Abundances and depletion}

The column density of different species can then be obtained under the optically-thin assumption. This assumption 
is only valid for weak lines. We again caution that even very strong lines in the optically-thick regime 
appear non-saturated at the BOSS spectral resolution. We estimate column densities 
for \NiII, \MnII, \TiII, \MgI\ and \CrII. \ZnII\ has two transitions at 2026 and 2062~{\AA} that are unfortunately 
badly blended with \MgI$\lambda$2026 and \CrII$\lambda$2062 \citep[see][for a discussion about these lines blending 
at SDSS spectral resolution]{York06}. 

We use the \MgI$\lambda$2852 line to derive the \MgI\ column density and estimate its contribution 
to the  2026~{\AA} feature, which we find amounts to about 10\% of the total equivalent width. 
Removing this contribution, we then get $\log N(\ZnII)=13.1 \pm 0.1$. 
We note that if \MgI$\lambda$2852 exceeds the optically thin regime, its column density 
and hence the contribution of \MgI\ to the 2026~{\AA} feature could be underestimated. In any case,  
\ZnII$\lambda2026$ provides the dominant component in this absorption.
Repeating the same procedure with 
\CrII\ and the 2062~{\AA} feature, we measure $\log N(\ZnII)=12.7 \pm 0.6$. The error here is larger 
because 
the equivalent width of this absorption is mainly due to \CrII. 

\begin{figure}
\centering
\addtolength{\tabcolsep}{-5pt}
\begin{tabular}{cc}
\includegraphics[bb = 70 175 315 570,clip=,height=0.8\hsize]{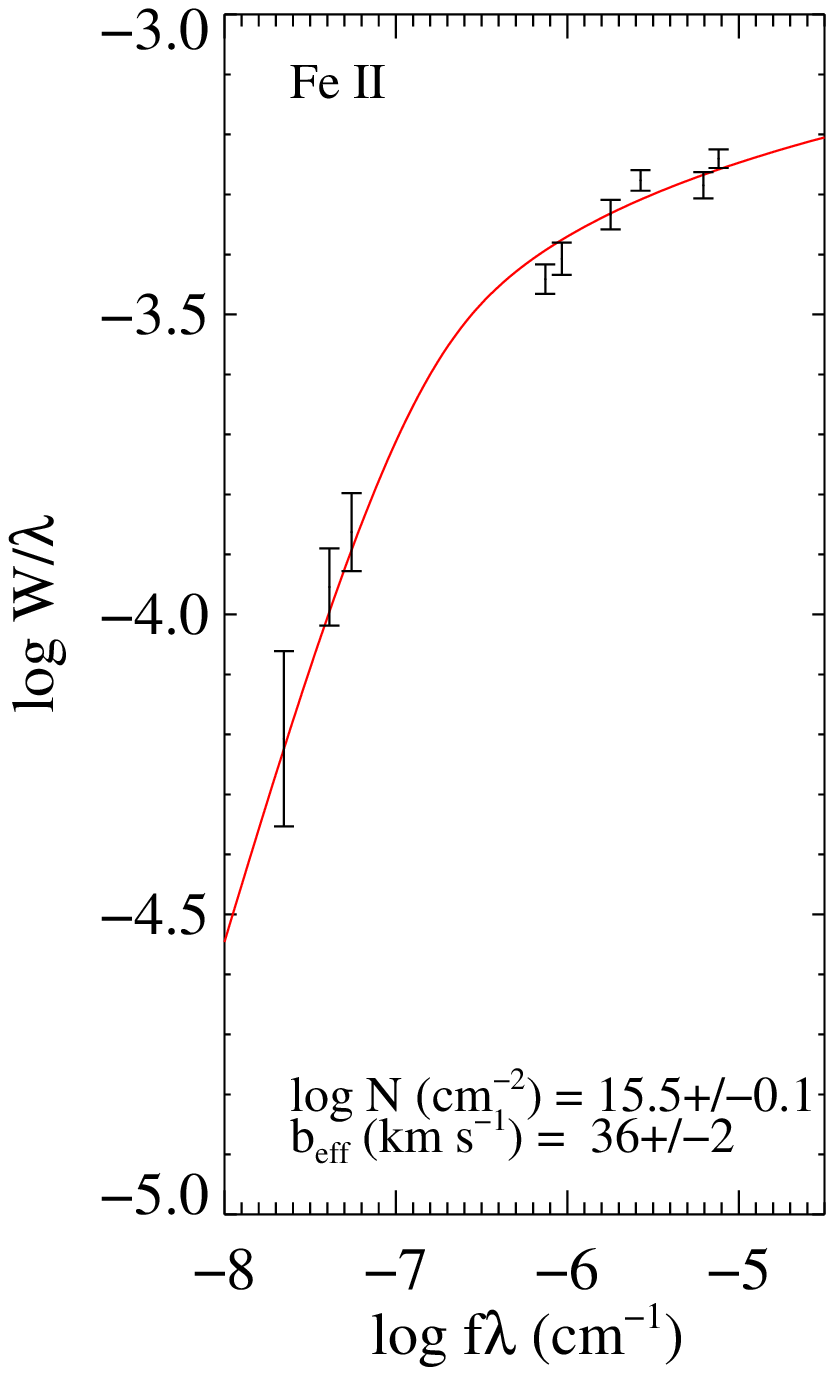} &
\includegraphics[bb = 130 175 315 570,clip=,height=0.8\hsize]{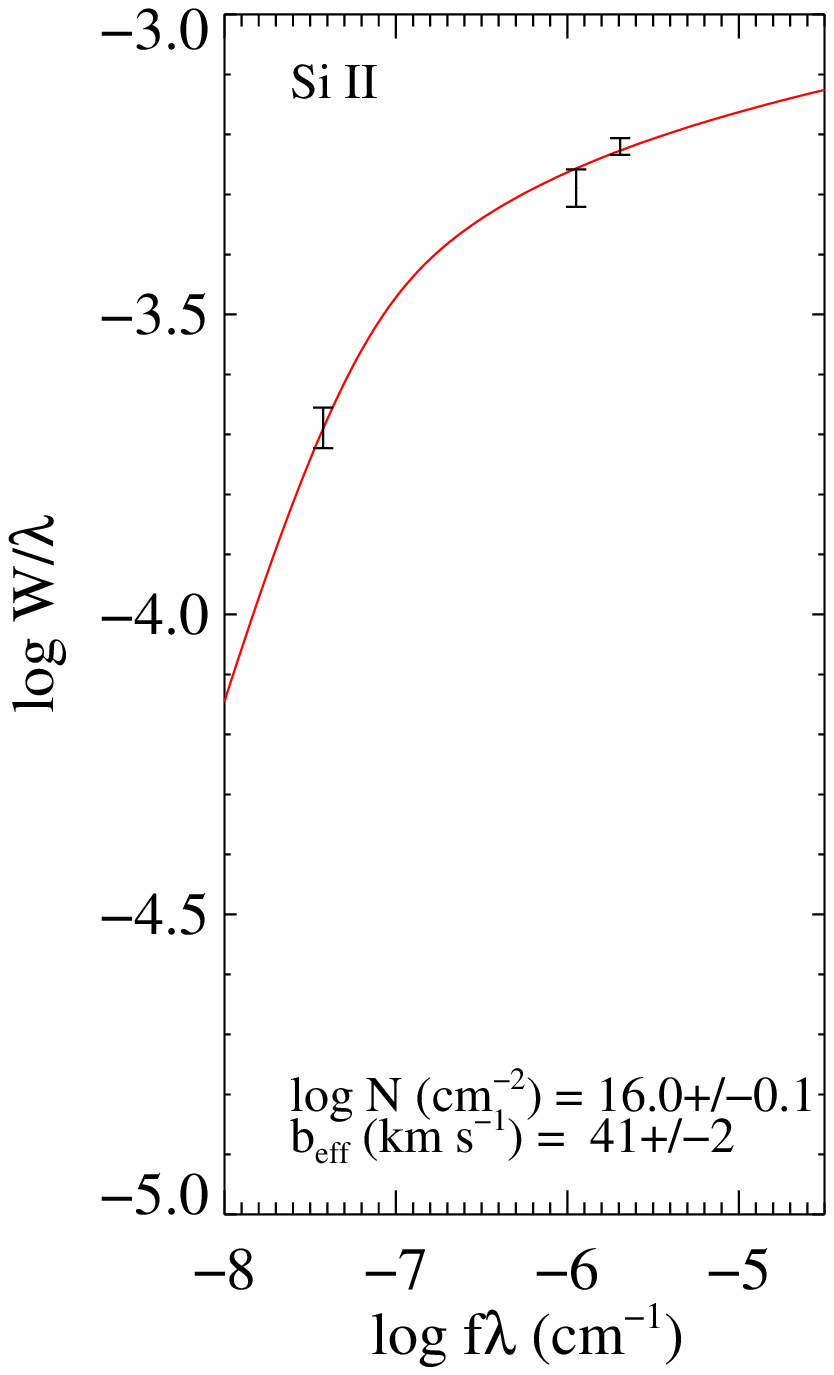} \\
\end{tabular}
\addtolength{\tabcolsep}{5pt}
\caption{Curve-of-growth for \FeII\ (left) and \SiII\ (right) in the stacked spectrum. 
\label{fig:cog}}
\end{figure}

Since several \FeII\ and \SiII\ lines spanning a wide range in oscillator 
strengths are available, it is possible to construct the curve-of-growth for these species and derive both the column density 
and the effective Doppler parameter (Fig.~\ref{fig:cog}). The column density is well constrained by weak 
lines but for stronger lines the 
equivalent widths 
reflect mostly the velocity extension of the profile 
\citep[e.g.][]{Nestor03,Ellison06}. We thus get the average $\log N(\SiII) = 16.0 \pm 0.1$ and $\log N(\FeII) = 15.5 \pm 0.1$.

\begin{figure}
\centering
\includegraphics[bb=74 178 491 397,width=\hsize]{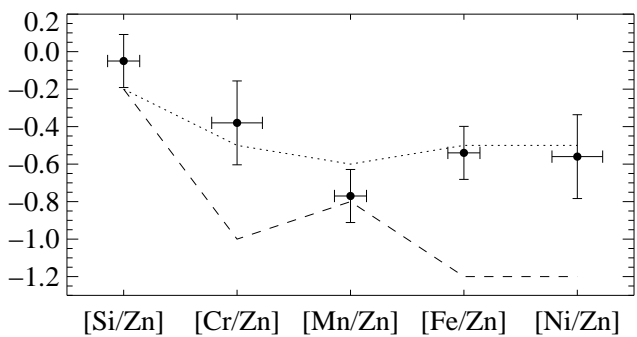}
\caption{Relative abundance pattern derived from measurements in the stacked absorption spectrum. The dotted 
(resp. dashed) line shows the typical patterns found in the Halo (resp. warm disc) of the Galaxy \citep[see][]{Welty99}.
\label{fig:deple}}
\end{figure}

Using the \ZnII\ column density derived from the stacked spectrum and the median hydrogen column density in our 
sample of ESDLAs, $\log N(\HI)= 21.8$, we estimate the average ESDLA metallicity to be about 1/20$^{\rm th}$ solar. 
This is consistent with the mean metallicity found for the overall DLA population across the same redshift 
range \citep[$\avg{Z} = (-0.22 \pm 0.03)\,z - (0.65 \pm 0.09)$,][]{Rafelski12}. 
We observe that iron is depleted by about a factor of three to four compared to zinc, which is slightly higher 
than the mean for the overall DLA population. Fig.~\ref{fig:deple} shows the abundance relative to that of zinc for iron and other species. 
As with most DLAs, such an abundance pattern is similar to the mean Halo abundance pattern of the Galaxy. However, we note 
that the Small Magellanic Cloud also has a ``Halo-like'' depletion pattern \citep{Welty97}, although it is a gas-rich dwarf 
galaxy. 
It is therefore hazardous to rely only on the depletion pattern to derive information about the physical origin 
of the gas, the depletion being more dictated by the metallicity than by the location of the gas in the galaxy. 
The dust-to-metal ratio measured following \citet{DeCia13}, ${\cal DTM} \simeq 0.75$, 
is similar to that measured in DLAs associated with long-duration GRB afterglows (GRB-DLAs). 

\subsection{Velocity extent}
The effective Doppler parameters that 
we derive independently from the curve-of-growth of \FeII\ and \SiII\ (Fig.~\ref{fig:cog}) 
are consistent with a single value of $b_{\rm eff} \sim 40$~\kms, which reflects the absorption line kinematics. 
Deconvolving the line width from the SDSS spectral profile
would be uncertain because of the insufficient SDSS resolution and smoothing resulting from redshift 
uncertainties when co-adding the spectra. 
Before comparing our results with previous studies, we note that DLA kinematics are more commonly quantified by their 
velocity width, $\Delta v$, defined as the velocity interval comprising 
5\%-95\% of the line optical depth \citep[see][]{Prochaska97}. 

\begin{figure}
\centering
\includegraphics[bb=74 178 520 397,width=\hsize]{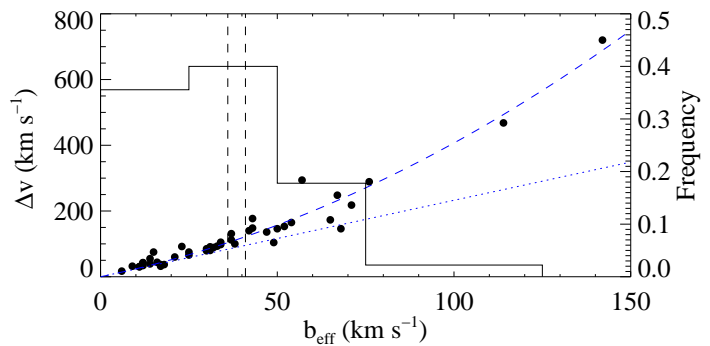}
\caption{Velocity width, $\Delta v$ measured by \citet{Ledoux06a} from VLT/UVES spectra 
as a function of the effective Doppler parameter derived from the equivalent widths of 
\FeII\ and/or \SiII\ lines. As expected, the theoretical relation for a single unsaturated component, 
$\Delta v = 2.33 b$ 
(dotted line), encompasses the data. The dashed line 
shows the empirical fitted relation $\Delta v = 2.21\,b_{\rm eff} + 0.02\,{b_{\rm eff}}^2$. 
The histogram shows the distribution of $b_{\rm eff}$ measured in the sample of \citet{Ledoux06a}, 
with the scale on the right axis. The vertical dashed lines marks the values independently 
found in ESDLAs from \SiII\ and \FeII.
 \label{fig:beff_dv}}
\end{figure}

We empirically derive the relation between $b_{\rm eff}$ 
and $\Delta v$ by applying a curve-of-growth analysis based on \SiII\ and/or \FeII\ equivalent widths measured 
for systems 
observed at high spectral resolution, for which $\Delta v$ is measured accurately \citep{Ledoux06a}. 
We 
derive $\Delta v = 2.21\,b_{\rm eff} + 0.02\,{b_{\rm eff}}^2$ (Fig.~\ref{fig:beff_dv}), which departs from 
the linear theoretical relation in the Gaussian regime, $\Delta v = 2.33 b$, at large values. 
Interestingly, this 
could indicate that satellite components increase $\Delta v$ while not changing 
much the curve-of-growth (which is derived from integrated equivalent widths). In this view, $b_{eff}$ could 
be an alternative method to quantify absorption-line kinematics, being less sensitive to 
satellite components than $\Delta v$ and applicable to data with any spectral resolution as soon as lines with a range of 
oscillator strength are covered. Our method also removes the degeneracy between column density and kinematics 
which could affect studies based on single line equivalent width \citep[e.g.][]{Prochaska08}.
Using this relation, the $b_{\rm eff}$ value derived 
above translates to $\Delta v \simeq 120$~\kms\ for the ESDLAs 
studied here. The mean velocity width of ESDLAs is again consistent with what is seen in the overall DLA population  
and agrees well with the linear relation between metallicity and velocity extent \citep{Ledoux06a}, within errors.

We emphasise that the mean abundances and depletion factors in ESDLAs are typical of 
the overall DLA population at this redshift. This suggests that ESDLAs probe the same underlying population 
of galaxies (i.e. same chemical enrichment history) as most DLAs and that the higher integrated column densities
observed  along the lines of sight (metal and neutral hydrogen) are a consequence of the hypothesised 
small impact parameter. 
The small average velocity extent ($\sim$120~km~s$^{-1}$) suggests that ESDLAs do not 
arise in distant gas ejected from or falling into the host galaxy, contrary to what 
has been invoked for DLA galaxies with large impact parameters \citep[e.g.][]{Bouche13,Krogager13}. 

\subsection{Note on molecules}
According to models by \citet{Schaye04}, the high column density gas collapses into cold and 
molecular gas \citep[see also][]{Schaye01} and eventually forms stars. We therefore 
anticipate the presence of H$_2$ in high column density systems. 
In the disc of the Milky Way, the molecular fraction increases sharply 
beyond $\log N(H) > 20.7$ \citep[][]{Savage77}. 
High-latitude Galactic lines of sight show a similar transition 
\citep{Gillmon06}, as do those towards 
the Magellanic Clouds, albeit at higher column densities \citep[$\log N(H)=21.3$ and 22 for the LMC and SMC, respectively;][]{Tumlinson02}. 
However, 
while there is a possible dependence on the H$_2$ fraction with $N(\HI)$ in high-$z$ DLAs, a sharp 
transition has not yet been observed \citep{Ledoux03,Noterdaeme08}. 
This is likely 
due to the very small cross-sections of clouds with large $N($H$_2)$ \citep{Zwaan06} as well as 
induced extinction. 
In addition, the transition could occur at higher H\,{\sc i} column densities than probed 
to date. Our sample is therefore 
a unique opportunity to identify the transition. Although the BOSS spectral resolution and S/N 
are far too poor to allow for detecting typical H$_2$ lines, it is 
nonetheless possible to detect H$_2$ if the column density is high enough to produce 
damping wings. 
About 20 DR9 ESDLAs from our sample are among the $\sim$10\,000 DLAs that \citet{Balashev14} 
inspected for very strong H$_2$ systems.
Two of them (towards SDSS\,J004349$-$025401 and SDSS\,J084312+022117J) belong to their sample of 23 confident strong H$_2$ systems (with $\log N($H$_2) \ga 19$).
Among the remaining ESDLAs, the $\log N(\HI)\sim 22$ DLA detected toward 
J\,0816$+$1446 has been observed with UVES and presents H$_2$ lines 
with $N($H$_2) \sim 5 \times 10^{18}$~\cmsq\ \citep{Guimaraes12}. 
Finally, although it is not in 
SDSS and hence not part of our sample, the well-known H$_2$-bearing DLA toward HE\,0027$-$1836 
with $N($H$_2) \sim 2-3 \times 10^{17}$ \citep{Noterdaeme07lf, Rahmani13} and $\log N(\HI)=21.75$ 
also classifies 
as an ESDLA according to our definition. 
Molecular hydrogen may be conspicuous in ESDLAs, but observations 
at higher spectral resolution would be required to investigate the overall sample.

\subsection{High-ionisation species}

\begin{figure}
\centering
\includegraphics[bb=74 178 491 397,width=\hsize]{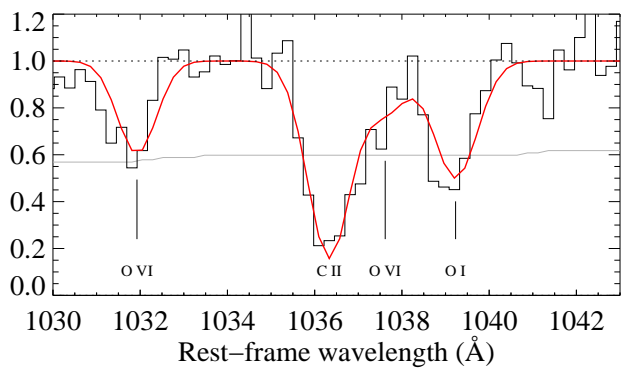}
\caption{Portion of the normalised stacked spectrum around the \OVI\ doublet. \label{fig:o6}}
\end{figure}

Analysing the \lya\ forest of a stacked spectrum is challenging because of uncertainties 
in the continuum placement and random blending with intervening \HI\ lines 
\citep{Pieri10}. Furthermore, because of the 
different DLA absorption redshifts, only a fraction of the sample contributes to a given 
rest wavelength in the stacked spectra. 
Although the \OVI\ doublet is always located in the forest, these transitions are useful to probe hot gas that 
is potentially related to galactic outflows originating from star-formation activity. 
Thanks to 
the small separation between the two lines of the \OVI\ doublet, the above-mentionned difficulties 
are mostly avoided for this species. 
We measured rest equivalent widths of 0.44$\pm$0.22 and 0.27$\pm$0.15 for \OVI\,$\lambda$1031 and 1037 respectively from the corresponding portion of the normalised stacked spectrum in Fig.~\ref{fig:o6}.
Two points do not allow any consistency check when building a curve-of-growth (that has two parameters), 
but it is interesting to note that these equivalent widths indicate a column density of about 
$\log N(\OVI) \sim 14.8$ and $b_{\rm eff} \sim 80$~\kms. The effective Doppler parameter
is larger than what is observed for low-ionisation metal lines and indicates a 
velocity extent of $\Delta v \sim$300~\kms\ (see Fig.~\ref{fig:beff_dv}). This is among the high  
values seen in DLAs \citep{Fox07b}, while the \OVI\ column density and metallicity of the neutral gas 
match the metallicity-$N(\OVI)$ correlation presented by these authors. We caution however that 
these values are indicative only, due to the large error on the \OVI\ EWs.
This may indicate that outflows of hot gas (best seen when the
line of sight has small impact parameters) could be present. 
Indeed, \citet{Fox07b} demonstrated that the
\OVI\ phase should be hot ($>$10$^5$~K) and collisionally ionised with a 
baryonic content of at least of similar order as that in the \HI\  phase. However, we caution again that 
the measurement of $\Delta v$ is highly uncertain here since \OVI$\lambda$1037 is blended 
with \CII$\lambda$1036 and \OI$\lambda$1040, in addition to the \lya\ forest. 
Furthermore, the product of oscillator strengths 
and wavelengths of the two lines differ only by a factor of two.

\subsection{Comparison with GRB-DLAs \label{GRBs}}

\begin{figure}
\centering
\includegraphics[bb =  30 175 500 570,clip=,width=\hsize]{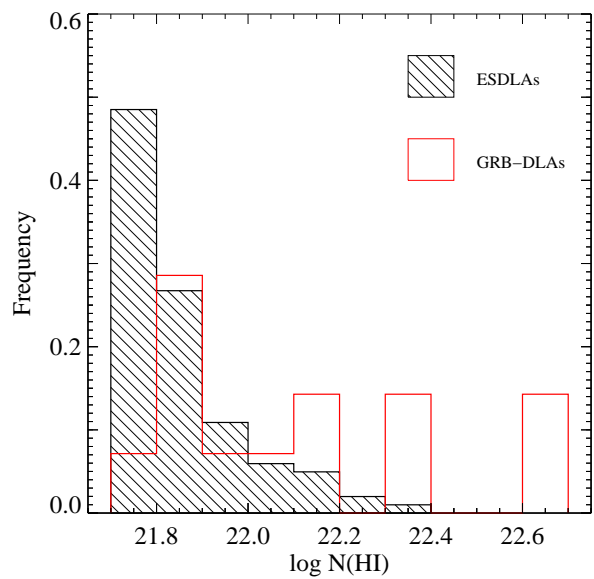}
\caption{\HI\ column density distribution for our intervening 
QSO-DLAs with $N(\HI)\ge 5\times10^{22}$~\cmsq\ (hashed histogram) compared to that of 
associated GRB-DLAs (red unfilled histogram, \citealt{Fynbo09,deUgartePostigo12}). \label{fig:nhi_dist}}
\end{figure}

In this Section, we compare the ESDLAs properties with 
those of GRB-DLAs 
in the same redshift range, $z=2-4$. We use the GRB-DLA measurements
from \citet{Fynbo09} with additional values from \citet{deUgartePostigo12}. 
Since GRBs originate from the collapse of massive stars \citep{Bloom99}, they are expected 
to occur in star-forming regions. A large fraction of GRB-DLAs have 
high $N$(\HI) values \citep[e.g.][]{Prochaska07,Fynbo09}, 
and we concentrate only on the high end of the QSO-DLAs and GRB-DLAs $N(\HI)$ distributions 
(Fig.~\ref{fig:nhi_dist}). The distribution for GRB-DLAs is flatter than 
that for QSO-DLAs and extends up to higher $N(\HI)$ values. Interestingly, in most 
cases, the gas that gives rise to GRB-DLAs is not directly associated with the 
dying star, but is instead located further away in the host galaxy \citep[e.g.][]{Vreeswijk07}. This 
implies that the difference between the two distributions is not related to the properties of
the immediate GRB environment but rather related to biases in the GRB sample  and 
inclination effects \citep{Fynbo09}:  Since QSOs randomly probe the intervening gas, the 
corresponding distribution reflects the average gas cross-section at different column 
densities and includes geometrical effects due to the galaxy inclination. 
In turn, GRB-DLAs uniformly probe their host galaxy, regardless of their inclination. 
In other words, face on discs would have a higher probability to appear in QSO 
surveys compared to edge-on discs, while this should not be the case for GRBs.

\begin{figure}
\centering
\begin{tabular}{c}
\includegraphics[bb =  40 215 500 570,clip=,width=\hsize]{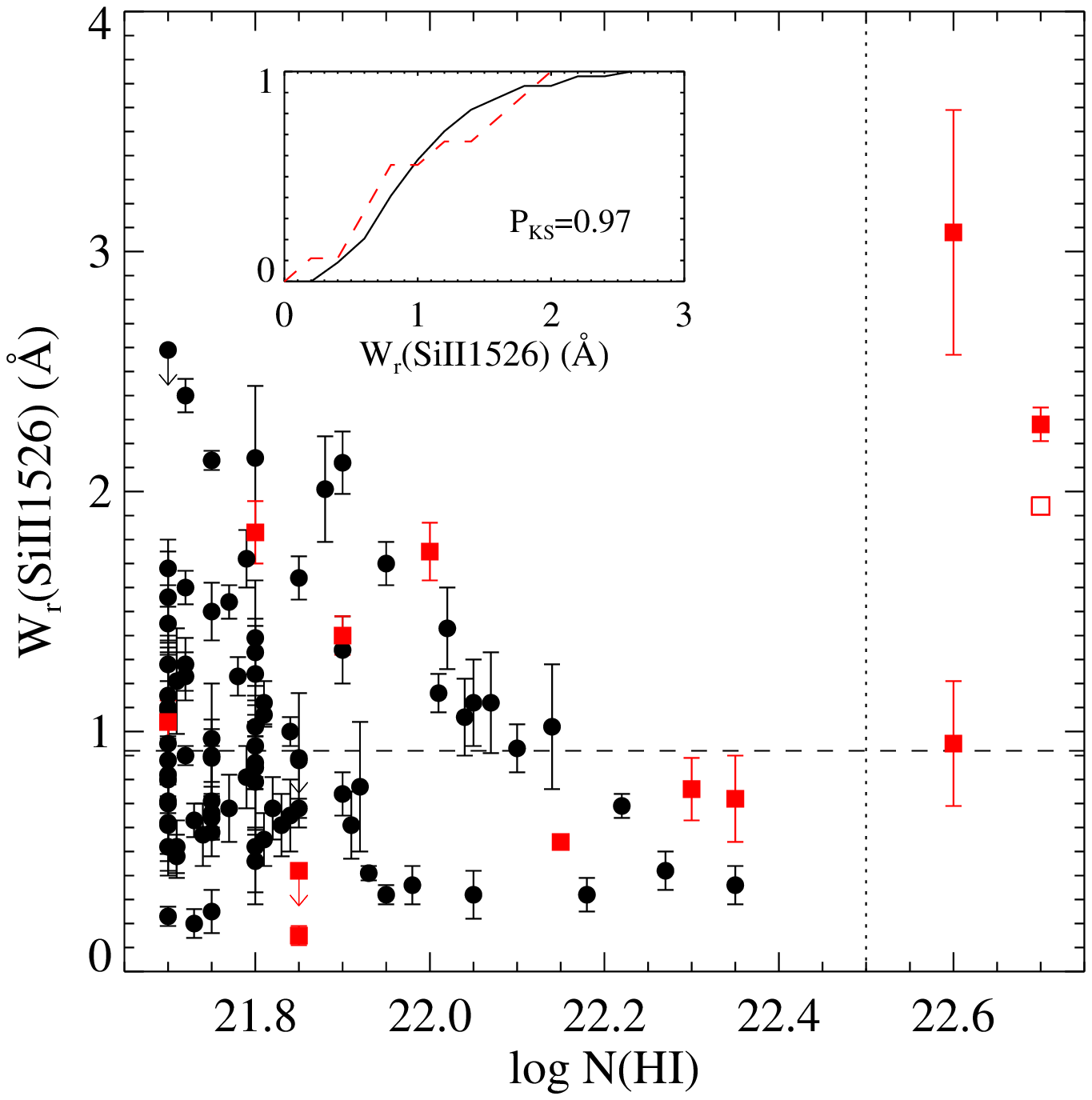}\\
\includegraphics[bb =  40 215 500 570,clip=,width=\hsize]{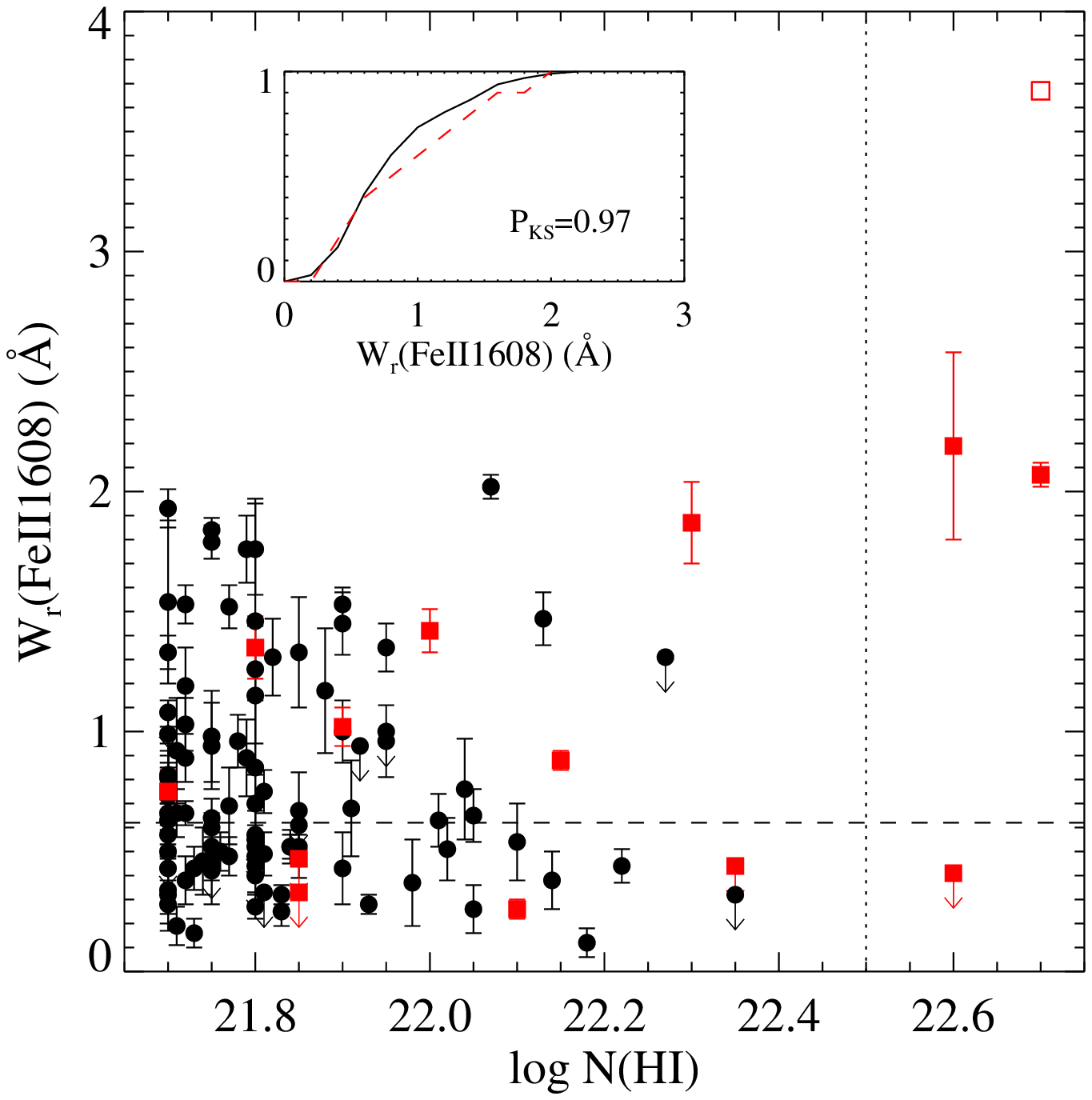}\\
\includegraphics[bb =  40 175 500 570,clip=,width=\hsize]{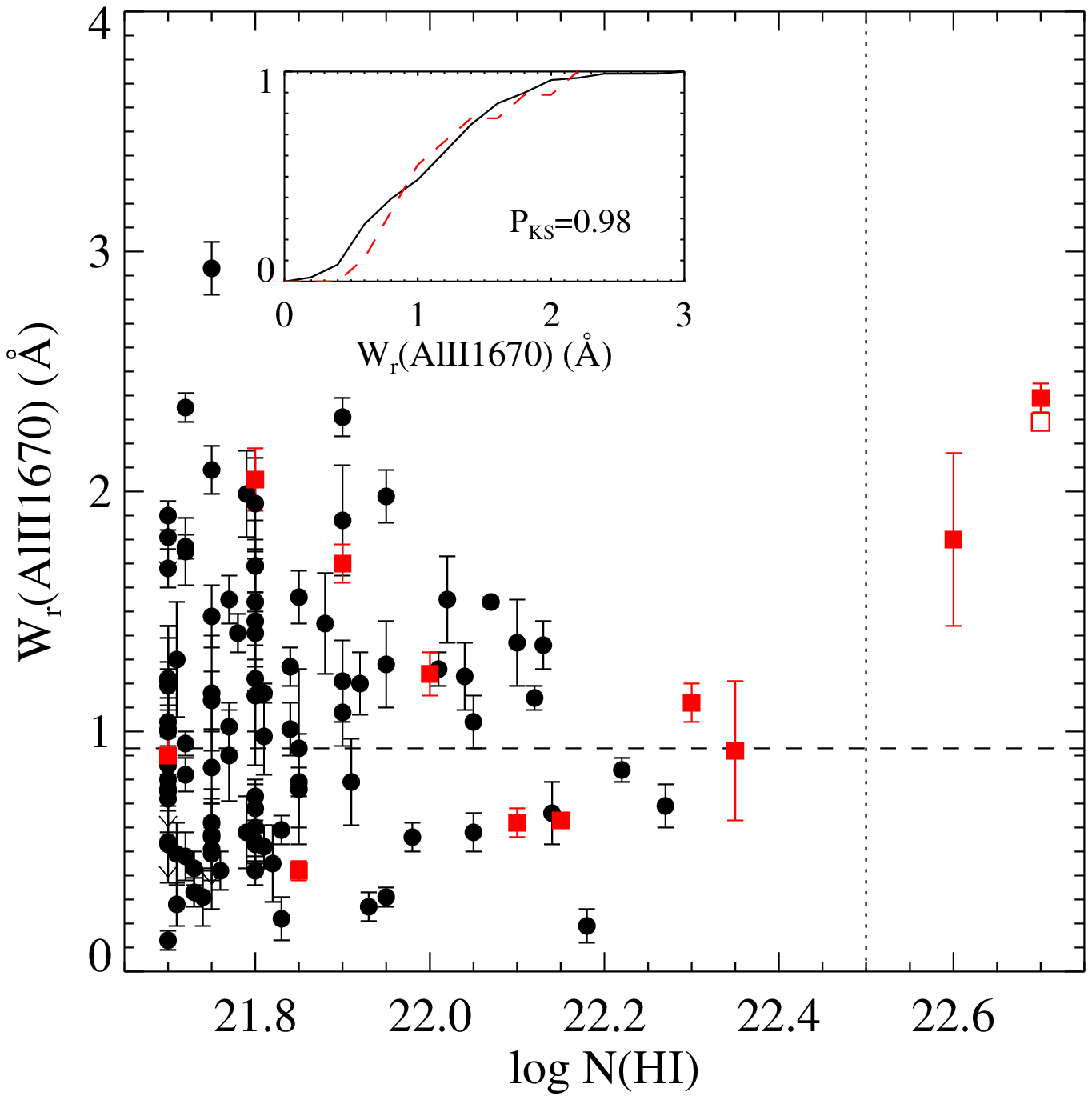}\\
\end{tabular}
\caption{Rest-frame equivalent widths of metal lines as a function of $\log N(\HI)$. 
Black are ESDLAs, red are GRB-DLAs from \citealt{Fynbo09} and \citealt{deUgartePostigo12}. 
The open/filled squares at $\log N(\HI)=22.8$ 
correspond to EW in the GRB-DLA 080607 measured at R=400 and R=1200 respectively. The horizontal dashed 
lines correspond to the average EWs measured from the stacked spectrum.
\label{fig:ew_grb}}
\end{figure}

We compare the equivalent widths of three metal lines for ESDLAs and GRB-DLAs 
(Fig.~\ref{fig:ew_grb}). We use  \SiII$\lambda$1526, \FeII$\lambda$1608 and 
\AlII$\lambda$1670 which are generally located in a clean portion of the spectra, 
redwards of the QSO \lya\ emission and bluewards of sky emission lines.
The range of \SiII\ and \AlII\ equivalent widths decreases with increasing $N(\HI)$ for ESDLAs. 
Conversely, EWs appear to increase at large $N(\HI)$ for GRB-DLAs, 
but this is mostly due to three 
GRB-DLAs that have column densities in a regime still unprobed by ESDLAs ($\log N(\HI)>22.5$). 
These also correspond to dark GRBs.
If we restrict the comparison to systems with column densities below $\log N(\HI)=22.5$, 
then the distributions of metal EWs for GRB-DLAs 
and ESDLAs are very similar, as the Kolmogorov-Smirnov tests indicates. 
Dust biasing could account for the decrease in the EW upper bound with increasing 
$N(\HI)$, since the extinction increases linearly with the metal column 
density (\citealt{Vladilo05}). 
In this case, the 
dependence on $N(\HI)$ arises 
indirectly from the fact that the EWs may more represent the velocity extent than the column density, 
and that this velocity extent increases with metallicity \citep{Ledoux06a}. In other words, systems 
with both high EW and high-$N(\HI)$ are therefore likely to have large metal column densities and 
hence produce more extinction.
However, as we will see in the next section, 
the average extinction per H atom appears to be relatively small.  
Another explanation is that at the highest column densities, the line of sight passes closer to the inner 
region of the host galaxy, where the dispersion in velocity could be less important.
It is also possible that, because both high-$N(\HI)$ systems and high EWs 
are uncommon, systems with these two characteristics are simply rarer and would 
require higher statistics to be represented in the figure.

It is possible that such a decrease in the range of EW does not apply for \FeII$\lambda$1608. 
This could be due the fact that this line is 
weaker, and therefore its equivalent width is less dominated by kinematics and more indicative 
of the column density. If so, the above explanation involving lower kinematics at higher $N(\HI)$ 
would be preferred. In 
addition, the decrease of $W_{\rm r}$(\SiII) and $W_{\rm r}$(\AlII) is mostly seen from the upper boundary, 
while the minimum 
EW for a given $N$(HI) bin seems rather to increase with $N$(\HI). This is expected if 
the metallicity does not depends on $N$(\HI), as the column density contributes more significantly 
for small EWs.
We note that, because GRB DLAs are generally observed at even lower spectral resolution, 
\FeII$\lambda$1611 and \FeII$^{\star}\lambda$1612 also contribute to 
the equivalent width 
measured for \FeII$\lambda$1608. Excited iron, \FeII$^{\star}$, which is absent in QSO-DLAs, 
is generally enhanced in GRB-DLAs because of strong excitation from the burst itself.
In the case of GRB\,080607, \citet{deUgartePostigo12} 
quote an \FeII\ EW measurement based on the $R = 400$ spectrum from \citet{Prochaska09} that 
surprisingly appears significantly higher than the \SiII$\lambda$1526 EW, which is an 
isolated line. The equivalent width measured from their $R=1200$ spectrum avoids contamination. 
As a result, the \FeII$\lambda$1608 EW is significantly lower 
and more consistent with \SiII$\lambda$1526 and \AlII$\lambda$1670 
(see Fig.~\ref{fig:ew_grb}). The quoted \FeII$\lambda$1608 equivalent widths in 
GRB-DLAs are thus probably overestimated.

\section{Induced colour distortions of the background QSO light. \label{colours}}

\begin{figure}
\begin{tabular}{c}
\includegraphics[bb=60 175 500 400,clip=,width=\hsize]{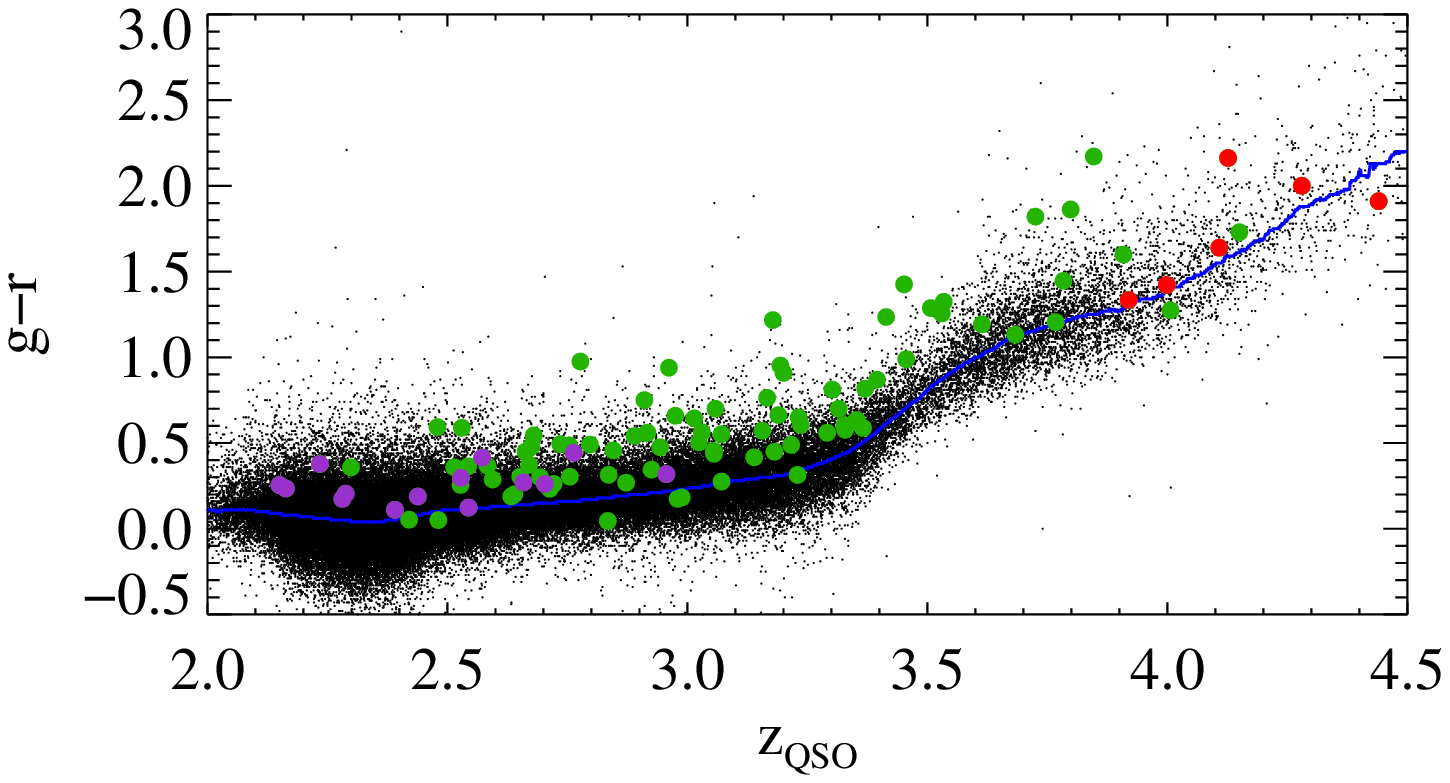}\\
\includegraphics[bb=60 175 500 400,clip=,width=\hsize]{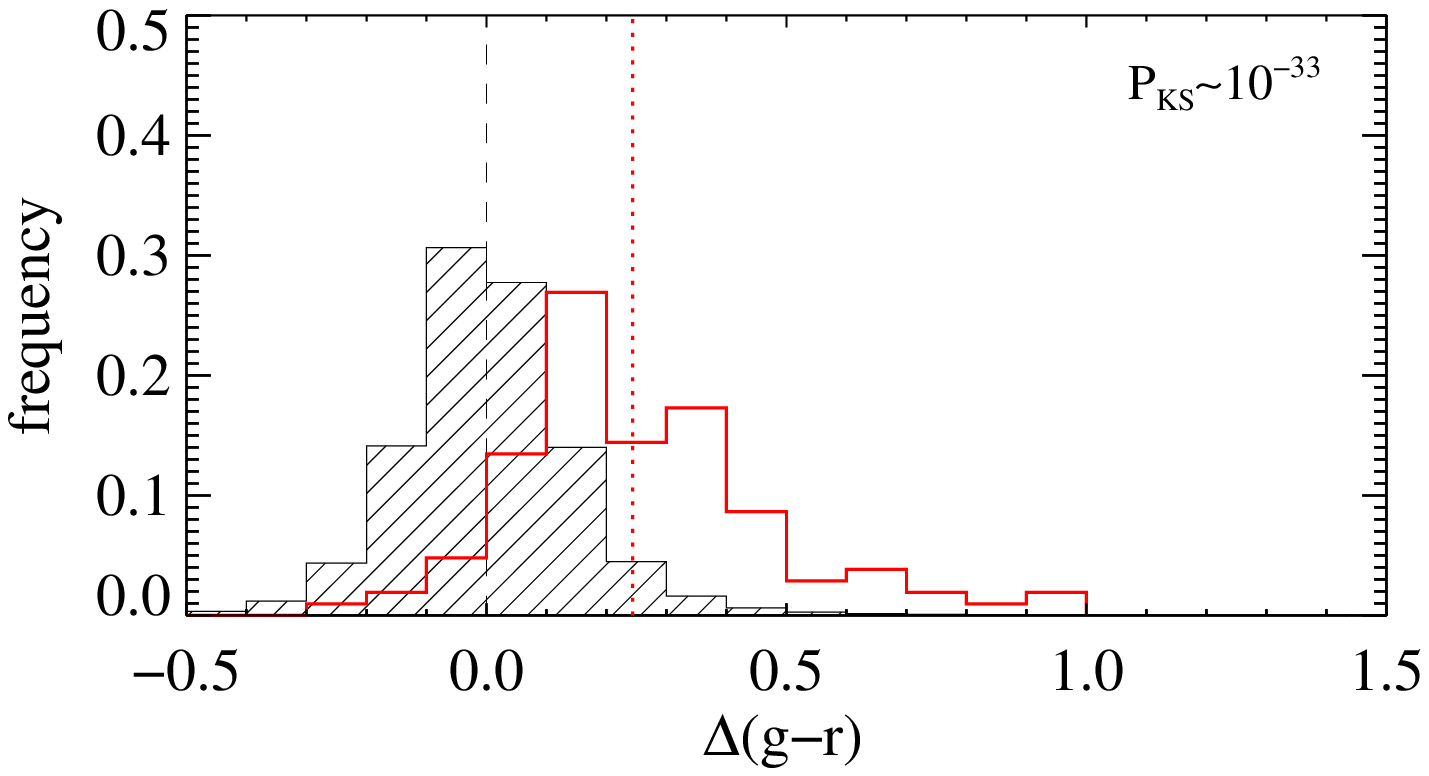}\\
\end{tabular}
\caption{{\sl Top}: ($g-r$) colour of the background QSOs as a function of redshift. The black points represent 
the DR11 QSO sample. QSOs with foreground ESDLA that have a \lya\ centroid that falls in the $u$, $g$, 
and $r$-band are represented by purple, green and red points respectively. The blue line shows the change 
in the median colour of the overall QSO sample as a function of redshift.
{\sl Bottom}: Normalised distributions of colour excess $\Delta (g-r)=(g-r)-\avg{g-r}_z$ for the DR11 QSO 
sample (black hashed histogram) 
and the QSO with foreground ESDLAs (red unfilled histogram). The vertical lines mark the median of the two distributions. \label{fig:gr}}
\end{figure}

In this section, we investigate whether the presence of an ESDLA has an effect on the the background 
quasar colour.
The top panels of Figs.~\ref{fig:gr} and \ref{fig:iz} represent respectively the $(g-r)$ and $(i-z)$ colours 
of the quasars as a function of their redshift. 

\subsection{\lya\ absorption}
The $g-r$ values for QSOs with foreground ESDLAs are 
systematically higher than the median colour for the BOSS DR11 QSO population at a given redshift ($\avg{g-r}_z$, blue line in the figure). 
This trend is better seen in the distribution of the colour excess $\Delta (g-r)=(g-r)-\avg{g-r}_z$ in the bottom 
panel, which indicates a systematic difference of about 0.2~mag between the two populations, with an 
almost zero probability that this 
is due to chance coincidence. 
The presence of a damped \lya\ absorption can explain the 0.2 mag difference, because the centroid often 
falls in the $g$-band. Even when it does not, the extended wings of the absorption profile also 
affect this band.

\subsection{Dust extinction}

\begin{figure}
\begin{tabular}{c}
\includegraphics[bb=60 175 500 400,clip=,width=\hsize]{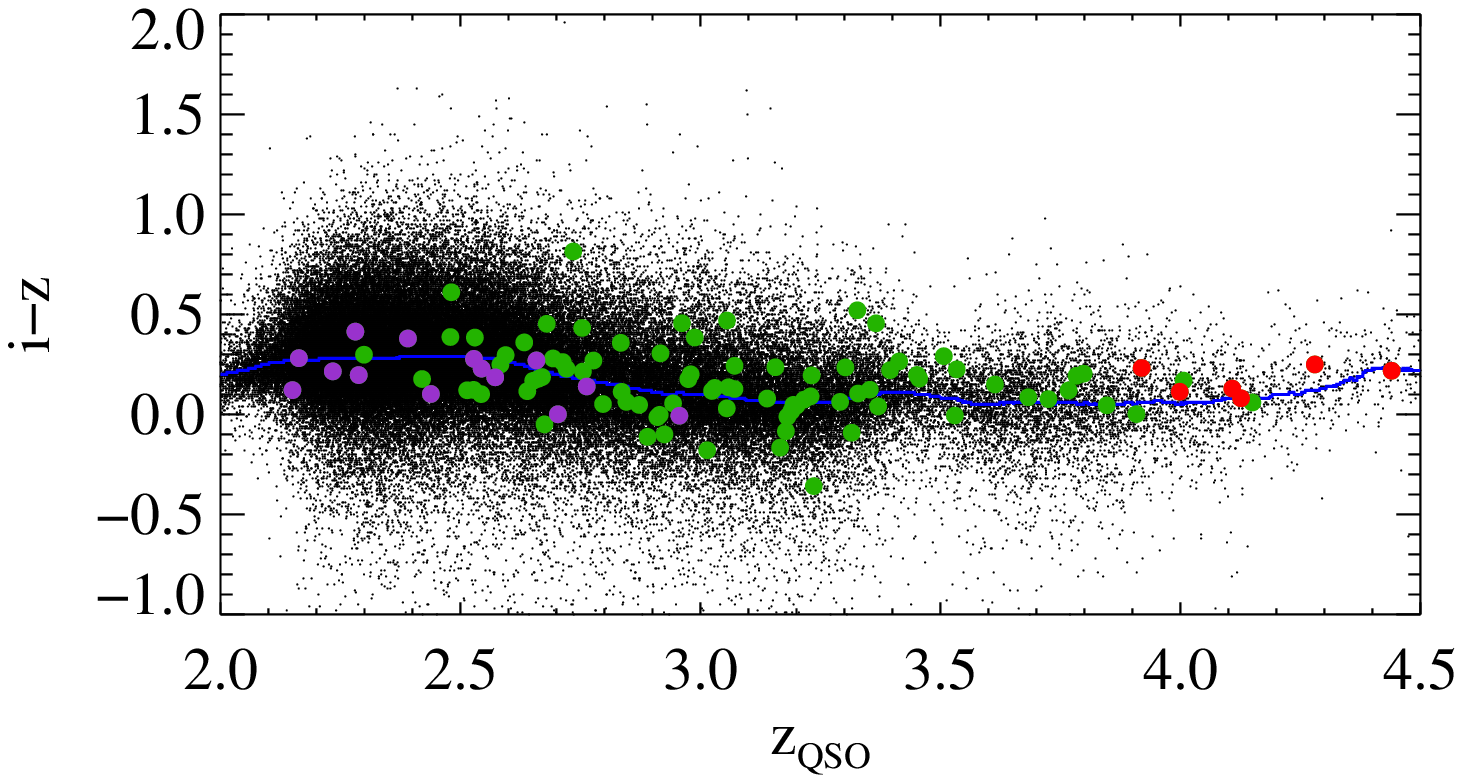}\\
\includegraphics[bb=60 175 500 400,clip=,width=\hsize]{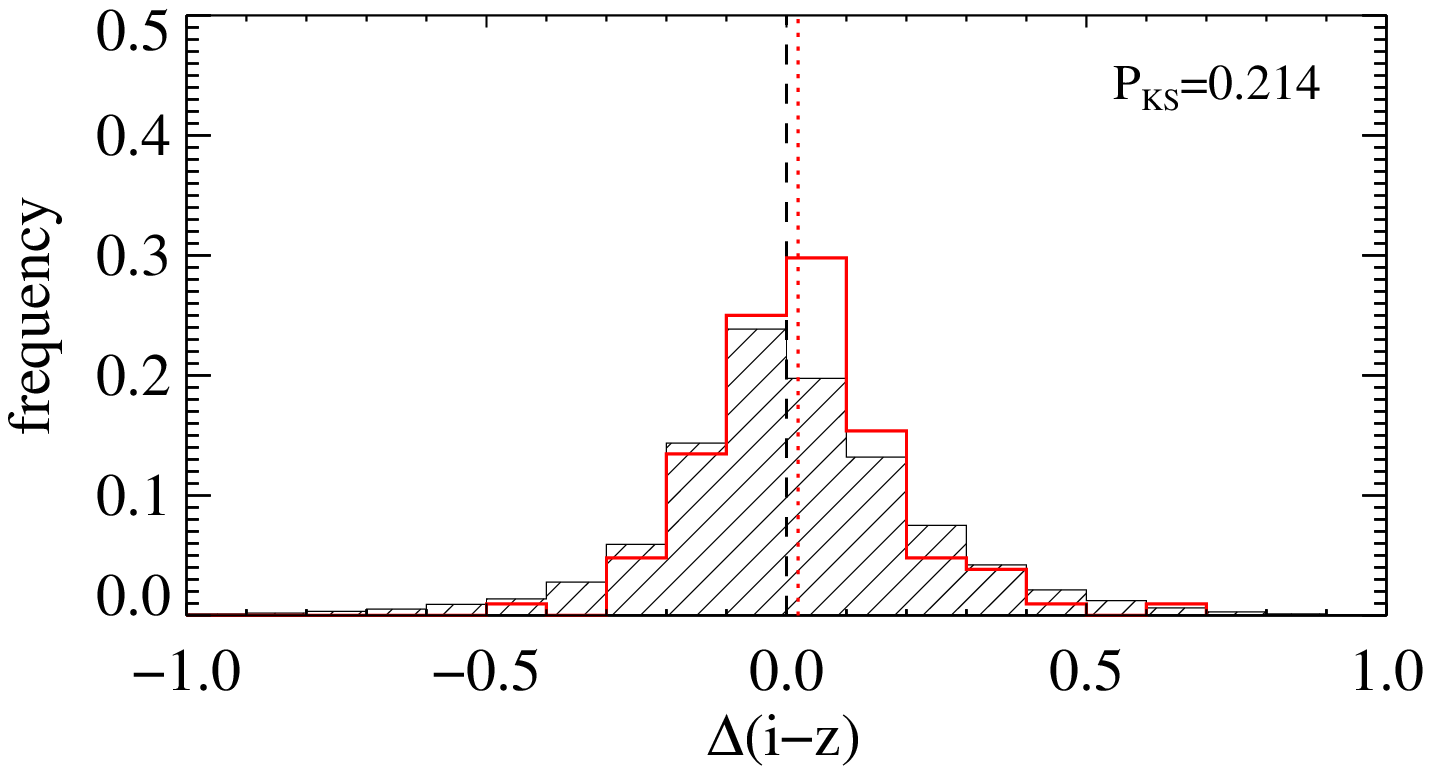}\\
\end{tabular}
\caption{Same as Fig.~\ref{fig:gr} for the $(i-z)$ colours. \label{fig:iz}}
\end{figure}

Potential continuum absorption by dust must be investigated at wavelength ranges unaffected by 
Lyman absorption lines, i.e., the $i$ and $z$ bands. 
The colour excess is hard to see directly in the $(i-z)$ vs $\zqso$ plot (Fig.~\ref{fig:iz}), but 
produces a 0.02~mag difference 
in the median of the two distributions in the bottom panel. The Kolmogorov-Smirnov test indicates a 20\% probability that the 
two samples are drawn from the same parent distribution.

\begin{figure*}
\centering
\includegraphics[bb = 125 24 452 735,clip=,angle=90,width=0.9\hsize]{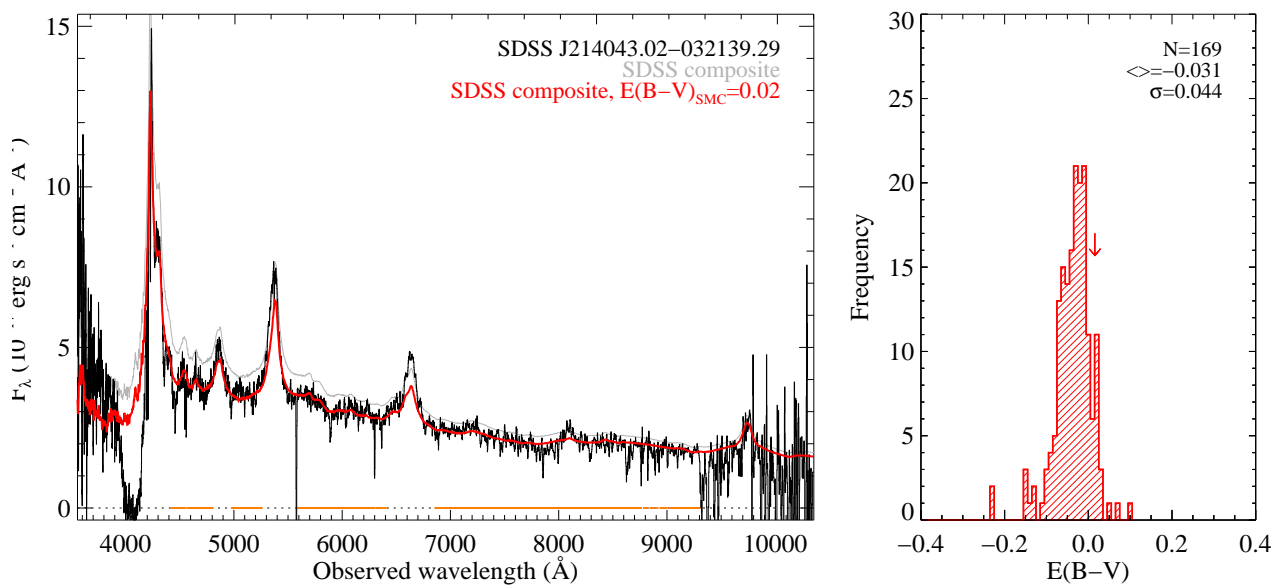}
\caption{{\sl Left}: ESDLA BOSS spectrum (black) with the SDSS composite spectrum (grey) reddened by the SMC extinction-law 
at \zabs (red) with $E(B-V)=0.02$. The orange segments indicate the regions used for the fit. {\sl Right}: Distribution of E(B-V)-values for the corresponding 
control sample. In this example, the ``zero-point'' (resp. dispersion) of the distribution is $-0.03$ (resp 0.044). The value retained for the DLA here is thus $E(B-V)=0.05\pm0.05$.\label{fig:sed}}
\end{figure*}

In order to quantify the reddening effect of intervening ESDLAs on the QSO light, 
we apply the technique 
described in \citet{Srianand08bump} and \citet{Noterdaeme10co}. We match each spectrum with a QSO 
composite spectrum from \citet{VandenBerk01} redshifted to the same $\zqso$ and reddened by a 
SMC-extinction law at $\zabs$.
As for most QSO absorbers \citep{York06}, this is the preferred extinction law for all but one ESDLA, 
(towards J\,104054$+$250709) which is best fitted with a LMC-extinction law featuring a 2175-{\AA} UV bump.
In the fitting process, we ignore the emission line regions as well as wavelengths bluewards of the 
QSO \lya\  emission.
For each QSO with intervening ESDLA in our sample, we repeat the same procedure on a control 
sample drawn from the same original DR11 QSO sample that we searched for DLAs with an emission redshift 
close to that of the ESDLA-bearing QSO. We restrict the redshift difference to $\Delta z=0.001$, yielding a typical 
control sample size of over 100 QSO spectra. For about 25\% of the systems we needed to increase this range 
to obtain at least 
50 QSOs for the control sample. The maximum redshift interval 
in the sample is then $\Delta z = 0.04$, which is certainly still small enough to avoid any 
redshift-dependent differences in the QSO colours. An example of the fitting is shown in Fig.~\ref{fig:sed}.

\begin{figure}
\centering
\includegraphics[bb = 60 175 500 570,clip=,width=\hsize]{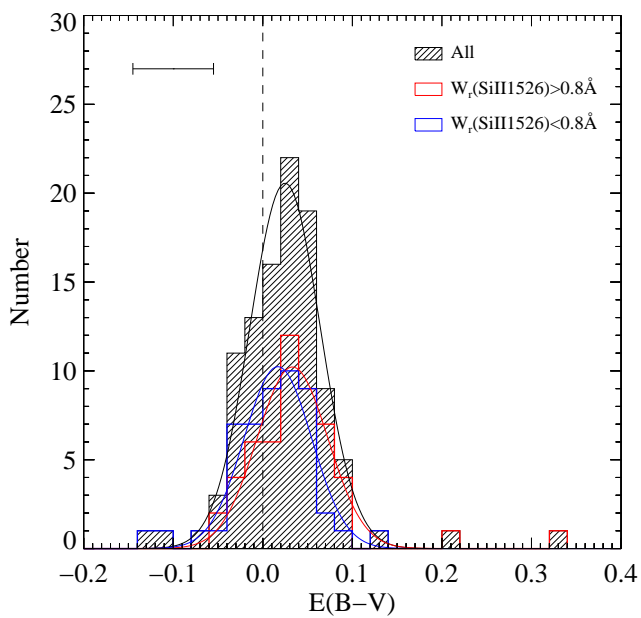} 
\caption{Distribution of $E(B-V)$ values (full sample as hashed histogram, subsample with $W_{\rm r}(SiII\lambda1526)$ above (resp. below) 0.8~{\AA} in red (resp. blue) unfilled histogram ), fitted with Gaussian functions.
The horizontal error bar in the upper-left corner indicates the typical error on $E(B-V)$, obtained from 
the standard deviation of the values in each control sample.  \label{fig:ebv}}
\end{figure}

We measure $E(B-V)$ by subtracting the median of each control sample (our ``zero-point''). 
The control sample dispersion provides the total error on $E(B-V)$ due to fitting 
uncertainties and intrinsic shape variations. We note that although unrelated absorbers 
could contribute to the measured reddening in individual ESDLAs, this has no consequence on 
our statistical result as the control sample is affected exactly the same way.
The distribution of $E(B-V)$ is shown in Fig.~\ref{fig:ebv}, and is well modelled 
by a Gaussian centred at $E(B-V)=0.025$. The dispersion around the mean is 0.05~mag, matching the mean 
error on $E(B-V)$, shown as a horizontal error bar. 
While it is hard to conclusively detect reddening in any individual ESDLA, on average these systems 
present a small but statistically significant reddening $\avg{E(B-V)} \approx 0.02-0.03$, which 
explains the typical 0.02~mag $(i-z)$ excess measured above.

This corresponds to a specific extinction of the order of $A_{\rm V}/N(\HI) \sim 10^{-23}$~mag\,cm$^2$, which is 
similar\footnote{We caution however that 
the derivation of $E(B-V)$ here and in \citet{Vladilo08} are quite different, the former being based on SED 
fitting with a template and the later based on photometric colour excess.} to the median value for the overall ($\log N(\HI)\ge 20.3$) 
DLA population \citep{Vladilo08}.

Next we divide our sample into two subsamples with \SiII\,$\lambda$1526 rest-frame equivalent width 
above and below 0.8~{\AA}. The reddening is statistically 
stronger in systems with higher $W_{\rm r}(\SiII\,\lambda 1526)$, as observed by \citet{Khare12} for the overall population of DLAs. 

\section{Lyman $\alpha$ emission from the host galaxy  \label{Lya}}

The background QSO light is completely absorbed at the position of the DLA trough, enabling 
us to search for 
\lya\ emission from star-formation activity in the vicinity of the neutral gas 
\citep[see e.g.][]{Rahmani10}. However, searches for DLA galaxy 
counterparts based on \lya\ emission have resulted mostly in non-detections 
\citep[e.g.][as well as numerous unpublished searches]{Lowenthal95,Moller04}. Indeed, because 
of their cross-section selection, DLAs should arise mostly from galaxies with low star-formation rates 
\citep[e.g.][]{Cen12}. The impact parameters can also be larger than the fibre radius, which means that the 
\lya\ emission will not necessarily be detected in the quasar spectrum 
\citep[such an example is shown in][]{Fynbo11,Krogager13}, although simulations indicate 
that the impact parameters should be of the order of a few kpc on average \citep{Pontzen08,Rahmati14}. 
Even in cases where oxygen or Balmer 
emission lines are detected, the \lya\ escape fraction can be far from unity, putting the 
line flux well below the detection threshold. 
Finally, resonant scattering widens the \lya\ line far beyond the virial velocity of the 
star-forming region. The \lya\ line can therefore be spread over the whole DLA core, 
making it difficult to distinguish the emission from residuals in the zero-flux level. 
This, together with fibre losses, explains in part why interpreting the residual flux in a stacked DLA 
spectrum can be challenging \citep{Rahmani10,Rauch11}.

Extremely strong DLAs allow us to avoid most of the observational obstacles. If very strong 
column densities truly arise from gas located within a galaxy, then we expect that the light 
from the galaxy will fall well within the radius of the BOSS fibre ($r=1\arcsec$ or equivalently, 
$\sim$8 kpc at $z \sim 2.5$). Furthermore, the light from the background quasar is completely 
absorbed across more than 10~{\AA} (rest-frame) for $\log N(\HI)\ge 21.7$, isolating any possible \lya\ 
emission from the wings of the damped profile.

\subsection{Stacking procedure}
\begin{figure}
\includegraphics[bb=60 175 500 400,width=\hsize]{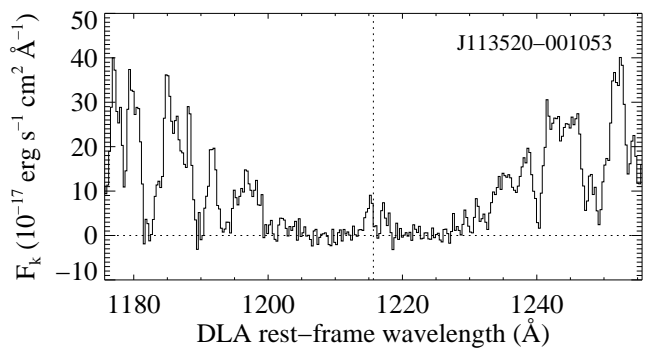}
\caption{Portion of the BOSS spectrum of SDSS~J\,113520$-$001053 in which the double peaked 
\lya\ emission is clearly seen in the DLA trough. \label{fig:j1135}}
\end{figure}

In order to detect faint \lya\ emission, we stack quasar spectra in the DLA rest-frame, following 
the technique described in \citet{Rahmani10}. 
We restrict our sample to ESDLAs with $2 < \zabs <3.56$ to ensure that the expected \lya\ emission always falls on the blue CCD. We thereby avoid both the very blue end of the spectrum and the region that overlaps with the red CCD, 
where spurious spikes are frequently seen. 
We also exclude the ESDLA toward \jonze, for which strong \lya\ emission is detected (see Fig.~\ref{fig:j1135}) 
\footnote{\citet{Noterdaeme12a} analysed this system and derived SFR~$\sim 25$~M$_{\sun}$\,yr$^{-1}$ 
from the detection of \Ha\ emission.} and a few systems that have very noisy spectra (CNR<2)  
where $\zabs$ and $N(\HI)$-measurements are highly uncertain.

Our sample for stacking includes 95 flux-calibrated spectra.
Each spectrum is shifted to the DLA-rest frame and then rebined to a common grid with the same 
velocity-constant pixel size as the original BOSS data, conserving the flux per unit wavelength interval. 
We next convert each spectrum into luminosity per unit wavelength using the luminosity distance at the 
DLA's redshift. We produce 
an average composite spectra using: 
$(i)$ median, $(ii)$ mean with iterative 3\,$\sigma$ rejection of deviant values (``3\,$\sigma$-clipped 
mean''), and $(iii)$ weighted mean. The latter is produced by weighting each spectrum (in luminosity 
units) by the squared inverse of the corresponding error ($\sigma_{\rm dark}$ in Table~\ref{tab:esdla}). 

\begin{figure*}
\centering
\begin{tabular}{ccc}
~~ ~~ \,median & 3\,$\sigma$-clipped mean & weighted mean \\
\includegraphics[bb =  90 212 480 570,height=0.3\hsize]{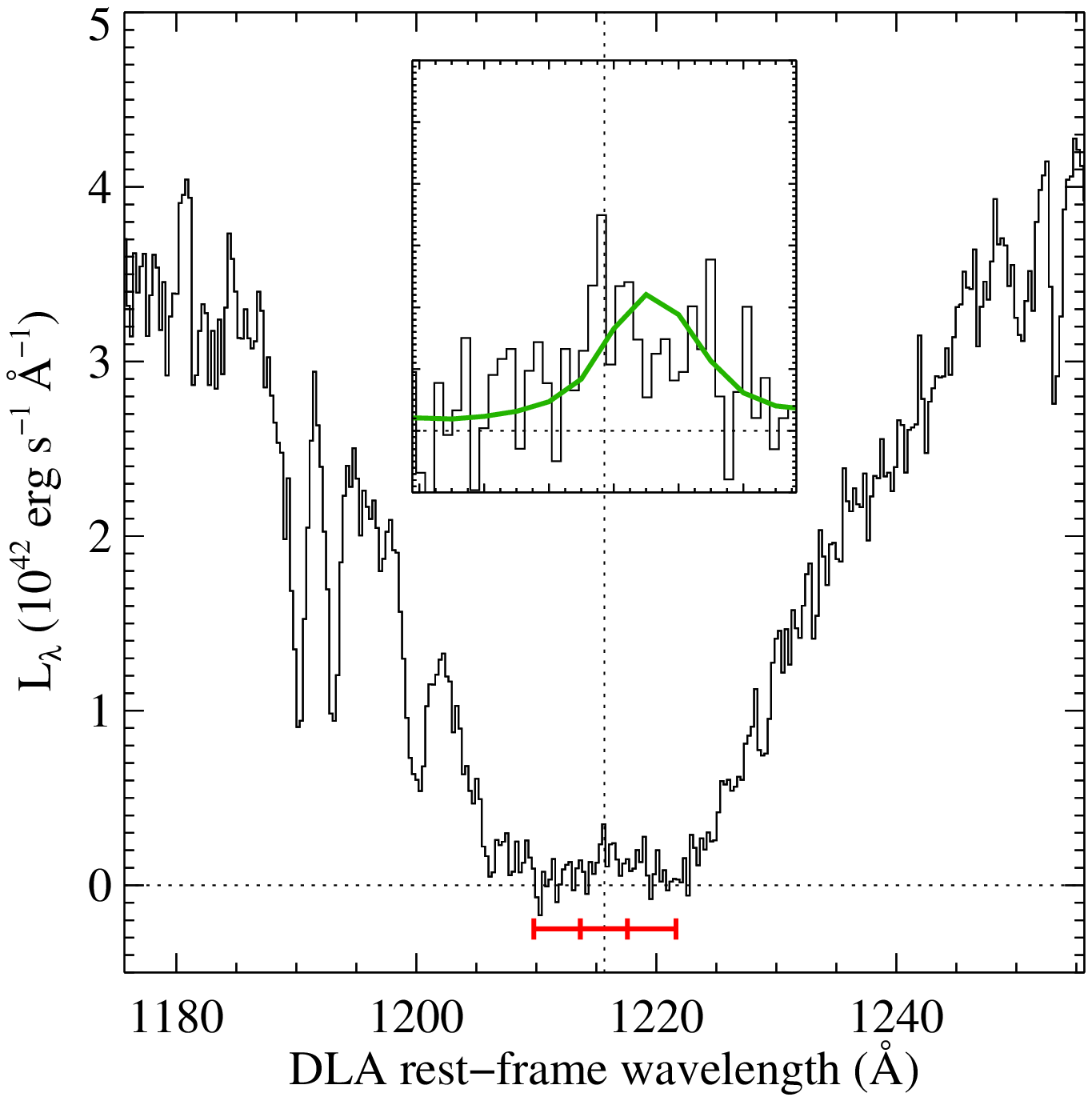} &
\includegraphics[bb = 132 212 480 570,height=0.3\hsize]{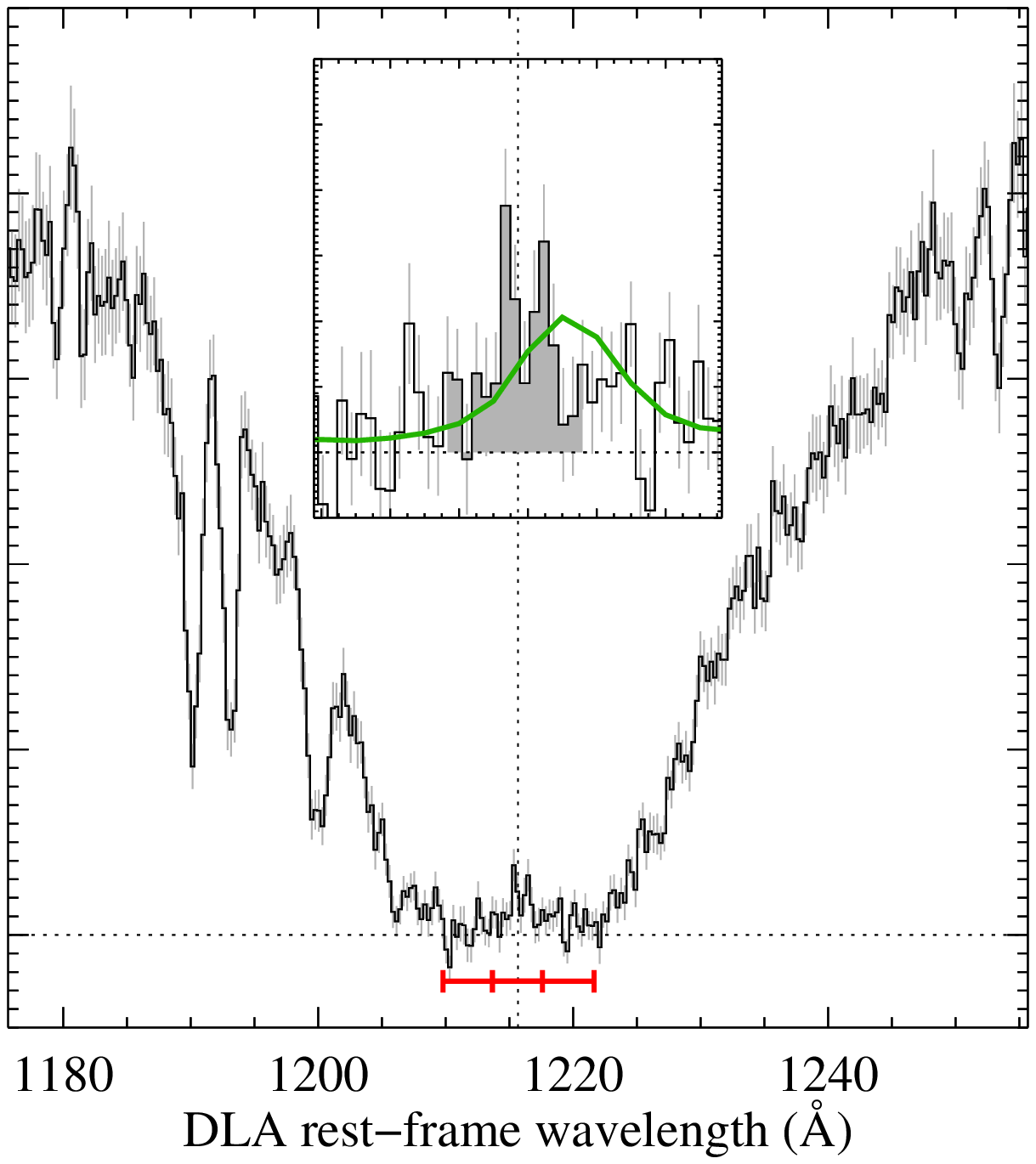} &
\includegraphics[bb = 132 212 480 570,height=0.3\hsize]{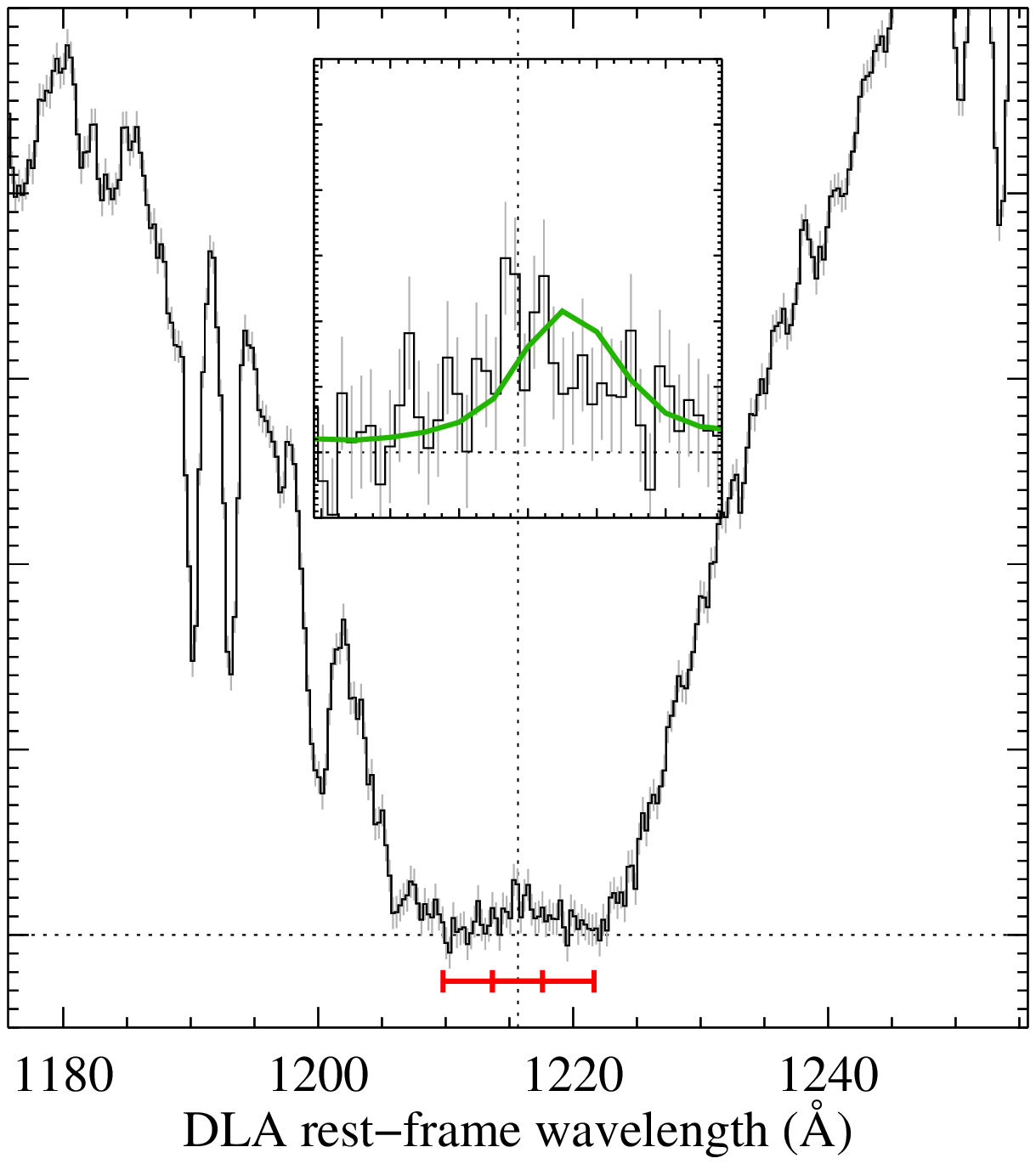} \\
{\Large \strut}  & & \\
\end{tabular}
\caption{Results from spectral stacking. 
From left to right, the stacked spectrum corresponds to the median values, 3\,$\sigma$-clipped mean 
and weighted mean. 
The long red segment shows the DLA core region over which $\tau>6$ for $\log N(\HI)\ge 21.7$, ensuring no residual 
flux from the quasar. This region is highlighted in the inset figures. 
The short red segment (inner tick marks) indicates the 1000~km\,s$^{-1}$ central region over which the \lya\ luminosity is integrated. The green spectrum is the 
composite Lyman-break galaxy spectrum from \citet{Shapley03}, scaled to match the same luminosity in the \lya\ region. 
\label{fig:stack}}
\end{figure*}

Fig.~\ref{fig:stack} shows the results of stacking 
with the three different averaging methods. The \lya\ emission appears clearly in the three composite 
spectra as positive flux in the central pixels. We overplot the scaled LBG composite from 
\citet{Shapley03} for comparison.

\subsection{Robustness of the detection and uncertainties}

By integrating the emission line seen over the central 1000~\kms in the 3\,$\sigma$-clipped composite, 
we measure $\left< L_{\rm ESDLA}(\lya) \right> \simeq (0.6 \pm 0.1) \times 10^{42}$~erg\,s$^{-1}$, where 
the error is derived from the noise in the stacked spectrum. Using the median or the weighted 
composite provides very similar results, within less than 7\%.

\begin{figure}
\centering
\includegraphics[bb =  60 175 500 570,clip=,width=\hsize]{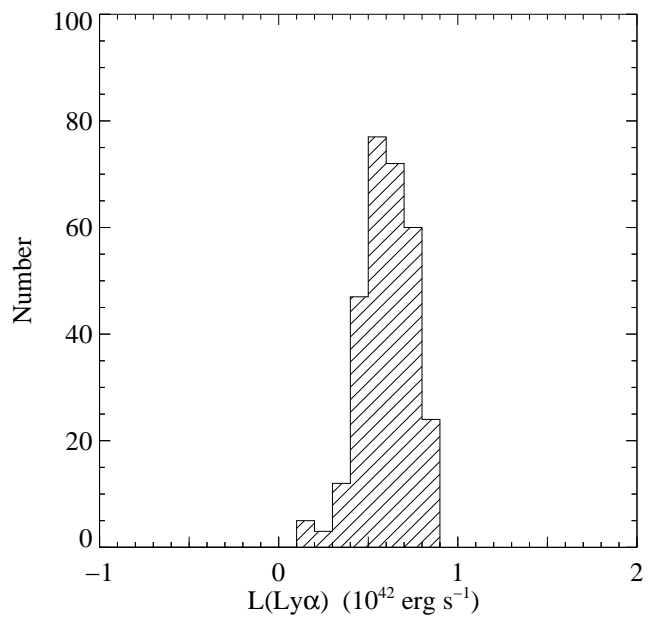}
\caption{Results from bootstrapping: Distribution of the measured mean \lya\ luminosities using 300 random subsamples 
with half the size of the original sample. \label{fig:boot}}
\end{figure}

We apply bootstrapping to further test the robustness of the detection and compare 
the statistical error obtained 
from the noise spectrum and from the data itself. We repeat the spectral stacking for 300 subsamples obtained by randomly keeping only half of the sample. The distribution of measured luminosities is shown on Fig.~\ref{fig:boot}. The distribution is clearly shifted from zero, centred 
at the same value as derived above with a standard deviation $\sigma = 0.15 \times 10^{42}$~\ergs. 
This implies a statistical error of $(0.15/\sqrt{2}) \times 10^{42}$~\ergs, which is in good agreement 
with that derived previously from the noise in the stacked spectrum.
Using different bootstrap sample sizes (keeping only a fraction $1/n$ of the total sample with $n \ge 2$) and scaling the error 
accordingly by $n^{-1/2}$ provides also very similar results.

All this shows that the detection of emission is robust. We caution however that the associated 
uncertainty represents only the statistical error and not possible systematics.

We indeed observe a non-zero emission on both sides of the \lya\ peak 
which, in addition to light from the DLA galaxy, could also result from residual 
sky light emission \citep{Paris12} or FUV light from the galaxy hosting the background QSO 
\citep[see][]{Cai14,Finley13,Zafar11}.
By coadding QSO spectra with high-redshift Lyman-breaks, \citeauthor{Cai14} put an upper-limit 
to the sky residual to $F_{\lambda}<3\times10^{-19}$~\ergsa\ in the wavelength range of interest for us, 
which could explain the continuum emission observed here. This translates to a contribution 
to the \lya\ luminosity of $L<0.25\times 10^{42}$~erg\,s$^{-1}$. FUV emission from the QSO host that 
would leak through the DLA galaxy because of non-unity covering factor would amount to about the same quantity. 

However, we observe that the continuum emission is not flat, decreasing from the line centre towards 
each side of the DLA core, with a possibly slightly higher emission on the red side, as seen in the LBG 
composite. This can result from averaging \lya\ emission lines with different shifts, as discussed  
by \citet{Rauch11} to explain the tilt in the core flux in previous studies \citep{Rahmani10}. In this 
case, the excess flux would also come from \lya\ photons and should be included. A contribution from 
stellar UV emission in the DLA host galaxy is also not excluded, although this is expected to be 
small. 

While the origin of the continuum emission is hard to establish, we note that subtracting the mean 
continuum observed in the composite spectrum before integrating the emission line results in a \lya\ 
luminosity $0.2\times 10^{42}$~erg\,s$^{-1}$ lower. In turn, integrating over a twice-wider velocity 
range, we obtain a $0.2\times 10^{42}$~erg\,s$^{-1}$ higher luminosity. This should be considered as 
very conservative upper-limit however. 
 
 In summary, we get 
$\left< L_{\rm ESDLA}(\lya) \right> \simeq (0.6 \pm 0.1 ({\rm stat}) \pm 0.2 ({\rm syst})) \times 10^{42}$~erg\,s$^{-1}$. 
We caution however that this result depends on the absolute flux-calibration. Comparison 
with photometric data shows these are usually of the order of 5\% \citep[][Schlegel et al., in prep.]{Dawson13} 
and can thus be neglected compared to the uncertainties discussed above. While larger flux-calibration 
errors are expected in a few cases, our statistical measurement is insensitive to possible outliers.

\subsection{Comparison with emission-selected \lya\ emitters}

If ESDLAs truly probe the population of emission-selected Lyman-$\alpha$ emitting galaxies,  
then we can expect the \lya\ luminosity distribution in ESDLAs hosts to follow the LAE luminosity function, 
which is well described by a Schechter function \citep{Schechter76}:

\begin{equation}
\Phi(L)dL = \Phi^{\star}(L/L^{\star})^{\alpha} e^{-L/L^{\star}} d(L/L^{\star})
\end{equation}

We use parameters derived at $z\sim 2-3$ from the VIMOS VLT Deep Survey 
\citep{Cassata11} ($\Phi^{\star}=7.1^{+2.4}_{-1.8} \times 10^{-4}$~Mpc$^{-3}$,  
$L^{\star}=5 \times10^{42}$~erg\,s$^{-1}$ and $\alpha=-1.6$), which probe the 
faint end of the luminosity 
function down to $L$(\lya) $\sim 10^{41}$~\ergs. We note that although 
the Subaru/XMM-Newton Deep Survey does not reach such faint luminosities, using 
the corresponding parameters \citep{Ouchi08} does not significantly change our 
results. 
The average luminosity of LAEs is then given by

\begin{equation}
\avg{L_{\rm LAE}} = {{\int_{L_{\rm min}}^\infty\!L \Phi(L) dL} \over {\int_{L_{\rm min}}^\infty\!\Phi(L) dL}},
\end{equation}

\begin{figure}
\includegraphics[bb=60 175 500 400,width=\hsize]{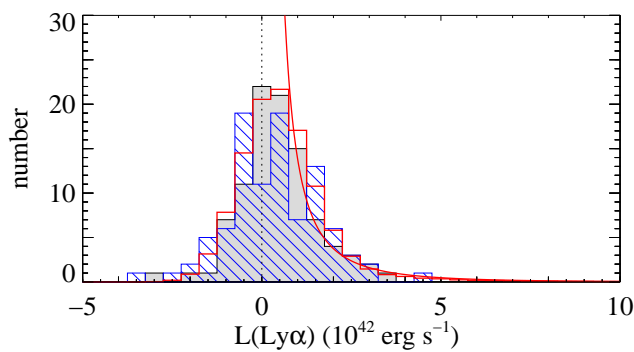}
\caption{Distribution of \lya\ luminosities estimated in individual ESDLAs (grey filled histogram). The blue hashed histogram 
represents the distribution after subtracting the mean luminosity observed on both sides of the \lya\ region. The solid 
red curve represents the LAE luminosity function from \citet{Cassata11} (arbitrary scaling), and the red unfilled histogram gives the 
expected distribution of $L \ge 10^{41}$~\ergs LAEs when Gaussian noise is added to mimic measurement uncertainties.
 \label{fig:ltest}}
\end{figure}

We find that the average luminosity of ESDLAs matches that of LAEs for $L_{\rm min} \sim 10^{41}~\ergs$. 
 In order to further 
test whether the \lya\ luminosities\footnote{To simplify the writing, here and in the following, ``$L$'' implicitly stands 
for \lya\ luminosity, i.e. ``$L(\lya)$''} of ESDLA hosts follows that of LAEs with $L> 10^{41} \ergs$, 
we estimate $L$ in each individual ESDLA by integrating the luminosity within $\Delta v = \pm 300$~\kms\ 
from the systemic redshift. The corresponding luminosity distribution is shown as grey histogram 
in Fig.~\ref{fig:ltest}. 
As expected due to the large uncertainties in individual measurements, the distribution is wide 
and nearly Gaussian. However, we do find that the mean is offset from zero and observe a tail at positive 
values. We note that the distribution does not change significantly (but gets noisier) when we subtract 
the mean luminosity per unit 
wavelength on both sides of the \lya\ region ($\Delta v = \pm [700-1000]~\kms$ from the systemic redshift) before 
integrating over the central region (blue hashed histogram). 
This again shows that a possible systematic positive zero-flux offset has no significant 
effect on the results.
We perform a Monte-Carlo analysis by generating a population 
of 100,000 LAEs with $L > 10^{41} \ergs$ that follows the \citet{Cassata11} luminosity function (solid red curve 
in Fig.~\ref{fig:ltest}). We randomly 
added Gaussian noise with $\sigma = 0.9 \times 10^{42}$~\ergs\ to mimic typical measurement uncertainties\footnote{This corresponds to the $0.09 \times 10^{42}~\ergs$ statistical uncertainty on the mean, scaled up by the square root of the number of systems used for the stack.}.
The resulting 
distribution (red unfilled histogram) follows the observed one very well. LAEs with $L> 5\times 10^{42} \ergs$ 
are expected to occur about a hundred times less frequently than those with $L> 10^{41} \ergs$, consistently with the fact that only 
one ESDLA in our sample has a \lya\ line with $L> 5\times 10^{42} \ergs$ (see Fig.~\ref{fig:j1135}).
{The \lya\ luminosity measured from follow-up observations on Magellan/MagE 
and VLT/X-shooter of this system is $L \sim 6\times 10^{42} \ergs$ \citep{Noterdaeme12a}.}. 

\subsection{\lya\ profile}

As it is subject to resonant scattering, the \lya\ line has a complex structure in absorption and emission. 
Often
the two are kinetically displaced. 
Where possible, \citet{Shapley03} measured a mean difference of about 650~\kms\ between 
the absorption from metal lines and the peak of the \lya\ emission in LBGs, which suggests that large-scale outflows are common in these objects. 
In the case of LAEs, spectroscopic observations indicate a velocity difference of about 150~\kms\ between 
\lya\ in emission and non-resonant lines \citep{McLinden11, Finkelstein11, Hashimoto13}, i.e. much less than the 
400~\kms\ derived for LBGs \citep{Steidel10,Shapley03}.

\begin{figure}
\centering
\begin{tabular}{c}
\includegraphics[bb=60 175 500 395,width=\hsize,clip=]{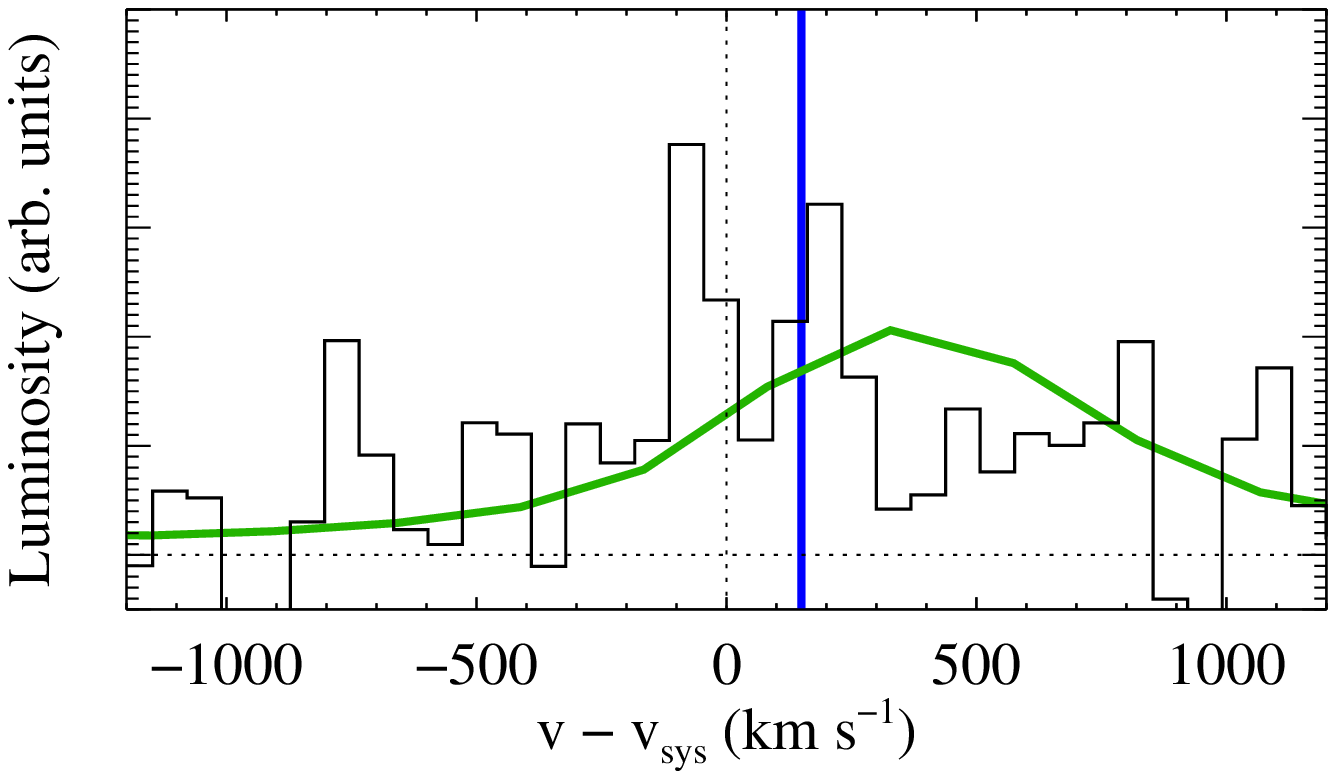}\\
\includegraphics[bb=60 175 500 395,width=\hsize]{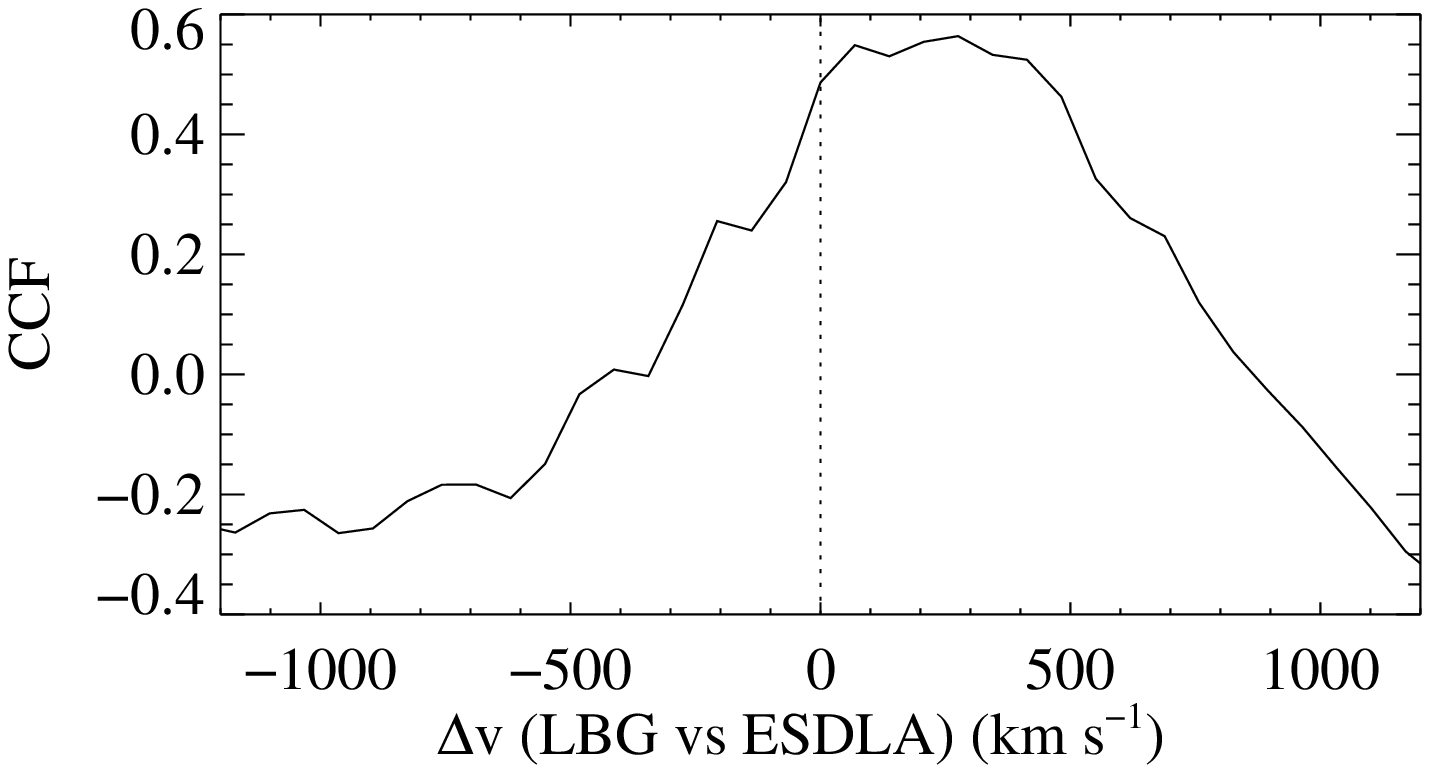}\\
\end{tabular}
\caption{{\sl Top}: Velocity offset of the \lya\ profile compared to the systemic redshift. The black histogram 
shows the stacked ESDLA spectrum, the green line represents the LBG composite \citep[][scaled down for illustration 
purposes]{Shapley03} and the vertical blue line marks the typical velocity shift observed in LAEs 
\citep{McLinden11, Finkelstein11, Hashimoto13}.
{\sl Bottom}: Cross-correlation function obtained by correlating the Lyman-break galaxy and the ESDLA (3,$\sigma$-clipped mean luminosity) composites at different velocities.
The \lya\ emission in the LBG composite is redshifted compared to that of ESDLAs by about 250~\kms. The 
ESDLA \lya\ emission redshift is much closer to the systemic redshift from the low-ionisation metal lines. 
\label{dv_esdla_lbg}.
}
\end{figure}

In the present case, our spectra do not cover NIR emission lines. However, because the line of sight passes 
through the galaxy, the systemic redshift can be well estimated from the absorption lines, unlike LBGs 
that have ISM absorption lines arising only from the gas located between the central star-forming region and the observer. 
For each system in our sample, we carefully measured the redshift from 
low-ionisation metal lines, cross-correlating the spectrum with a metal template to derive a first guess. 
We estimate the accuracy to be better than the pixel size, i.e., a few 10~\kms.
This shows that the \lya\ profile is slightly shifted toward the red relative to the systemic redshift 
(see top panel of Fig.~\ref{dv_esdla_lbg}). This shift is comparable to what is seen in LAEs and 
significantly ($\sim 250$~\kms) less than what is typical of LBGs (bottom panel of Fig.~\ref{dv_esdla_lbg}). 
There is also a hint of a double or multiple peaked profile \citep[as described in][]{Kulas12}, that 
is seen in the three different composite spectra, but the S/N achieved is still too low to be conclusive. 

It is initially somewhat surprising that such high \HI\ column densities 
do not introduce larger velocity offsets, as the \lya\ photons need to scatter away from the line centre 
to escape \citep[e.g.][]{Neufeld90}. However, the column density measured along the line of sight is 
the {\sl total} column density, and an inhomogeneous ISM could decrease the number of scattering 
required by \lya\ photons to escape the medium \citep[e.g.][]{Finkelstein11}. 
Indeed, \lya\ transfer is a complex process that depends on ISM clumpiness, kinematics, 
dust attenuation and geometry \citep[e.g.][]{Haiman99}. In the case of \jonze, an escape fraction as high as 0.20 
is observed, with a double-peaked profile and little velocity offset, in spite of a very high associated
\HI\ column density. 
In this particular case, the model that successfully reproduces all observable includes anisotropic 
galactic winds and distributes the total column density across numerous low-$N(\HI)$ clouds.
However, broadly speaking, the small shift of the \lya\ profile in our stacked spectrum indicates 
kinematic fields with velocities relatively low compared to LBGs and probably 
lower mass systems on average \citep{Zheng10}.

\subsection{Incidences of ESDLAs and LAEs}

In the previous subsections, we found that the distribution, mean value, and shape of the \lya\ 
emission agree with that of emission-selected LAEs with $L>10^{41}$~\ergs. Here, we compare 
the incidence of this population of LAEs with that of ESDLAs along the QSO lines of sight.

Integrating  \fhi\ over $N(\HI) \ge 0.5\times10^{22}$~\cmsq, we derive the number of ESDLA per unit 
absorption distance $dN_{\rm ESDLA}/d\chi \approx 7\times 10^{-4}$. 
Using 

\begin{equation}
{{d\chi} \over {dz}} \equiv {{(1+z)^2} \over {E(z)}}, 
\end{equation}

\noindent where $E(z)=\sqrt{\Omega_{\rm m} (1+z)^3 + \Omega_{\rm k} (1+z)^2 + \Omega_{\Lambda}}$, and the co-moving distance per unit redshift, 

\begin{equation}
{{dl_c} \over {dz}} = {c \over {\Ho E(z)}},
\end{equation}

\noindent we get

\begin{equation}
{{dN_{\rm ESDLA}} \over {dl_c}} = {{dN_{\rm ESDLA}} \over {d\chi}} {\Ho (1+z)^2 \over c} \approx 2\times10^{-6}$~Mpc$^{-1}.
\label{eq:dNesdla}
\end{equation}

\noindent On the other hand, the co-moving incidence of LAEs that give rise to ESDLAs can be written as 

\begin{equation}
dN_{\rm LAE}(L > L_{\rm min})/dl_c = \sigma_{\rm gas} (1+z)^2 d\Omega \int_{L_{\rm min}}^\infty\!\Phi(L) dL,
\label{eq:dNlae}
\end{equation}

\noindent where $\sigma_{\rm gas} = \pi r_{\rm gas}^2$ and $r_{\rm gas}$ is the mean physical projected extent of the gas 
with $N(\HI) \ge 5\times10^{21}$~\cmsq. By equating Eq.~\ref{eq:dNesdla} and Eq.~\ref{eq:dNlae} 
at $\avg{z} = 2.5$, we derive $r_{\rm gas} \approx $2.5~kpc. In other words, the expected number density of 
LAEs within an impact parameter of 2.5~kpc from the quasar lines of sight accounts for 
the incidence of ESDLAs, in very good agreement with our hypothesis. We also emphasise  
that this value is significantly smaller than the SDSS fibre radius (1~$\arcsec$ corresponding 
to $\sim$8~kpc at $z \sim 2.5$), indicating that fibre losses are likely negligible and not a 
problem for our study\footnote{See discussions in \citet{Lopez12} about the dangers of 
associating a population of 
extended absorbers with emitting galaxies in aperture-limited surveys.}. Finally, 
we note that if our measured \lya\ luminosity is overestimated due to zero-flux offset, 
then the luminosity function should be integrated down to lower $L_{\rm min}$, and hence 
we would derive an even smaller high-column density gas radius (or equivalently impact 
parameter).

\subsection{Star formation rate}

Assuming case B recombination, the \lya\ to H\,$\alpha$ ratio is theoretically 8.7. 
Since this does not take into account dust and escape fraction corrections, the \lya\ luminosity 
provides a lower-limit on the star-formation rate. Using the \Ha-SFR calibration from \citet{Kennicutt98b}, 
\begin{equation}
{\rm SFR} (\msyr)~= 7.9~L({\rm H}\,\alpha)~(10^{42}~\ergs) , 
\end{equation}
we get SFR (\msyr)~$=0.9~L($\lya$)/f_{\rm esc}$, where 
$L($\lya) is the observed \lya\ luminosity in units of $10^{42}~\ergs$. This gives 

\begin{equation}
\avg{\rm SFR} (\msyr) \approx 0.6/f_{\rm esc} , 
\end{equation}

\noindent in good agreement with what is expected for most DLAs 
from cosmological hydrodynamic simulations in a standard cold dark matter model \citep{Cen12}. 
At low redshift, \citet{Kennicutt98a} derives the best fit relation:

\begin{equation}
\Sigma_{\rm SFR}~(\msyr\,{\rm kpc}^{-2}) = (2.5 \pm 0.7) \times 10^{-4} \left({\Sigma_{\rm gas}} \over {1\,{\rm M_{\odot}\,pc^{-2}}}\right)^{1.4 \pm 0.15}
\end{equation}

\noindent Applying this to the mean gas surface density, $N(\HI)\approx 10^{21.8}$~\cmsq\ in our sample, we expect 
$\Sigma_{\rm SFR} \approx 0.08~{\rm M_{\odot}\,yr^{-1}\,kpc^{-2}}$. 
From the integrated SFR derived above, we can write:

\begin{equation}
\Sigma_{\rm SFR} \pi r_{\rm gal}^2 = 0.6/f_{\rm esc}~\msyr,
\end{equation}

\noindent where $r_{\rm gal}$ is the radius of the system. 
In the limiting case, $f_{\rm esc} =1$, the expected surface star-formation rate will match the observed 
integrated SFR if it remains constant over an effective galaxy radius of $r_{\rm gal}\sim 1.5$\,kpc. In the real 
situation, the escape fraction will be smaller than unity. For the global population of high-$z$ galaxies, 
\citet{Hayes10} estimate $f_{\rm esc}=0.05$, which would imply $r_{\rm gal}\sim 6~$kpc, i.e., a significant 
fraction of the galaxy light could easily fall outside the BOSS fibre. However, the typical escape fraction for 
high-$z$ LAEs is significantly higher, about $f_{\rm esc}\approx 0.30$ \citep{Blanc11,Nakajima12}\footnote{The follow-up 
observations of the ESDLA towards \jonze\ indicate $f_{\rm esc} \sim 0.2$.}, implying 
$r_{\rm gal}\sim 2.5$\,kpc. 
While the $\sim$35\% uncertainty on the \lya\ luminosity propagates to 
17\% on $r_{gas}$, we caution that this remains a simple model ignoring projection effects 
and where the most important unknown remains the escape fraction. With these limitations in mind, we can 
say that our overall picture is satisfactorally consistent. Indeed, the emission radius 
($r_{\rm gal}$) matches the high column density radius estimated above ($r_{\rm gas}$), further indicating that ESDLAs 
do arise from lines of sight passing through the ISM which feeds star-formation. This result also matches expectations 
from hydrodynamical simulations \citep[e.g.][]{Altay13}.

\subsection{Dependence on dust}

In this section, we wish to study the effect of dust on the \lya\ emission as 
attenuation by dust is known to have a significant impact on the \lya\ escape fraction.
High-redshift LAEs, like ESDLAs have a generally low dust content, with 
E(B-V)~$\sim 0-0.07$ at $z\sim 3$ \citep{Nilsson07, Ono10}, 
although it evolves significantly from $z \sim 3$ to $z \sim 2$ \citep{Guaita11, Nakajima12}.
However, due to resonant scattering, the optical path \lya\ photons follow can be very different 
and longer than that of photons from the UV continuum or non-resonant lines like H$\alpha$ or H$\beta$. 
Consequently, even a small amount of dust can affect the \lya\ escape fraction. 

\citet{Blanc11} report a correlation between the \lya\ equivalent width and reddening in high-$z$ LAEs. 
Similarly, \citet{Atek14} observe a clear dependence of the \lya\ escape fraction   
on the dust extinction in nearby galaxies. 

 Moreover, we note that galaxies hosting high-metallicity, dust-rich DLAs generally 
have no detectable \lya\ emission, despite their high star-formation rates \citep{Fynbo13}.
As an additional difficulty, here, we have access only to the extinction along the QSO line of sight. Because 
the impact parameter is small, we can, however, expect this extinction to be similar to that in the star-forming region 
as evidenced in e.g. \citet[][]{Gupta13}. 

\begin{figure}
\centering
\begin{tabular}{cc}
\includegraphics[bb =  85 215 490 570,clip=,width=0.47\hsize]{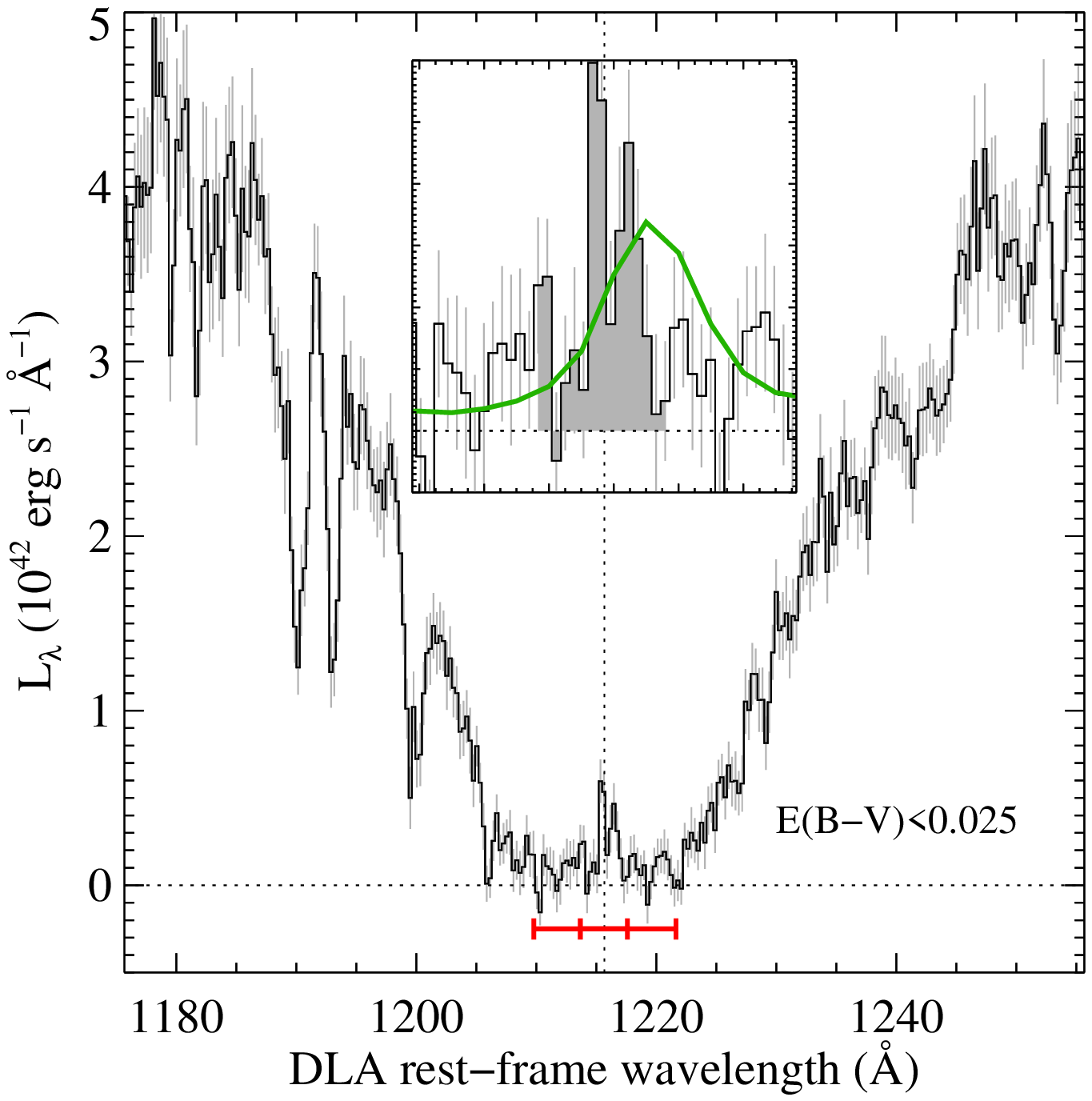} &
\includegraphics[bb =  85 215 490 570,clip=,width=0.47\hsize]{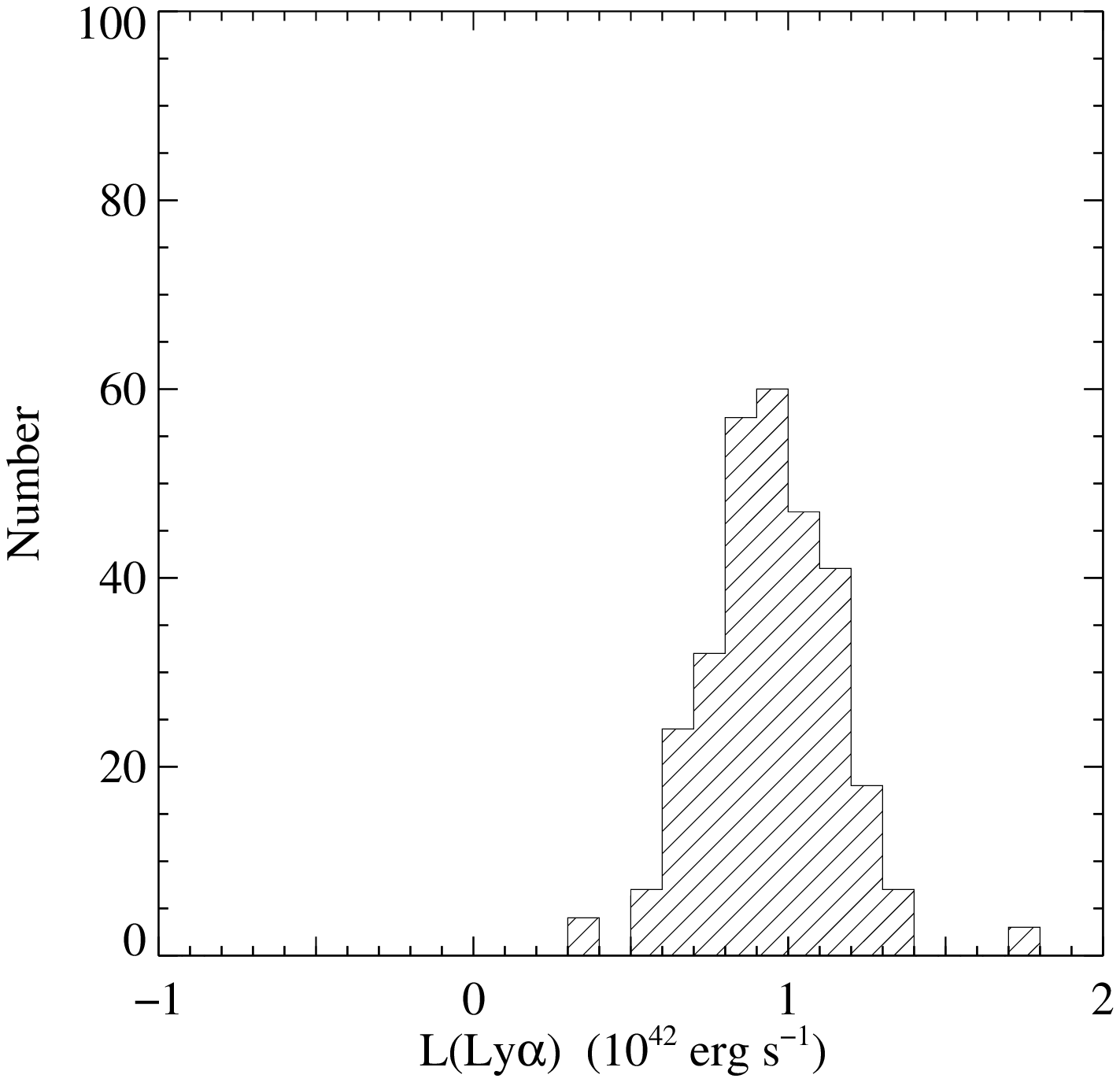} \\
\includegraphics[bb =  85 175 490 570,clip=,width=0.47\hsize]{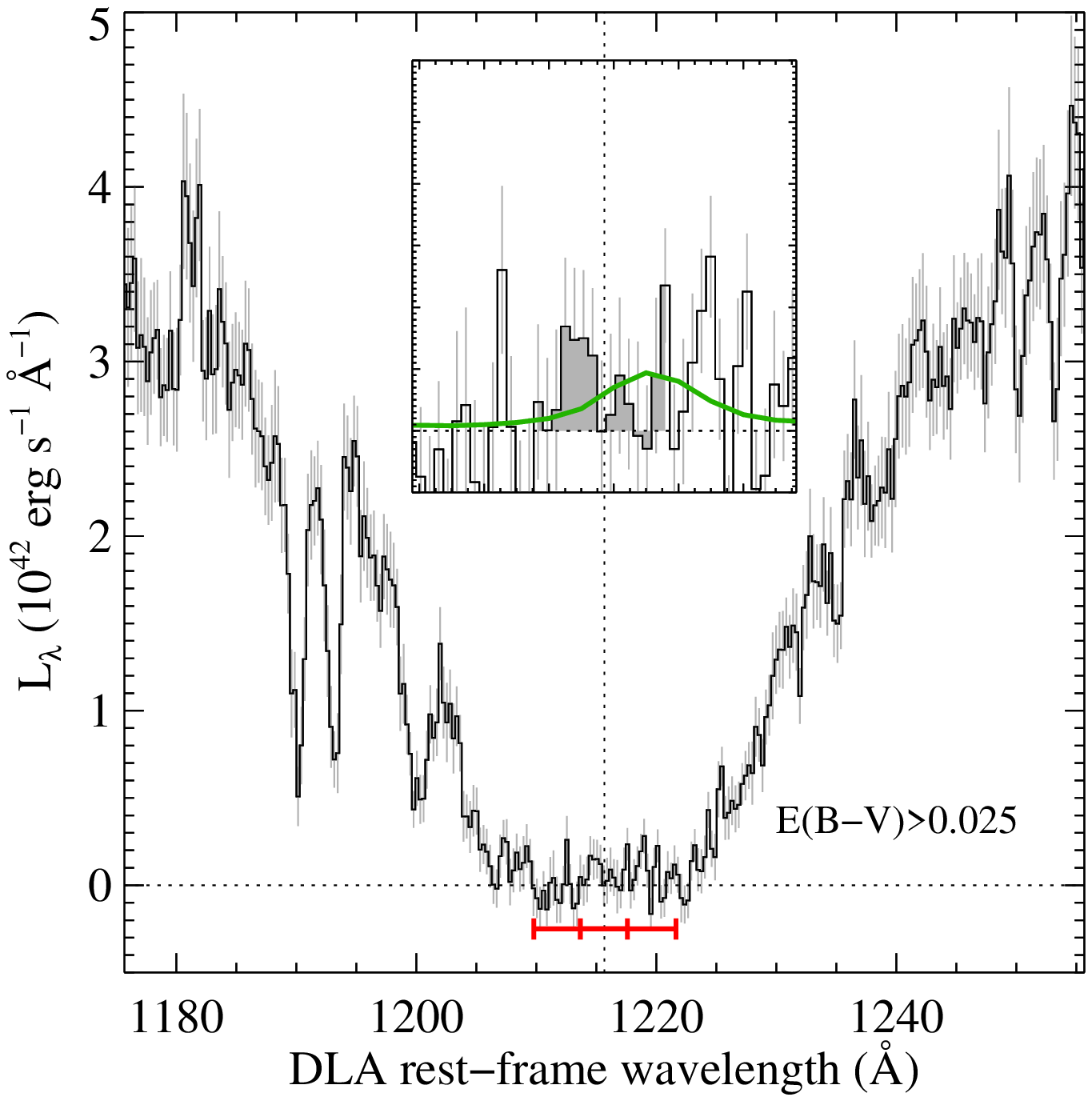} &
\includegraphics[bb =  85 175 490 570,clip=,width=0.47\hsize]{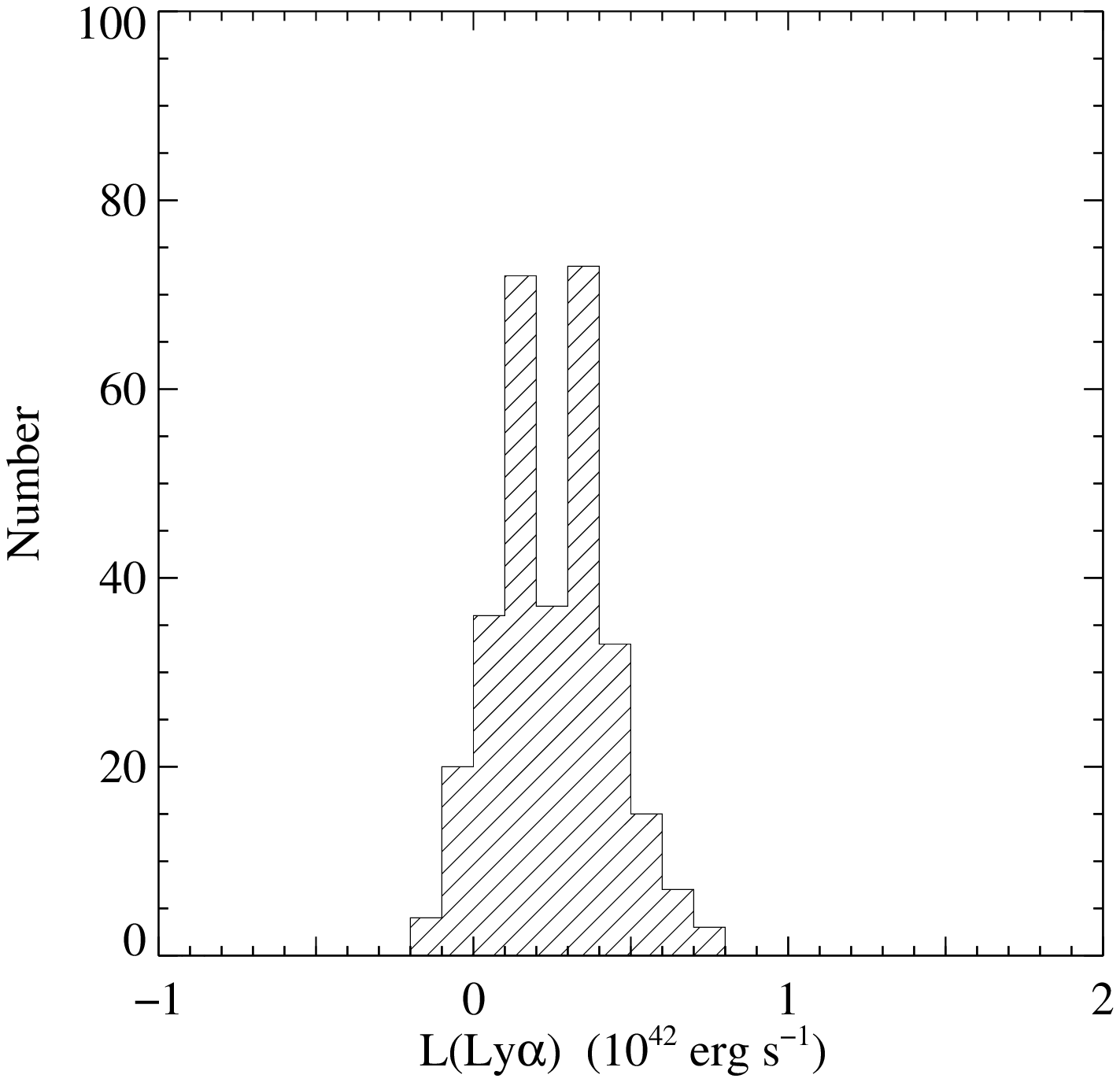} \\
\end{tabular}
\caption{Effect of dust on the \lya\ emission. The 3\,$\sigma$-clipped mean luminosity are represented on the left panels 
and the corresponding bootstrap analysis on the right panels. The top (resp. bottom) panels correspond to the ESDLA 
subsample with $E(B-V)<0.025$ (resp. $E(B-V) \ge 0.025$).
\label{fig:dust_lya}}
\end{figure}

We thus divide our sample into two subsamples with $E(B-V)$ below and above the median $E(B-V)$ value 
and perform two independent stacks with the accompanying bootstrap analysis (using random samples half the 
size of each subsample), see Fig.~\ref{fig:dust_lya}. The \lya\ luminosity is four times higher in the less dusty subsample, with
$\avg{L(\lya)}~\sim 10^{42}$~\ergs, compared to $\avg{L(\lya)}~\sim 0.25 \times 10^{42}$~\ergs\ in the 
dustier subsample. 
This is consistent with larger 
escape fractions for less dusty systems.

\section{Conclusion}

The historical $N(\HI)$-threshold for DLAs, $\log N(\HI) \ge 20.3$, was originally chosen 
mostly for observational reasons \citep{Wolfe86} and 
was found similar to \HI\ column densities measured in the disks of local spiral galaxies. 
However, it is becoming 
more and more clear that a significant fraction of DLAs at high redshift probe gas on the outskirts of a galaxy. 
Recent 21-cm observations of nearby galaxies have shown that higher \HI\ column 
densities are mostly found at small impact parameters \citep[$\sim$~80\% probability of being located at less 
than 5~kpc for $\log N(\HI)\ge 21.7$,][]{Zwaan05}. At high redshift, simulations also indicate that only the 
highest column density absorptions probe ISM gas that feeds star formation \citep[e.g.][]{Altay13,Rahmati13,Rahmati14}.
Here, we have studied an elusive population of extremely strong DLAs detected in BOSS. 
The small incidence of these systems reflects their small 
cross-section. Confirming our previous result in \citep[see][]{Noterdaeme12c}, the high column density end 
of the $N(\HI)$-distribution function has a moderate power-law slope, similar to that of the local Universe. 
We find that the metallicities and dust-depletion of ESDLAs are similar to those of the overall DLA 
population and thus indicate that they are related to a similar population of galaxies. Their higher column densities are mainly the consequence of small impact parameters.
Indeed, we found that the absorption characteristics are very similar to what is seen in DLAs 
associated with GRB afterglows, which are known to be intimately related to star-formation regions. 

Using stacking techniques, we detect \lya\ emission in the core of ESDLAs with 
a mean luminosity    
$\avg{L_{Ly\,\alpha}} \simeq (0.6 \pm 0.1 ({\rm stat}) \pm 0.2 ({\rm syst})) \times 10^{42}\,\ergs$ 
which corresponds to that of $L_{Ly\,\alpha} \ge 10^{41}$~\ergs\ Lyman-$\alpha$ 
emitting galaxies. We also show that the distribution of luminosities measured 
in individual spectra, although noisy, is also consistent with that of the above LAE 
population.

The incidences 
of ESDLAs and LAEs indicates impact parameters $b <2.5$~kpc.
The properties of the Ly$\alpha$ emission 
in both populations are very similar. All of this strongly suggests that the ESDLA host galaxies 
are actually LAEs that emit most of their light well within the area 
covered by the the BOSS fibre (8~kpc radius) and obey the Schmidt-Kennicutt law. 
We caution however that the measured \lya\ luminosity may be overestimated by $\sim 35\%$ 
due to sky light residuals and/or FUV emission from the QSO host and that we have neglected 
flux-calibration uncertainties. However, this has little consequence on our 
overall picture. 
Indeed, a lower \lya\ luminosity would imply a fainter but more numerous LAE population (hence a 
smaller extent of gas to match the incidence of ESDLAs) and at the same time a smaller galactic size 
according to the Schmidt-Kennicutt law.

\citet{Hashimoto13} recently suggested that LAEs should have small neutral hydrogen 
column densities. However, this suggestion arises from considerations based 
on homogeneous expanding shell models \citep{Verhamme06}, while the true configuration is probably 
much more complicated \citep[e.g.][]{Kulas12}. Indeed, recent works have highlighted the 
importance of ISM clumpiness and geometry in allowing \lya\ photons to escape from  
star-forming regions \citep[e.g.][]{Laursen13}, even at high integrated \HI\ column densities 
\citep[e.g.][]{Noterdaeme12a}. Finally we note that the viewing angle seems to play an 
important role in anisotropic configurations \citep{Zheng13}.

Interestingly, using very deep (92\,h of VLT/FORS2) long-slit spectroscopy, \citet{Rauch08}  
revealed a population of faint LAEs with $L$(\lya)~$\sim 10^{41}$~\ergs\ that have a total 
cross-section consistent with that of DLAs \citep[see also][]{Barnes09}. 
Targeting high-metallicity DLAs has successfully produced a number of host galaxy detections with 
higher SFR, but the host galaxies are frequently at large impact parameters and either have no \lya\ emission 
or a suppressed blue peak \citep[see][]{Fynbo10,Fynbo11,Fynbo13,Krogager12,Krogager13}.
This indicates that high metallicity DLAs could be associated with massive and luminous galaxies, but 
their cross-section selection increases the probability that the DLAs will probe the galaxy outskirts. 
This is in line with other studies suggesting that the large cross-section of gas around massive galaxies 
is responsible for higher metallicities in sub-DLAs on average \citep{Khare07,Kulkarni10}.
ESDLAs, however, are selected solely on the basis of high \HI\ column densities. They should arise in more 
typical galaxies that have not yet converted their gas reservoirs into stars and thereby produced little 
metals, as seen from the low metallicities. 

Follow-up studies of ESDLAs and their host galaxies will contribute important clues for understanding  
galaxy formation at high redshift and constrain crucial parameters for numerical simulations such as 
the amount of stellar feedback and the gas consumption rate. In particular, deep multi-wavelength 
spectroscopy, covering both \lya\ and nebular emission lines (redshifted in the near-infrared) 
are required to measure accurately the star-formation rate and hence the \lya\ escape fraction as well 
as bringing constraints on the \lya\ transfer.

\begin{acknowledgements}

We thank the referee for careful reading as well as constructive comments 
  and suggestions and Susanna Vergani for helpful discussions.
The SDSS-III French participation group was supported by the Agence Nationale de la
Recherche under grants ANR-08-BLAN-0222 and ANR-12-BS05-0015.

Funding for SDSS-III has been provided by the Alfred P. Sloan Foundation, the Participating 
Institutions, the National Science Foundation, and the U.S. Department of Energy Office of Science. 
The SDSS-III web site is \url{http://www.sdss3.org/}.

SDSS-III is managed by the Astrophysical Research Consortium for the Participating Institutions 
of the SDSS-III Collaboration including the University of Arizona, the Brazilian Participation 
Group, Brookhaven National Laboratory, University of Cambridge, Carnegie Mellon University, 
University of Florida, the French Participation Group, the German Participation Group, Harvard 
University, the Instituto de Astrof\'isica de Canarias, the Michigan State/Notre Dame/JINA Participation 
Group, Johns Hopkins University, Lawrence Berkeley National Laboratory, Max Planck Institute for 
Astrophysics, Max Planck Institute for Extraterrestrial Physics, New Mexico State University, New 
York University, Ohio State University, Pennsylvania State University, University of Portsmouth, 
Princeton University, the Spanish Participation Group, University of Tokyo, University of Utah, 
Vanderbilt University, University of Virginia, University of Washington, and Yale University.

\end{acknowledgements}

\bibliographystyle{aa}

\onecolumn
\small
\begin{longtab}
\begin{longtable}{c c c c c c c c}
\caption{Intervening ESDLA sample \label{tab:esdla}}\tabularnewline
\hline \hline 
{\large \strut} QSO   &MJD-plate-fiber &$z_{\rm QSO}$ &CNR &$z_{\rm abs}$ & $\log N(\HI)$ &$\sigma_{dark}$ \tablefootmark{a}       & used in stack? \tabularnewline
{\large \strut} J2000 &                &            &    &           &[cm$^{-2}$]      & [$10^{42}$~erg\,s$^{-1}$\,{\AA}$^{-1}$] & \tabularnewline
\hline
\endfirsthead
\caption{continued.}\tabularnewline
\hline \hline
{\large \strut} QSO   &MJD-plate-fiber &$z_{\rm QSO}$ &CNR &$z_{\rm abs}$ & $\log N(\HI)$ &$\sigma_{dark}$ \tablefootmark{a}       & used in stack? \tabularnewline
{\large \strut} J2000 &                &            &    &           &[cm$^{-2}$]      & [$10^{42}$~erg\,s$^{-1}$\,{\AA}$^{-1}$] & \tabularnewline
\hline
\endhead
\hline
\endfoot
   J\,001743.88+130739.84 &56267-6184-0444 &2.594 & 11.8 &2.326 &21.70 & 0.61 &y  \tabularnewline
   J\,002503.03+114547.80 &56237-6189-0912 &2.961 &  7.5 &2.304 &21.75 & 0.70 &y  \tabularnewline
   J\,004349.39-025401.91 &55534-4370-0422 &2.956 & 15.7 &2.013 &22.12 & 1.14 &y  \tabularnewline
   J\,004810.37+213818.32 &56217-6200-0644 &3.232 &  4.0 &2.941 &21.92 & 0.89 &y  \tabularnewline
   J\,004953.46+012217.40 &55584-4306-0208 &2.651 &  1.8 &2.433 &21.70 & 0.72 &n  \tabularnewline
   J\,005954.31+045941.27 &55531-4307-0945 &2.754 &  4.2 &2.205 &21.79 & 0.98 &y  \tabularnewline
   J\,010029.31+290100.44 &56274-6257-0084 &2.544 &  3.1 &2.157 &21.75 & 0.62 &y  \tabularnewline
   J\,010153.03+335746.05 &56270-6593-0706 &2.670 & 16.3 &2.424 &22.22 & 0.56 &y  \tabularnewline
   J\,014005.40-010333.38 &55444-4231-0328 &3.919 &  8.8 &3.631 &21.95 & 1.28 &n  \tabularnewline
   J\,014858.31+141235.41 &55591-4657-0502 &3.178 &  8.2 &2.874 &21.70 & 1.20 &y  \tabularnewline
   J\,015445.22+193515.89 &55925-5117-0113 &2.530 &  5.5 &2.252 &21.77 & 0.91 &y  \tabularnewline
   J\,022759.79+000947.46 &55455-4238-0700 &2.675 &  3.7 &2.261 &21.77 & 0.60 &y  \tabularnewline
   J\,023011.30-033450.07 &55540-4386-0544 &2.872 & 13.1 &2.503 &21.84 & 0.81 &y  \tabularnewline
   J\,072059.38+391955.97 &55240-3655-0314 &3.783 &  9.4 &2.740 &21.77 & 0.70 &y  \tabularnewline
   J\,074344.26+142134.89 &55564-4497-0283 &2.281 & 12.3 &2.045 &21.93 & 1.46 &y  \tabularnewline
   J\,074700.26+345301.65 &55234-3751-0665 &3.216 & 14.1 &2.258 &21.83 & 0.82 &y  \tabularnewline
   J\,074815.54+225838.03 &55589-4473-0634 &3.189 &  6.6 &2.394 &21.90 & 0.79 &y  \tabularnewline
   J\,075330.07+252000.88 &55533-4459-0212 &2.527 &  3.5 &2.464 &21.75 & 0.92 &y  \tabularnewline
   J\,081206.74+105738.85 &55574-4509-0394 &3.328 &  4.1 &2.945 &21.80 & 0.83 &y  \tabularnewline
   J\,081634.39+144612.47 &55571-4504-0748 &3.846 & 10.5 &3.287 &22.01 & 1.49 &y  \tabularnewline
   J\,082532.46+424033.14 &55511-3807-0064 &2.755 &  2.3 &2.428 &21.80 & 0.88 &y  \tabularnewline
   J\,084312.72+022117.35 &56015-3810-0725 &2.917 & 10.0 &2.787 &21.80 & 0.88 &y  \tabularnewline
   J\,084533.05-000919.81 &55513-3812-0494 &3.229 &  8.9 &2.300 &21.72 & 0.70 &y  \tabularnewline
   J\,084646.06+211257.36 &56245-5177-0428 &3.507 &  3.1 &3.266 &21.85 & 0.98 &y  \tabularnewline
   J\,085201.02+050659.23 &55924-4867-0254 &2.150 &  4.6 &2.046 &22.27 & 1.25 &y  \tabularnewline
   J\,090203.10+222732.20 &56010-5776-0334 &2.543 & 15.0 &2.376 &21.95 & 0.56 &y  \tabularnewline
   J\,090227.15+411753.73 &55983-4604-0026 &2.797 &  6.5 &2.389 &21.85 & 0.83 &y  \tabularnewline
   J\,091334.76+164506.31 &55987-5301-0154 &3.071 &  6.3 &2.649 &21.72 & 1.01 &y  \tabularnewline
   J\,092233.41+395518.54 &55947-4641-0058 &2.639 &  3.7 &2.441 &21.79 & 0.86 &y  \tabularnewline
   J\,092515.06+071354.11 &55929-5309-0362 &3.166 &  3.4 &3.027 &22.14 & 0.81 &y  \tabularnewline
   J\,093033.36+202311.74 &56245-5767-0950 &2.163 &  3.6 &2.041 &21.70 & 1.19 &y  \tabularnewline
   J\,093653.82+040600.92 &55662-4797-0273 &2.658 &  8.4 &2.180 &21.71 & 0.96 &y  \tabularnewline
   J\,094849.87+272435.03 &56298-5795-0016 &2.925 &  6.1 &2.453 &21.75 & 1.15 &y  \tabularnewline
   J\,100645.60+462717.36 &56364-6662-0500 &4.440 &  2.9 &4.255 &21.95 & 3.40 &n  \tabularnewline
   J\,102420.45+370321.74 &55570-4564-0742 &2.420 &  4.6 &2.212 &21.74 & 0.92 &y  \tabularnewline
   J\,102657.74+222220.91 &56272-6424-0069 &3.024 & 38.5 &2.688 &21.75 & 0.83 &y  \tabularnewline
   J\,103035.67+445033.49 &55651-4691-0692 &2.846 & 10.3 &2.749 &21.70 & 0.77 &y  \tabularnewline
   J\,103049.13+262926.80 &56330-6457-0426 &3.139 & 12.4 &2.719 &22.05 & 0.89 &y  \tabularnewline
   J\,103508.64+175306.06 &56036-5885-0254 &2.481 &  3.6 &2.353 &21.70 & 0.72 &y  \tabularnewline
   J\,103729.89+010711.60 &55290-3833-0832 &3.181 &  6.6 &2.861 &21.75 & 0.75 &y  \tabularnewline
   J\,104054.61+250709.50 &56358-6439-0160 &2.735 &  6.8 &2.240 &21.70 & 0.61 &y  \tabularnewline
   J\,104629.64-025114.59 &55563-3786-0332 &2.391 &  6.1 &2.131 &21.73 & 0.80 &y  \tabularnewline
   J\,104803.88+184350.10 &56038-5875-0326 &3.366 &  2.6 &3.014 &21.80 & 1.05 &y  \tabularnewline
   J\,111252.25+375910.25 &55629-4622-0704 &3.999 &  7.1 &3.821 &21.70 & 3.22 &n  \tabularnewline
   J\,111743.20+124554.99 &55945-5365-0044 &2.438 &  4.4 &2.075 &21.70 & 1.21 &y  \tabularnewline
   J\,112444.88+100235.60 &55976-5371-0708 &3.236 &  3.8 &2.600 &21.70 & 0.67 &y  \tabularnewline
   J\,113421.08+035200.86 &55944-4768-0027 &4.150 &  7.3 &2.993 &21.75 & 0.96 &y  \tabularnewline
   J\,113520.40-001053.56 &55574-3840-0148 &2.915 & 26.5 &2.207 &22.07 & 0.58 &n  \tabularnewline
   J\,113959.21+221930.96 &56311-6431-0978 &3.055 &  2.8 &2.945 &21.70 & 0.96 &y  \tabularnewline
   J\,114252.88+322619.58 &55617-4616-0046 &3.058 &  2.3 &2.652 &21.80 & 0.80 &y  \tabularnewline
   J\,114347.21+142021.60 &56009-5381-0604 &2.583 &  3.0 &2.324 &21.90 & 0.61 &y  \tabularnewline
   J\,114638.95+074311.29 &55982-5382-0482 &3.030 &  6.7 &2.840 &21.80 & 1.12 &y  \tabularnewline
   J\,120705.64+031637.16 &55631-4748-0756 &2.910 &  2.3 &2.722 &21.71 & 0.82 &y  \tabularnewline
   J\,120716.58+221117.40 &56067-5973-0794 &3.530 &  7.9 &3.133 &21.80 & 0.88 &y  \tabularnewline
   J\,122923.90+373128.89 &55302-3965-0894 &2.988 &  3.8 &2.820 &21.85 & 0.86 &y  \tabularnewline
   J\,123248.44+070830.78 &55927-5402-0140 &2.234 &  8.8 &1.984 &21.86 & 1.30 &y  \tabularnewline
   J\,123816.04+162042.47 &56013-5404-0984 &3.451 & 11.9 &3.208 &21.72 & 1.16 &y  \tabularnewline
   J\,124817.38+010120.52 &55274-3849-0942 &2.975 &  2.8 &2.825 &21.75 & 0.85 &y  \tabularnewline
   J\,125336.36-022807.81 &55222-3779-0140 &4.007 & 22.4 &2.783 &21.80 & 0.77 &y  \tabularnewline
   J\,125855.41+121250.21 &55983-5419-0568 &3.055 & 12.6 &2.444 &21.98 & 0.62 &y  \tabularnewline
   J\,130150.97+460633.46 &56401-6618-0152 &3.014 &  4.7 &2.721 &21.75 & 1.06 &y  \tabularnewline
   J\,130504.55+405713.82 &55681-4704-0092 &2.980 &  5.8 &2.407 &21.71 & 0.85 &y  \tabularnewline
   J\,133707.41+305205.24 &56363-6496-0384 &2.663 &  3.3 &2.376 &21.70 & 0.85 &y  \tabularnewline
   J\,134508.82+365214.88 &55243-3852-0468 &2.288 &  5.2 &2.149 &22.18 & 0.80 &y  \tabularnewline
   J\,134910.45+044819.91 &55659-4785-0722 &3.353 &  5.1 &2.481 &21.70 & 0.79 &y  \tabularnewline
   J\,135316.83+095636.73 &55987-5445-0327 &3.614 & 16.0 &3.333 &21.73 & 1.37 &y  \tabularnewline
   J\,141120.51+122935.96 &56001-5453-0109 &2.713 & 11.2 &2.545 &21.83 & 0.87 &y  \tabularnewline
   J\,143047.09+060201.05 &55691-4860-0180 &4.108 &  7.0 &3.981 &21.90 & 3.12 &n  \tabularnewline
   J\,143107.52+342730.93 &55269-3860-0290 &4.280 &  7.7 &4.180 &21.81 & 2.13 &n  \tabularnewline
   J\,143121.31+401544.04 &56038-5171-0652 &3.291 &  7.1 &2.509 &21.80 & 0.64 &y  \tabularnewline
   J\,143703.74+315742.10 &55360-3868-0656 &3.908 &  4.8 &3.326 &21.81 & 1.04 &y  \tabularnewline
   J\,143725.16+173444.41 &56037-5469-0644 &3.200 &  6.9 &2.892 &22.05 & 0.99 &y  \tabularnewline
   J\,143746.91+501245.59 &56390-6725-0432 &2.528 & 12.3 &2.076 &21.81 & 1.23 &y  \tabularnewline
   J\,144250.63+385414.83 &56046-5173-0914 &3.370 &  7.0 &2.904 &22.02 & 0.99 &y  \tabularnewline
   J\,145258.80+252554.51 &56088-6024-0358 &3.767 & 14.1 &3.453 &21.72 & 1.40 &y  \tabularnewline
   J\,145646.48+160939.32 &56030-5477-0974 &3.683 & 18.3 &3.352 &21.85 & 1.76 &y  \tabularnewline
   J\,150731.89+435429.69 &56072-6048-0548 &4.126 &  3.3 &3.616 &22.10 & 1.84 &n  \tabularnewline
   J\,151203.52+205547.67 &55656-3956-0827 &2.943 & 12.0 &2.230 &21.84 & 0.74 &y  \tabularnewline
   J\,151349.52+035211.68 &55652-4776-0060 &2.680 &  2.4 &2.464 &21.80 & 1.04 &y  \tabularnewline
   J\,153906.70+325007.62 &56033-4723-0950 &3.725 &  3.0 &3.308 &22.04 & 1.19 &y  \tabularnewline
   J\,154235.24+360003.36 &56038-4974-0940 &2.836 & 10.7 &2.303 &21.70 & 0.85 &y  \tabularnewline
   J\,155125.64+083545.73 &56003-5210-0792 &3.395 &  2.5 &3.083 &21.70 & 1.10 &y  \tabularnewline
   J\,155556.90+480015.14 &56425-6730-0850 &3.302 & 10.1 &2.391 &21.90 & 0.95 &y  \tabularnewline
   J\,160311.35+212158.41 &55335-3929-0492 &3.456 & 14.3 &2.414 &21.80 & 0.75 &y  \tabularnewline
   J\,161838.17+192110.44 &55365-4069-0944 &3.534 & 10.7 &2.514 &21.76 & 0.71 &y  \tabularnewline
   J\,162629.25+274921.07 &55706-5006-0904 &2.633 &  2.9 &2.314 &21.91 & 0.67 &y  \tabularnewline
   J\,162717.19+231932.00 &55450-4184-0808 &2.763 &  8.1 &2.110 &21.72 & 0.94 &y  \tabularnewline
   J\,165426.78+320602.29 &55723-4992-0478 &2.777 &  6.7 &2.652 &22.13 & 0.81 &y  \tabularnewline
   J\,165434.56+180751.47 &55682-4176-0188 &2.572 &  4.2 &2.170 &21.80 & 0.76 &y  \tabularnewline
   J\,165645.24+300306.58 &55720-4996-0732 &2.834 &  5.0 &2.477 &21.80 & 0.94 &y  \tabularnewline
   J\,171102.04+313507.60 &55738-4997-0872 &2.721 &  4.5 &2.545 &21.82 & 0.94 &y  \tabularnewline
   J\,171200.18+262716.06 &55717-5014-0154 &3.156 &  9.8 &2.800 &21.78 & 0.81 &y  \tabularnewline
   J\,214043.02-032139.29 &55883-4374-0401 &2.479 &  5.6 &2.340 &22.35 & 0.82 &y  \tabularnewline
   J\,220525.56+102118.68 &55739-5065-0812 &3.414 &  8.3 &3.256 &21.70 & 1.15 &y  \tabularnewline
   J\,220536.70+242242.39 &56102-5951-0750 &2.890 &  1.9 &2.724 &21.80 & 0.77 &n  \tabularnewline
   J\,221122.53+133451.24 &55749-5041-0374 &3.071 &  8.6 &2.838 &21.81 & 0.77 &y  \tabularnewline
   J\,222338.81+070246.04 &56189-4428-0714 &3.798 &  2.5 &3.011 &21.88 & 0.78 &y  \tabularnewline
   J\,223250.98+124225.29 &56187-5043-0168 &2.299 & 15.6 &2.228 &21.70 & 0.75 &y  \tabularnewline
   J\,224327.99+220312.54 &56181-6119-0945 &3.315 &  3.7 &2.463 &21.95 & 0.76 &y  \tabularnewline
   J\,224621.14+132821.32 &56186-5044-0020 &2.514 & 15.1 &2.214 &21.70 & 0.86 &y  \tabularnewline
   J\,231624.74+213237.59 &56209-6114-0348 &3.327 &  3.3 &2.860 &21.75 & 0.66 &y  \tabularnewline
   J\,232207.30+003348.99 &55446-4211-0506 &2.693 &  5.1 &2.477 &21.75 & 0.95 &y  \tabularnewline
   J\,233035.50+005842.36 &55446-4211-0946 &2.703 &  4.9 &2.152 &21.70 & 0.81 &y  \tabularnewline
   J\,235854.43+014953.65 &55505-4278-0010 &3.194 & 11.6 &2.979 &21.72 & 0.91 &y  \tabularnewline
\hline
\end{longtable}
\tablefoot{
\tablefoottext{a}{1\,$\sigma$ noise level in the DLA trough (luminosity at DLA rest-frame)}}
\end{longtab}

\begin{figure*}
\centering
\setlength{\tabcolsep}{1pt}
\begin{tabular}{c c c c c c}
\includegraphics[bb=190 265 435 535,clip=,angle=90,width=0.145\hsize]{/home/pnoterda/Science/BIGDLAs/DR11/FITS/56267-6184-0444.ps}&
\includegraphics[bb=190 265 435 535,clip=,angle=90,width=0.145\hsize]{/home/pnoterda/Science/BIGDLAs/DR11/FITS/56237-6189-0912.ps}&
\includegraphics[bb=190 265 435 535,clip=,angle=90,width=0.145\hsize]{/home/pnoterda/Science/BIGDLAs/DR11/FITS/55534-4370-0422.ps}&
\includegraphics[bb=190 265 435 535,clip=,angle=90,width=0.145\hsize]{/home/pnoterda/Science/BIGDLAs/DR11/FITS/56217-6200-0644.ps}&
\includegraphics[bb=190 265 435 535,clip=,angle=90,width=0.145\hsize]{/home/pnoterda/Science/BIGDLAs/DR11/FITS/55584-4306-0208.ps}&
\includegraphics[bb=190 265 435 535,clip=,angle=90,width=0.145\hsize]{/home/pnoterda/Science/BIGDLAs/DR11/FITS/55531-4307-0945.ps}\\
\includegraphics[bb=190 265 435 535,clip=,angle=90,width=0.145\hsize]{/home/pnoterda/Science/BIGDLAs/DR11/FITS/56274-6257-0084.ps}&
\includegraphics[bb=190 265 435 535,clip=,angle=90,width=0.145\hsize]{/home/pnoterda/Science/BIGDLAs/DR11/FITS/56270-6593-0706.ps}& 
\includegraphics[bb=190 265 435 535,clip=,angle=90,width=0.145\hsize]{/home/pnoterda/Science/BIGDLAs/DR11/FITS/55444-4231-0328.ps}& 
\includegraphics[bb=190 265 435 535,clip=,angle=90,width=0.145\hsize]{/home/pnoterda/Science/BIGDLAs/DR11/FITS/55591-4657-0502.ps}& 
\includegraphics[bb=190 265 435 535,clip=,angle=90,width=0.145\hsize]{/home/pnoterda/Science/BIGDLAs/DR11/FITS/55925-5117-0113.ps}& 
\includegraphics[bb=190 265 435 535,clip=,angle=90,width=0.145\hsize]{/home/pnoterda/Science/BIGDLAs/DR11/FITS/55455-4238-0700.ps}\\
\includegraphics[bb=190 265 435 535,clip=,angle=90,width=0.145\hsize]{/home/pnoterda/Science/BIGDLAs/DR11/FITS/55540-4386-0544.ps}& 
\includegraphics[bb=190 265 435 535,clip=,angle=90,width=0.145\hsize]{/home/pnoterda/Science/BIGDLAs/DR11/FITS/55240-3655-0314.ps}&
\includegraphics[bb=190 265 435 535,clip=,angle=90,width=0.145\hsize]{/home/pnoterda/Science/BIGDLAs/DR11/FITS/55564-4497-0283.ps}&
\includegraphics[bb=190 265 435 535,clip=,angle=90,width=0.145\hsize]{/home/pnoterda/Science/BIGDLAs/DR11/FITS/55234-3751-0665.ps}& 
\includegraphics[bb=190 265 435 535,clip=,angle=90,width=0.145\hsize]{/home/pnoterda/Science/BIGDLAs/DR11/FITS/55589-4473-0634.ps}& 
\includegraphics[bb=190 265 435 535,clip=,angle=90,width=0.145\hsize]{/home/pnoterda/Science/BIGDLAs/DR11/FITS/55533-4459-0212.ps}\\
\includegraphics[bb=190 265 435 535,clip=,angle=90,width=0.145\hsize]{/home/pnoterda/Science/BIGDLAs/DR11/FITS/55574-4509-0394.ps}& 
\includegraphics[bb=190 265 435 535,clip=,angle=90,width=0.145\hsize]{/home/pnoterda/Science/BIGDLAs/DR11/FITS/55571-4504-0748.ps}& 
\includegraphics[bb=190 265 435 535,clip=,angle=90,width=0.145\hsize]{/home/pnoterda/Science/BIGDLAs/DR11/FITS/55511-3807-0064.ps}&
\includegraphics[bb=190 265 435 535,clip=,angle=90,width=0.145\hsize]{/home/pnoterda/Science/BIGDLAs/DR11/FITS/56015-3810-0725.ps}& 
\includegraphics[bb=190 265 435 535,clip=,angle=90,width=0.145\hsize]{/home/pnoterda/Science/BIGDLAs/DR11/FITS/55513-3812-0494.ps}& 
\includegraphics[bb=190 265 435 535,clip=,angle=90,width=0.145\hsize]{/home/pnoterda/Science/BIGDLAs/DR11/FITS/56245-5177-0428.ps}\\
\includegraphics[bb=190 265 435 535,clip=,angle=90,width=0.145\hsize]{/home/pnoterda/Science/BIGDLAs/DR11/FITS/55924-4867-0254.ps}& 
\includegraphics[bb=190 265 435 535,clip=,angle=90,width=0.145\hsize]{/home/pnoterda/Science/BIGDLAs/DR11/FITS/56010-5776-0334.ps}& 
\includegraphics[bb=190 265 435 535,clip=,angle=90,width=0.145\hsize]{/home/pnoterda/Science/BIGDLAs/DR11/FITS/55983-4604-0026.ps}& 
\includegraphics[bb=190 265 435 535,clip=,angle=90,width=0.145\hsize]{/home/pnoterda/Science/BIGDLAs/DR11/FITS/55987-5301-0154.ps}&
\includegraphics[bb=190 265 435 535,clip=,angle=90,width=0.145\hsize]{/home/pnoterda/Science/BIGDLAs/DR11/FITS/55947-4641-0058.ps}& 
\includegraphics[bb=190 265 435 535,clip=,angle=90,width=0.145\hsize]{/home/pnoterda/Science/BIGDLAs/DR11/FITS/55929-5309-0362.ps}\\
\includegraphics[bb=190 265 435 535,clip=,angle=90,width=0.145\hsize]{/home/pnoterda/Science/BIGDLAs/DR11/FITS/56245-5767-0950.ps}& 
\includegraphics[bb=190 265 435 535,clip=,angle=90,width=0.145\hsize]{/home/pnoterda/Science/BIGDLAs/DR11/FITS/55662-4797-0273.ps}& 
\includegraphics[bb=190 265 435 535,clip=,angle=90,width=0.145\hsize]{/home/pnoterda/Science/BIGDLAs/DR11/FITS/56298-5795-0016.ps}& 
\includegraphics[bb=190 265 435 535,clip=,angle=90,width=0.145\hsize]{/home/pnoterda/Science/BIGDLAs/DR11/FITS/56364-6662-0500.ps}&
\includegraphics[bb=190 265 435 535,clip=,angle=90,width=0.145\hsize]{/home/pnoterda/Science/BIGDLAs/DR11/FITS/55570-4564-0742.ps}&
\includegraphics[bb=190 265 435 535,clip=,angle=90,width=0.145\hsize]{/home/pnoterda/Science/BIGDLAs/DR11/FITS/56272-6424-0069.ps}\\ 
\includegraphics[bb=190 265 435 535,clip=,angle=90,width=0.145\hsize]{/home/pnoterda/Science/BIGDLAs/DR11/FITS/55651-4691-0692.ps}& 
\includegraphics[bb=190 265 435 535,clip=,angle=90,width=0.145\hsize]{/home/pnoterda/Science/BIGDLAs/DR11/FITS/56330-6457-0426.ps}& 
\includegraphics[bb=190 265 435 535,clip=,angle=90,width=0.145\hsize]{/home/pnoterda/Science/BIGDLAs/DR11/FITS/56036-5885-0254.ps}&
\includegraphics[bb=190 265 435 535,clip=,angle=90,width=0.145\hsize]{/home/pnoterda/Science/BIGDLAs/DR11/FITS/55290-3833-0832.ps}& 
\includegraphics[bb=190 265 435 535,clip=,angle=90,width=0.145\hsize]{/home/pnoterda/Science/BIGDLAs/DR11/FITS/56358-6439-0160.ps}&
\includegraphics[bb=190 265 435 535,clip=,angle=90,width=0.145\hsize]{/home/pnoterda/Science/BIGDLAs/DR11/FITS/55563-3786-0332.ps}\\ 
\includegraphics[bb=190 265 435 535,clip=,angle=90,width=0.145\hsize]{/home/pnoterda/Science/BIGDLAs/DR11/FITS/56038-5875-0326.ps}& 
\includegraphics[bb=190 265 435 535,clip=,angle=90,width=0.145\hsize]{/home/pnoterda/Science/BIGDLAs/DR11/FITS/55629-4622-0704.ps}&
\includegraphics[bb=190 265 435 535,clip=,angle=90,width=0.145\hsize]{/home/pnoterda/Science/BIGDLAs/DR11/FITS/55945-5365-0044.ps}& 
\includegraphics[bb=190 265 435 535,clip=,angle=90,width=0.145\hsize]{/home/pnoterda/Science/BIGDLAs/DR11/FITS/55976-5371-0708.ps}& 
\includegraphics[bb=190 265 435 535,clip=,angle=90,width=0.145\hsize]{/home/pnoterda/Science/BIGDLAs/DR11/FITS/55944-4768-0027.ps}&
\includegraphics[bb=190 265 435 535,clip=,angle=90,width=0.145\hsize]{/home/pnoterda/Science/BIGDLAs/DR11/FITS/55574-3840-0148.ps}\\
\includegraphics[bb=155 265 435 535,clip=,angle=90,width=0.145\hsize]{/home/pnoterda/Science/BIGDLAs/DR11/FITS/56311-6431-0978.ps}&
\includegraphics[bb=155 265 435 535,clip=,angle=90,width=0.145\hsize]{/home/pnoterda/Science/BIGDLAs/DR11/FITS/55617-4616-0046.ps}& 
\includegraphics[bb=155 265 435 535,clip=,angle=90,width=0.145\hsize]{/home/pnoterda/Science/BIGDLAs/DR11/FITS/56009-5381-0604.ps}& 
\includegraphics[bb=155 265 435 535,clip=,angle=90,width=0.145\hsize]{/home/pnoterda/Science/BIGDLAs/DR11/FITS/55982-5382-0482.ps}& 
\includegraphics[bb=155 265 435 535,clip=,angle=90,width=0.145\hsize]{/home/pnoterda/Science/BIGDLAs/DR11/FITS/55631-4748-0756.ps}&
\includegraphics[bb=155 265 435 535,clip=,angle=90,width=0.145\hsize]{/home/pnoterda/Science/BIGDLAs/DR11/FITS/56067-5973-0794.ps}\\
\end{tabular}                                                                                                 
\setlength{\tabcolsep}{6pt}
\caption{Normalised flux (black, 3 pixel boxcar smoothed) and error (orange) around the damped \lya\ line. 
The Voigt-profile fit is overplot in red, with the shaded area corresponding to an uncertainty of 0.2~dex. 
The QSO name, DLA resdshift and \HI\ column density are indicated above each panel. \label{fig:esdla}}
\end{figure*}

\newpage

\begin{figure*}
\centering
\setlength{\tabcolsep}{1pt}
\begin{tabular}{c c c c c c}
\includegraphics[bb=190 265 435 535,clip=,angle=90,width=0.145\hsize]{/home/pnoterda/Science/BIGDLAs/DR11/FITS/55302-3965-0894.ps}&
\includegraphics[bb=190 265 435 535,clip=,angle=90,width=0.145\hsize]{/home/pnoterda/Science/BIGDLAs/DR11/FITS/55927-5402-0140.ps}& 
\includegraphics[bb=190 265 435 535,clip=,angle=90,width=0.145\hsize]{/home/pnoterda/Science/BIGDLAs/DR11/FITS/56013-5404-0984.ps}& 
\includegraphics[bb=190 265 435 535,clip=,angle=90,width=0.145\hsize]{/home/pnoterda/Science/BIGDLAs/DR11/FITS/55274-3849-0942.ps}& 
\includegraphics[bb=190 265 435 535,clip=,angle=90,width=0.145\hsize]{/home/pnoterda/Science/BIGDLAs/DR11/FITS/55222-3779-0140.ps}&
\includegraphics[bb=190 265 435 535,clip=,angle=90,width=0.145\hsize]{/home/pnoterda/Science/BIGDLAs/DR11/FITS/55983-5419-0568.ps}\\ 
\includegraphics[bb=190 265 435 535,clip=,angle=90,width=0.145\hsize]{/home/pnoterda/Science/BIGDLAs/DR11/FITS/56401-6618-0152.ps}& 
\includegraphics[bb=190 265 435 535,clip=,angle=90,width=0.145\hsize]{/home/pnoterda/Science/BIGDLAs/DR11/FITS/55681-4704-0092.ps}&
\includegraphics[bb=190 265 435 535,clip=,angle=90,width=0.145\hsize]{/home/pnoterda/Science/BIGDLAs/DR11/FITS/56363-6496-0384.ps}& 
\includegraphics[bb=190 265 435 535,clip=,angle=90,width=0.145\hsize]{/home/pnoterda/Science/BIGDLAs/DR11/FITS/55243-3852-0468.ps}&
\includegraphics[bb=190 265 435 535,clip=,angle=90,width=0.145\hsize]{/home/pnoterda/Science/BIGDLAs/DR11/FITS/55659-4785-0722.ps}&
\includegraphics[bb=190 265 435 535,clip=,angle=90,width=0.145\hsize]{/home/pnoterda/Science/BIGDLAs/DR11/FITS/55987-5445-0327.ps}\\ 
\includegraphics[bb=190 265 435 535,clip=,angle=90,width=0.145\hsize]{/home/pnoterda/Science/BIGDLAs/DR11/FITS/56001-5453-0109.ps}& 
\includegraphics[bb=190 265 435 535,clip=,angle=90,width=0.145\hsize]{/home/pnoterda/Science/BIGDLAs/DR11/FITS/55691-4860-0180.ps}& 
\includegraphics[bb=190 265 435 535,clip=,angle=90,width=0.145\hsize]{/home/pnoterda/Science/BIGDLAs/DR11/FITS/55269-3860-0290.ps}&
\includegraphics[bb=190 265 435 535,clip=,angle=90,width=0.145\hsize]{/home/pnoterda/Science/BIGDLAs/DR11/FITS/56038-5171-0652.ps}& 
\includegraphics[bb=190 265 435 535,clip=,angle=90,width=0.145\hsize]{/home/pnoterda/Science/BIGDLAs/DR11/FITS/55360-3868-0656.ps}&
\includegraphics[bb=190 265 435 535,clip=,angle=90,width=0.145\hsize]{/home/pnoterda/Science/BIGDLAs/DR11/FITS/56037-5469-0644.ps}\\ 
\includegraphics[bb=190 265 435 535,clip=,angle=90,width=0.145\hsize]{/home/pnoterda/Science/BIGDLAs/DR11/FITS/56390-6725-0432.ps}& 
\includegraphics[bb=190 265 435 535,clip=,angle=90,width=0.145\hsize]{/home/pnoterda/Science/BIGDLAs/DR11/FITS/56046-5173-0914.ps}&
\includegraphics[bb=190 265 435 535,clip=,angle=90,width=0.145\hsize]{/home/pnoterda/Science/BIGDLAs/DR11/FITS/56088-6024-0358.ps}& 
\includegraphics[bb=190 265 435 535,clip=,angle=90,width=0.145\hsize]{/home/pnoterda/Science/BIGDLAs/DR11/FITS/56030-5477-0974.ps}&
\includegraphics[bb=190 265 435 535,clip=,angle=90,width=0.145\hsize]{/home/pnoterda/Science/BIGDLAs/DR11/FITS/56072-6048-0548.ps}&
\includegraphics[bb=190 265 435 535,clip=,angle=90,width=0.145\hsize]{/home/pnoterda/Science/BIGDLAs/DR11/FITS/55656-3956-0827.ps}\\ 
\includegraphics[bb=190 265 435 535,clip=,angle=90,width=0.145\hsize]{/home/pnoterda/Science/BIGDLAs/DR11/FITS/55652-4776-0060.ps}&
\includegraphics[bb=190 265 435 535,clip=,angle=90,width=0.145\hsize]{/home/pnoterda/Science/BIGDLAs/DR11/FITS/56033-4723-0950.ps}& 
\includegraphics[bb=190 265 435 535,clip=,angle=90,width=0.145\hsize]{/home/pnoterda/Science/BIGDLAs/DR11/FITS/56038-4974-0940.ps}& 
\includegraphics[bb=190 265 435 535,clip=,angle=90,width=0.145\hsize]{/home/pnoterda/Science/BIGDLAs/DR11/FITS/56003-5210-0792.ps}& 
\includegraphics[bb=190 265 435 535,clip=,angle=90,width=0.145\hsize]{/home/pnoterda/Science/BIGDLAs/DR11/FITS/56425-6730-0850.ps}&
\includegraphics[bb=190 265 435 535,clip=,angle=90,width=0.145\hsize]{/home/pnoterda/Science/BIGDLAs/DR11/FITS/55335-3929-0492.ps}\\
\includegraphics[bb=190 265 435 535,clip=,angle=90,width=0.145\hsize]{/home/pnoterda/Science/BIGDLAs/DR11/FITS/55365-4069-0944.ps}& 
\includegraphics[bb=190 265 435 535,clip=,angle=90,width=0.145\hsize]{/home/pnoterda/Science/BIGDLAs/DR11/FITS/55706-5006-0904.ps}& 
\includegraphics[bb=190 265 435 535,clip=,angle=90,width=0.145\hsize]{/home/pnoterda/Science/BIGDLAs/DR11/FITS/55450-4184-0808.ps}& 
\includegraphics[bb=190 265 435 535,clip=,angle=90,width=0.145\hsize]{/home/pnoterda/Science/BIGDLAs/DR11/FITS/55723-4992-0478.ps}& 
\includegraphics[bb=190 265 435 535,clip=,angle=90,width=0.145\hsize]{/home/pnoterda/Science/BIGDLAs/DR11/FITS/55682-4176-0188.ps}&
\includegraphics[bb=190 265 435 535,clip=,angle=90,width=0.145\hsize]{/home/pnoterda/Science/BIGDLAs/DR11/FITS/55720-4996-0732.ps}\\
\includegraphics[bb=190 265 435 535,clip=,angle=90,width=0.145\hsize]{/home/pnoterda/Science/BIGDLAs/DR11/FITS/55738-4997-0872.ps}& 
\includegraphics[bb=190 265 435 535,clip=,angle=90,width=0.145\hsize]{/home/pnoterda/Science/BIGDLAs/DR11/FITS/55717-5014-0154.ps}& 
\includegraphics[bb=190 265 435 535,clip=,angle=90,width=0.145\hsize]{/home/pnoterda/Science/BIGDLAs/DR11/FITS/55883-4374-0401.ps}& 
\includegraphics[bb=190 265 435 535,clip=,angle=90,width=0.145\hsize]{/home/pnoterda/Science/BIGDLAs/DR11/FITS/55739-5065-0812.ps}&
\includegraphics[bb=190 265 435 535,clip=,angle=90,width=0.145\hsize]{/home/pnoterda/Science/BIGDLAs/DR11/FITS/56102-5951-0750.ps}&
\includegraphics[bb=190 265 435 535,clip=,angle=90,width=0.145\hsize]{/home/pnoterda/Science/BIGDLAs/DR11/FITS/55749-5041-0374.ps}\\ 
\includegraphics[bb=190 265 435 535,clip=,angle=90,width=0.145\hsize]{/home/pnoterda/Science/BIGDLAs/DR11/FITS/56189-4428-0714.ps}&
\includegraphics[bb=190 265 435 535,clip=,angle=90,width=0.145\hsize]{/home/pnoterda/Science/BIGDLAs/DR11/FITS/56187-5043-0168.ps}& 
\includegraphics[bb=190 265 435 535,clip=,angle=90,width=0.145\hsize]{/home/pnoterda/Science/BIGDLAs/DR11/FITS/56181-6119-0945.ps}&
\includegraphics[bb=190 265 435 535,clip=,angle=90,width=0.145\hsize]{/home/pnoterda/Science/BIGDLAs/DR11/FITS/56186-5044-0020.ps}& 
\includegraphics[bb=190 265 435 535,clip=,angle=90,width=0.145\hsize]{/home/pnoterda/Science/BIGDLAs/DR11/FITS/56209-6114-0348.ps}&
\includegraphics[bb=190 265 435 535,clip=,angle=90,width=0.145\hsize]{/home/pnoterda/Science/BIGDLAs/DR11/FITS/55446-4211-0506.ps}\\ 
\includegraphics[bb=155 265 435 535,clip=,angle=90,width=0.145\hsize]{/home/pnoterda/Science/BIGDLAs/DR11/FITS/55446-4211-0946.ps}&
\includegraphics[bb=155 265 435 535,clip=,angle=90,width=0.145\hsize]{/home/pnoterda/Science/BIGDLAs/DR11/FITS/55505-4278-0010.ps}&
&
&
\\
\end{tabular}
\setlength{\tabcolsep}{6pt}
\addtocounter{figure}{-1}
\caption{Continued}
\end{figure*}

\end{document}